\documentclass[11pt]{article}
\usepackage{hyperref,a4wide}
\usepackage{amssymb}
\usepackage{graphics}
\usepackage{epsfig}
\usepackage{a4wide}
\usepackage{latexsym}
\usepackage{graphicx}
\usepackage{amsmath}
\usepackage{multirow}
\usepackage{subfigure}
\usepackage{graphicx}

\allowdisplaybreaks[2]

\begin{document}

\begin{titlepage}

\begin{center}
{\Large \textbf{Convergence properties of \boldmath{$\eta\to 3\pi$} decays in chiral perturbation theory}}\\[1.5 cm%
] {\bf Marian Koles\'ar$^{\,1}$ and Ji\v{r}\'{\i} Novotn\'{y}$^{\,1}$}
\\[1.2 cm]

$\ ^{1}${\it Institute of Particle and Nuclear Physics, Faculty of
Mathematics and Physics,\\ Charles University in Prague, V
Hole\v{s}ovi\v{c}k\'ach 2, 18000
Prague, Czech Republic.}\\[0.4cm]
\end{center}

\vspace*{1.0cm}

\begin{abstract}
Theoretical efforts to describe and explain the $\eta $$\,\rightarrow $$3\pi$ decays reach far back in time. Even today, the convergence of the decay widths and some of the Dalitz plot parameters seems problematic in low energy QCD. In the framework of resummed $\chi$PT, we explore the question of compatibility of experimental data with a reasonable convergence of a carefully defined chiral series, where NNLO remainders are assumed to be small. By treating the uncertainties in the higher orders statistically, we numerically generate a large set of theoretical predictions, which are then confronted with experimental information. In the case of the decay widths, the experimental values can be reconstructed for a reasonable range of the free parameters and thus no tension is observed, in spite of what some of the traditional calculations suggest. The Dalitz plot parameters $a$ and $d$ can be described very well too. When the parameters $b$ and $\alpha$ are concerned, we find a mild tension for the whole range of the free parameters, at less than 2$\sigma$ C.L. This can be interpreted in two ways - either some of the higher order corrections are indeed unexpectedly large or there is a specific configuration of the remainders, which is, however, not completely improbable. Also, the distribution of the theoretical uncertainties is found to be significantly non-gaussian, so the consistency cannot be simply judged by the 1$\sigma$ error bars.

\end{abstract}

\end{titlepage}

\setcounter{footnote}{0}

\newpage 
\tableofcontents
\newpage

\section{Introduction}

Theoretical efforts to describe and explain the $\eta $$\,\rightarrow $$3\pi 
$ decays reach far back in time. From the very beginning it was known that this is an isospin breaking process, as
three isovectors can constitute an isoscalar state only through the fully
antisymmetric combination $\epsilon _{abc}\pi ^{a}\pi ^{b}\pi ^{c}$, which
together with Bose symmetry and charge conjugation invariance leads to zero
contribution to the amplitude.

Initially, the process was considered to be of electromagnetic origin \cite%
{Bose:1966,Bardeen:1967}, generated by the isospin breaking virtual photon
exchange

\begin{equation}
H_{QED}(x)=-\frac{e^{2}}{2}\int dyD^{\mu \nu }(x-y)T(j_{\mu }(x)j_{\nu }(y)).
\label{H_QED}
\end{equation}

\noindent Though calculations applying current algebra and PCAC obtained
correct order of magnitude values for the decay rates \cite%
{Bose:1966,Bardeen:1967}, it was soon pointed out that the decays are almost
forbidden in the framework of QED (the Sutherland theorem \cite%
{Sutherland:1966zz,Bell:1968mi}). The early works \cite%
{Bose:1966,Bardeen:1967} related the $\eta $-$\pi $ matrix elements to the
difference of squared kaon masses or kaon and pion masses, respectively, in
fact resembling the later Dashen's theorem, which cannot be justified by
electrodynamics \cite{Bell:1968mi}. Subsequently it became clear that there
has to be a source of isospin breaking beyond the term (\ref{H_QED}) \cite%
{Osborn:1970nn}. As we  know now, strong interactions break isospin via the
difference between the masses of the $u$ and $d$ quarks

\begin{equation}
H_{QCD}^{IB}(x)=\frac{m_{d}-m_{u}}{2}(\bar{d}(x)d(x)-\bar{u}(x)u(x)).
\end{equation}

\noindent The work \cite{Osborn:1970nn} collected all the relevant current
algebra terms contributing to the decays and thus can be considered to be
the first to provide the correct leading order calculation. However, the
obtained decay rates turned out to be significantly lower than the
experimental values, which were just becoming available.

When a systematic approach to low energy hadron physics was born in the form
of chiral perturbation theory ($\chi $PT) \cite%
{Weinberg:1978kz,Gasser:1983yg,Gasser:1984gg}, it was quickly applied to the 
$\eta$$\,\rightarrow$$3\pi $ decays \cite{Gasser:1984pr}. The one loop
corrections were very sizable, the result for the decay width of the charged
channel was 160$\pm $50 eV, compared to the current algebra prediction of 66
eV. However, already at that time there were hints that the experimental
value is still much larger. The current PDG value \cite{PDG:2014kda} is 
\begin{equation}
\Gamma^+_\mathrm{exp} = 300 \pm 12 \ \mathrm{eV}.
\end{equation}
For the neutral channel, the current average is \cite{PDG:2014kda}
\begin{equation}
\Gamma^0_\mathrm{exp} = 428 \pm 17 \ \mathrm{eV}.
\end{equation}

After the effective theory was extended to include virtual photon exchange
generated by (\ref{H_QED}) \cite{Urech:1994hd}, it was shown that the
next-to-leading electromagnetic corrections to the Sutherland's theorem are
still very small \cite{Baur:1995gc, Ditsche:2008cq}. Recently it was argued
that there is an indication this need not be true for the neutral channel 
\cite{Nehme:2011jp}, but that is a partial result which has not been
finalized yet.

The theory thus seems to converge really slowly for the decays. At last, the
two loop $\chi $PT calculation \cite{Bijnens:2007pr} has succeeded to
provide a reasonable prediction for the decay widths. 

Meanwhile, experimental data are being gathered with increasing precision in order to
make more detailed analysis of the Dalitz plot distribution possible.
Comparison of the recent experimental information with the NNLO $\chi $PT
results can be found in tables \ref{tab1} and \ref{tab2}, the conventionally
defined Dalitz plot parameters will be introduced in section \ref{kinematics}.
For the sake of brevity, we added the systematic a statistical uncertainties in squares.
As can be seen, a tension between $\chi $PT and experiments appears to
be in the charged decay parameter $b$ and the neutral decay parameter $\alpha$.

\begin{table}[t]
\begin{center}
{\small 
\begin{tabular}{|c|c|c|c|c|c|}
\hline
\rule[-0.2cm]{0cm}{0.6cm} $\eta\to\pi^+\pi^-\pi^0$ & $a$ & $b$ & $d$ & $f$ & $g$ \\ 
\hline
\rule[-0.2cm]{0cm}{0.6cm} Crystal Barrel '98 \cite{CrBarrel:1998yi} & 
$-1.22\pm0.07$ & $0.22\pm0.11$ & $0.06$ (input) &&\\ 
\rule[-0.2cm]{0cm}{0.5cm} KLOE '08 \cite{KLOE:2008ht} & $-1.090\pm0.020$ & $%
0.124\pm0.012$ & $0.057\pm0.017$ & $0.14\pm0.02$ &\\
\rule[-0.2cm]{0cm}{0.5cm} KLOE '16 \cite{KLOE:2016qvh} & $-1.095\pm0.004$ & $%
0.145\pm0.006$ & $0.081\pm0.007$ & $0.141\pm0.010$ & $-0.044\pm0.016$ \\
\rule[-0.2cm]{0cm}{0.5cm} BESIII '15 \cite{BESIII:2015cmz} & $-1.128\pm0.017$ & $%
0.153\pm0.017$ & $0.085\pm0.018$ & $0.173\pm0.035$ &\\
\rule[-0.2cm]{0cm}{0.5cm} WASA at COSY '14 \cite{Wasa:2014aks} & $-1.144\pm0.018$ & $%
0.219\pm0.042$ & $0.086\pm0.025$ & $0.115\pm0.037$ &\\ \hline
\rule[-0.2cm]{0cm}{0.6cm} NREFT '11 \cite{Schneider:2010hs} & $%
-1.213\pm0.014 $ & $0.308\pm0.023$ & $0.050\pm0.003$ & $0.083\pm0.019$ & $-0.039\pm0.002$ \\ 
\rule[-0.2cm]{0cm}{0.5cm} NNLO $\chi$PT '07 \cite{Bijnens:2007pr} & $%
-1.271\pm0.075$ & $0.394\pm0.102$ & $0.055\pm0.057$ & $0.025\pm0.160$ &\\ \hline

\end{tabular}
}
\end{center}
\caption{Recent experimental and theoretical results for the charged channel.}
\label{tab1}
\end{table}

\begin{table}[tbp]
\begin{center}
{\normalsize {\small 
\begin{tabular}{|c|c|}
\hline
\rule[-0.2cm]{0cm}{0.6cm} $\eta\to\pi^0\pi^0\pi^0$ & $\alpha$ \\ \hline
\rule[-0.2cm]{0cm}{0.6cm} Crystal Barrel '98 \cite{CrBarrel:1998yj} & $%
-0.052\pm0.020$ \\ 
\rule[-0.2cm]{0cm}{0.5cm} SND '01 \cite{Achasov:2001xi} & $-0.010\pm0.023$
\\ 
\rule[-0.2cm]{0cm}{0.5cm} Crystal Ball '01 \cite{CrBall:2001fm} & $%
-0.031\pm0.004$ \\ 
\rule[-0.2cm]{0cm}{0.5cm} CELSIUS/WASA '07 \cite{Wasa:2007aa} & $%
-0.026\pm0.014$ \\ 
\rule[-0.2cm]{0cm}{0.5cm} WASA at COSY '09 \cite{Wasa:2008vn} & $%
-0.027\pm0.009$ \\ 
\rule[-0.2cm]{0cm}{0.5cm} Crystal Ball at MAMI-B '09 \cite{CrBall:2008ny} & $%
-0.032\pm0.003$ \\ 
\rule[-0.2cm]{0cm}{0.5cm} Crystal Ball at MAMI-C '09 \cite{CrBall:2008ff} & $%
-0.0322\pm0.0025$ \\ 
\rule[-0.2cm]{0cm}{0.5cm} KLOE '10 \cite{KLOE:2010mj} & $-0.0301\pm0.0050$
\\ 
\rule[-0.2cm]{0cm}{0.5cm} PDG '14 \cite{PDG:2014kda} & $-0.0315\pm0.0015$ \\ 
\hline
\rule[-0.2cm]{0cm}{0.6cm} NREFT '11 \cite{Schneider:2010hs} & $%
-0.0246\pm0.0049 $ \\ 
\rule[-0.2cm]{0cm}{0.5cm} Prague disp.fit '11 \cite{Kampf:2011wr} & $%
-0.044\pm0.004$ \\ 
\rule[-0.2cm]{0cm}{0.5cm} Bern disp.fit '11 \cite{Colangelo:2011zz} & $%
-0.045\pm0.010$ \\ 
\rule[-0.2cm]{0cm}{0.5cm} NNLO $\chi$PT '07 \cite{Bijnens:2007pr} & $%
+0.013\pm0.032$ \\ \hline
\end{tabular}
} }
\end{center}
\caption{Recent experimental and theoretical results for the neutral
channel. }
\label{tab2}
\end{table}

Alternative approaches were developed in order to model the amplitudes more
precisely, namely dispersive approaches \cite%
{Kambor:1995yc,Anisovich:1996tx,Colangelo:2011zz,Kampf:2011wr} and
non-relativistic effective field theory \cite%
{Bissegger:2007yq,Gullstrom:2008sy,Schneider:2010hs}. These more or less
abandon strict equivalence to $\chi $PT and their success in reproducing a
negative sign for $\alpha $ (see table \ref{tab2}) can serve as a motivation
to ask what is the culprit of the failure of chiral perturbation theory to
do so.

There is a long standing suspicion that chiral perturbation theory might
posses slow or irregular convergence in the case of the three light quark
flavours \cite{Fuchs:1991cq,DescotesGenon:1999uh}, the $\eta $$\,\rightarrow $$%
3\pi $ decay rates might serve as a prime example. An alternative method,
now dubbed resummed $\chi $PT \cite%
{DescotesGenon:2003cg,DescotesGenon:2007ta}, was developed in order to
express these assumptions in terms of parameters and uncertainty bands. The
starting point is the realization that the standard approach to $\chi $PT,
as a usual treatment of perturbation series, implicitly assumes good
convergence properties and hides the uncertainties associated with a
possible violation of this assumption. The resummed procedure uses the same
standard $\chi $PT Lagrangian and power counting, but only expansions
derived directly from the generating functional are trusted. All subsequent
manipulations are carried out in a non-perturbative algebraic way. The
expansion is done explicitly to next-to-leading order and higher orders are
collected in remainders. These are not neglected, but retained as sources of
error, which have to be estimated.

In this paper, we concentrate on the technical details of the resummed $\chi
PT$ approach to $\eta$$\,\rightarrow$$3\pi $ decays and provide a first look at
numerical outputs of this formalism. Our goal is to use the resummed
framework  to analyze the problem from a theoretical point of view. We do
not aim to produce an alternative set of predictions, but rather to
understand whether the theory, by which we mean $\chi $PT as a low energy
representation of QCD, really does have difficulties explaining the data.
This is the aim for which we claim the formalism of resummed $\chi PT$ is
well suited.

The results of this paper form a basis for further applications, which will
follow in separate publications \cite{Kolesar:prep}. Namely, the resummed $\chi PT$
approach can be used as a tool for testing various scenarios of the QCD
chiral symmetry breaking; preliminary results are already available in \cite{Kolesar:2014zra}.
Also, by using complementary information both from inside and outside the $\chi
PT$, we can try to address the source of the problem of irregularities of the chiral
expansion (see \cite{Kolesar:2011wn} for first results).

The paper is organized as follows. In section \ref{kinematics}, we fix our notation
and provide a brief review of the kinematics of the decay. A concise summary of
the methods of resummed $\chi PT$ is presented in section \ref{formalism_section}, while a
more detailed discussion of the choice of safe observables, their properties
and safe manipulation with them is postponed to section \ref{good_observables_section}. The relation
between the amplitude and the corresponding safe observables in the presence
of $\eta-\pi^0$ mixing is given in section \ref{mixing_subsection}. Calculation of the  mixing angles,
an example of dangerous observables, is presented in section \ref{mixangles_section}. Sections \ref{strictexp_section}, \ref{matching section} and \ref{repar_section} are devoted to successive steps of the calculations within resummed $\chi PT$, namely the strict expansion, the matching with dispersive representation and
the reparameterization in terms of the masses and decay constants, respectively.
In section \ref{Remainders}, we comment on the treatment of free parameters and the role
of the higher order remainders. Numerical results are provided in section \ref{Numerical_analysis} and
we conclude with a summary in section \ref{discussion_section}.

An explicit form of the obtained formulae, as well as some other technical
details, are postponed to appendices. In appendix \ref{strict_expansion_appendix}, we present the strict
chiral expansion of all the relevant safe observables. Appendices \ref{reconstruction_appendix} and \ref{disp_poly_appendix} are
devoted to the application of the reconstruction theorem to the  $\eta$$\,\rightarrow$$3\pi $ decays and to the matching of the strict expansion with the
dispersive approach. Bare expansion of the safe observables under
consideration and its reparameterization are summarized in appendices \ref{A_D_bare_expansion_appendix} and
\ref{reparametrization_appendix}. More detailed discussion of the mixing in resummed $\chi PT$ is presented
in appendix \ref{Op4_mixing}.

\section{Notation and kinematics \label{kinematics}}

The S-matrix element of the charged decay $\eta \rightarrow \pi ^{+}\pi
^{-}\pi ^{0}$ can be expressed in terms of the invariant amplitude $A(s,t;u)$
as 
\begin{equation}
\langle \pi ^{+}(p_{+})\pi ^{-}(p_{-})\pi ^{0}(p_{0});\mathrm{out}|\eta (p);%
\mathrm{in}\rangle =\mathrm{i}(2\pi )^{4}\delta
(p-p_{+}-p_{-}-p_{0})A(s,t;u).  \label{A_charged}
\end{equation}%
The amplitude is a function of the Mandelstam variables 
\begin{eqnarray}
s &=&(p-p_{0})^{2}=(p_{+}+p_{-})^{2}  \notag \\
t &=&(p-p_{+})^{2}=(p_{0}+p_{-})^{2}  \notag \\
u &=&(p-p_{-})^{2}=(p_{0}+p_{+})^{2},  \label{Mandelstam_c}
\end{eqnarray}%
which satisfy the constraint 
\begin{equation}
s+t+u=3s_{0}=M_{\eta }^{2}+2M_{\pi ^{+}}^{2}+M_{\pi ^{0}}^{2}.
\end{equation}%
In what follows, we will work in the first order in the isospin breaking.
We will thus not make a difference between the charge and neutral pion
masses from now on, because their difference is of the second order in the
isospin breaking. In this case, the isospin symmetry and charge
conjugation invariance imply (we use the Condon-Shortley convention here) 
\begin{eqnarray}
A(s,t;u) &=&A(s,u;t)  \label{Bose_relation} \\
\overline{A}(s,t;u) &=&-A(s,t;u)-A(t,s;u)-A(u,t;s),  \label{Isospin_relation}
\end{eqnarray}%
where $\overline{A}(s,t;u)$ is the neutral channel amplitude. We can
therefore restrict ourselves to the investigation of the charged decay mode
only\footnote{%
However, for the numerical calculation of the decay widths we will hold a
distinction in the numerical values of the pion masses for the neutral and
the charged decay, and in the position of the Dalitz plot center as well. E.g. for
the neutral decay observables we put $M_{\pi }\rightarrow M_{\pi ^{0}}$ and 
\begin{equation}
3s_{0}=M_{\eta }^{2}+3M_{\pi ^{0}}^{2}.
\end{equation}%
For more details, see \cite{Kolesar:prep}.}.

The Mandelstam variables are bounded as follows 
\begin{equation}
4M_{\pi }^{2}\leq s,t,u\leq (M_{\eta }-M_{\pi })^{2}.
\end{equation}%
For fixed $s$, the bounds for $t$, $u$ are 
\begin{eqnarray}
t_{\max ,\min }(s) &=&\frac{1}{2}\left[ 3s_{0}-s\pm \sigma \left( s\right)
\lambda ^{1/2}\left( s,M_{\eta }^{2},M_{\pi }^{2}\right) \right]
\label{t_bounds} \\
u_{\max ,\min }(s) &=&\frac{1}{2}\left[ 3s_{0}-s\mp \sigma \left( s\right)
\lambda ^{1/2}\left( s,M_{\eta }^{2},M_{\pi }^{2}\right) \right],
\label{u_bounds}
\end{eqnarray}%
where $\sigma \left( s\right) $ means the velocity of the charged pions in
the $\pi ^{+}\pi ^{-}$ rest frame,\ i.e. 
\begin{equation}
\sigma \left( s\right) =\sqrt{1-\frac{4M_{\pi }^{2}}{s}}
\end{equation}%
and $\lambda $ is the Kallen triangle function%
\begin{equation}
\lambda \left( s,M_{\eta }^{2},M_{\pi }^{2}\right) =\left( s-\left( M_{\eta
}+M_{\pi }\right) ^{2}\right) \left( s-\left( M_{\eta }-M_{\pi }\right)
^{2}\right) .
\end{equation}%
For further convenience, we also denote
\begin{equation}
\Delta _{\pi \eta }=M_{\eta }^{2}-M_{\pi }^{2}\text{. }
\end{equation}%
The differential decay rate is then 
\begin{equation}
\frac{\mathrm{d}\Gamma }{\mathrm{d}s\mathrm{d}t}=\frac{1}{(2\pi )^{3}}\frac{%
|A(s,t;u)|^{2}}{32M_{\eta }^{3}}.
\end{equation}%
The usual phenomenological parametrization of $|A(s,t;u)|^{2}$ (known as the
Dalitz plot) is given in terms of the variables 
\begin{eqnarray}
x &=&\sqrt{3}\,\frac{T_{+}-T_{-}}{Q_{\eta }}=-\frac{\sqrt{3}}{2M_{\eta
}Q_{\eta }}\left( t-u\right)  \label{Dalitz_x} \\
y &=&3\,\frac{T_{0}}{Q_{\eta }}-1=-\frac{3}{2M_{\eta }Q_{\eta }}\left(
s-s_{0}\right),  \label{Dalitz_y}
\end{eqnarray}%
where $T_{0,\pm }$ are the kinetic energies of the final states $\pi ^{0,\pm
} $ and
\begin{equation}
Q_{\eta }=T_{0}+T_{+}+T_{-}=M_{\eta }-3M_{\pi }.
\end{equation}%
The parametrization then reads 
\begin{equation}
|A(s,t;u)|^{2}=|A(s_{0},s_{0};s_{0})|^{2}\left(
1+ay+by^{2}+dx^{2}+fy^{3}+gx^{2}y+\ldots \right) \label{Dalitz_charged}
\end{equation}%
and corresponds to the Taylor expansion at the center of the Dalitz plot%
\footnote{%
Let us note that beyond the first order in the isospin breaking, which
requires to take the neutral and charged pion masses as different, the point 
$x=y=0$ does not coincide with the point $s=t=u$ and we have the following
formula for $y$%
\begin{equation}
y=\frac{3}{2M_{\eta }Q_{\eta }}\left( (M_{\eta }-M_{\pi })^{2}-s\right) -1%
\text{.}
\end{equation}%
} $s=t=u=s_{0}$. Note that the charge conjugation invariance excludes terms
which are of odd powers in $x$.

In the case of the neutral decay, the amplitude is symmetric with respect to an
exchange of $s$, $t~\ $and $u$ and it is therefore more convenient to
introduce the variable 
\begin{equation}
z=x^{2}+y^{2}=\frac{3}{2M_{\eta }^{2}(M_{\eta }-M_{\pi })^{2}}\left(
s^{2}+t^{2}+u^{2}-3s_{0}^{2}\right)  \label{Dalitz_z}
\end{equation}%
and write the Dalitz plot parametrization in the form 
\begin{equation}
|\overline{A}(s,t;u)|^{2}=|A(s_{0},s_{0};s_{0})|^{2}\left( 1+2\alpha
z+2\beta y(3z-4y^{2})+\gamma z^{2}+\ldots \right) .  \label{Dalitz_neutral}
\end{equation}%

For reasons described bellow, the basic object of our investigation will be the quantity $G(s,t;u)$
\begin{equation}
G(s,t;u)\equiv F_{\pi }^{3}F_{\eta }A(s,t;u),  \label{G_definition}
\end{equation}%
where $F_{\pi }$, $F_{\eta }$ are the pion and eta decay constants. The
coefficients $A$, $B$, $C$, $D$ are defined by its expansion at the center of the
Dalitz plot  
\begin{equation}
G(s,t;u) = A+B(s-s_{0})+C(s-s_{0})^{2}+D\left[ (t-s_{0})^{2}+(u-s_{0})^{2}%
\right] + O((s-s_{0})^{3}).  \label{A_F_coefficients}
\end{equation}%
These coefficients are related to the Dalitz plot parameters $a$, $b$, $d$, $\alpha$ by means of nonlinear relations 
\begin{eqnarray}
a &=&-2R_{\eta }\mathrm{Re}\left( \frac{B}{A}\right)  \notag \\
b &=&R_{\eta }^{2}\left( \left\vert \frac{B}{A}\right\vert ^{2}+\mathrm{Re}%
\left( \frac{2C+D}{A}\right) \right)  \notag \\
d &=&3R_{\eta }^{2}\mathrm{Re}\left( \frac{D}{A}\right)  \notag \\
\alpha &=&\frac{1}{4}\left( d+b-R_{\eta }^{2}\left\vert \frac{B}{A}%
\right\vert ^{2}\right) ,  \label{ABCD_Dalitz}
\end{eqnarray}%
where 
\begin{equation}
R_{\eta }=\frac{2}{3}M_{\eta }Q_{\eta }.
\end{equation}
Note that the last relation for $\alpha $ holds only in the lowest order in
the isospin breaking.

\section{Resummed chiral perturbation theory - the formalism \label{formalism_section}}

In this section, we briefly review the formalism of resummed chiral
perturbation theory \cite{DescotesGenon:2003cg,DescotesGenon:2007ta}. The
general prescription can be summarized in the following points:

\begin{itemize}
\item The calculations are based on the standard $\chi PT$ Lagrangian and
standard chiral power counting given by the Weinberg formula \cite{Weinberg:1978kz}. In particular, the
quark masses $m_{q}$ are counted as $m_{q}=O(p^{2})$.

\item The crucial point is an identification of globally convergent
observables (named \emph{safe observables}, i.e. the chiral expansion of which can be trusted) related to
the amplitude and other physically relevant observables for the process under
consideration. As will be explained in more detail in the next section,
these safe observables are related to the Green functions of the quark bilinears by linear operations.

\item The next step consists of performing the \emph{strict chiral expansion} of
the safe observables, i.e. an expansion constructed in terms of the
parameters of the chiral Lagrangian and strictly respecting the chiral
orders. That means, e.g., that the propagators inside the loops carry the $%
O(p^{2})$ masses. The expansion is done up to the $O(p^{4})$ order
explicitly; the higher orders are collected implicitly in \emph{remainders},
which arise as additional parameters.

\item Then we construct a modified expansion (dubbed \emph{bare expansion}%
), which differs from the strict expansion by the location of the branching
points of the non-analytical unitarity part of the amplitudes - within the
bare expansion they are placed in their physical positions. This can be done either by means
of a matching with a dispersive representation or by hand.

\item After that we perform an algebraically exact nonperturbative
reparametrization of the bare expansion by expressing the $O(p^{4})$ LECs $%
L_{4},\ldots ,L_{8}$ in terms of physical values of experimentally well established safe observables
- the pseudoscalar decay constants and masses. The procedure generates additional higher order remainders. In what follows, we refer to these as \emph{indirect} remainders.

\item The physical amplitude and other relevant observables are then
obtained as algebraically exact nonperturbative expressions in terms of the
related safe observables and higher order remainders.

\item The higher order remainders are explicitly kept and carefully treated
by using various information stemming from both inside and outside $\chi PT$
(order of magnitude estimates, explicit higher order calculations, resonance saturation, etc.).
\end{itemize}

In the presence of particle mixing, which is the case of the $\pi ^{0}-\eta $
sector treated at the first order in the isospin breaking, the
implementation of the procedure is a little bit more complicated. We will therefore give
a more detailed explanation of the above points in the following
sections.

\section{Safe observables\label{good_observables_section}}

The starting point of the formalism of resummed $\chi PT$ is the
generating functional $Z[v,a,p,s]$ of the correlators of the quark bilinears
defined as 
\begin{equation}
Z[v,a,p,s]=\langle 0|T\exp \left( \frac{\mathrm{i}}{\hbar }\int d^{4}x\left( 
\overline{q}\gamma \cdot (v+a\gamma ^{5})q+\overline{q}(s+ip\gamma
^{5})q\right) \right) |0\rangle ,
\end{equation}%
where $v$, $a\,$, $p$, $s$ are the external classical sources and $q$ stands
for the $SU(N_{f})$ multiplet of the quark fields. Pseudo-Goldstone boson (PGB) fields are the only
relevant degrees of freedom at energies up to the
hadronic scale $E\ll \Lambda _{H}\sim 1GeV$. The low energy
representation of $Z[v,a,p,s]$ can thus be expressed in terms of the functional
integral over the PGB fields 
\begin{equation}
Z[v,a,p,s]=\int \mathcal{D}U\exp \left( \frac{i}{\hbar }\left(
S^{(2)}[v,a,p,s;U]+\hbar S^{(4)}[v,a,p,s;U]+\ldots \right) \right) .
\label{Z_int}
\end{equation}%
In this expression, the field $U$ corresponds to the $N_{f}\times N_{f}$
unitary matrix, which can be written, for $N_{f}=3$, in terms of the
pseudoscalar octet fields $\phi ^{a}\equiv \pi ,K,\eta $ as 
\begin{equation}
U=\exp \frac{\mathrm{i}}{F_{0}}\phi ^{a}\lambda ^{a},
\end{equation}%
with $\lambda ^{a}$ being the Gell-Mann matrices and 
\begin{equation}
\phi ^{a}\lambda ^{a}=\left( 
\begin{array}{lll}
\pi ^{3}+\frac{1}{\sqrt{3\,}}\eta ^{8} & \sqrt{2}\pi ^{+} & \sqrt{2}K^{+} \\ 
-\sqrt{2}\pi ^{-} & -\pi ^{3}+\frac{1}{\sqrt{3\,}}\eta ^{8} & \sqrt{2}K^{0}
\\ 
-\sqrt{2}K^{-} & \sqrt{2}\overline{K^{0}} & -\frac{2}{\sqrt{3\,}}\eta ^{8}%
\end{array}%
\right) .
\end{equation}%
$S^{(n)}[v,a,p,s;U]$ is the action functional of the chiral order $%
O(p^{n})$. The systematic chiral expansion of $Z[v,a,p,s]$ is then obtained
by means of a loop expansion of (\ref{Z_int}), which is correlated with the
chiral expansion by means of the Weinberg formula \cite{Weinberg:1978kz}. In practice, this
means integrating out the quantum fluctuations around the classical solution 
$\phi ^{i}[v,a,p,s]$ of the lowest order equation of motion (i.e. those
derived from the lowest order action $S^{(2)}[v,a,p,s;U]$) order by order in 
$\hbar $. The result can be then written as (we put $%
\hbar =1$ in what follows) 
\begin{eqnarray}
Z[v,a,p,s] &=&\sum_{n}\frac{F_{0}^{-n}}{n!}\int \mathrm{d}^{4}x_{1}\ldots 
\mathrm{d}^{4}x_{n}Z_{i_{1}i_{2}\ldots i_{n}}[v,a,p,s](x_{1},\ldots ,x_{n}) 
\notag \\
&&\times \phi ^{i_{1}}[v,a,p,s](x_{1})\ldots \phi ^{i_{n}}[v,a,p,s](x_{n}),
\label{Z_i_definition}
\end{eqnarray}%
where the chiral expansion of the coefficient functionals $%
Z_{i_{1}i_{2}\ldots i_{n}}[v,a,p,s](x_{1},\ldots ,x_{n})$ symbolically reads
\begin{equation}
Z_{i_{1}i_{2}\ldots i_{n}}=Z_{i_{1}i_{2}\ldots
i_{n}}^{(2)}+Z_{i_{1}i_{2}\ldots i_{n}}^{(4)}+Z_{i_{1}i_{2}\ldots
i_{n}}^{(6)}+\ldots  \label{Z_expansion}
\end{equation}%
(where $Z_{i_{1}i_{2}\ldots i_{n}}^{(k)}=O\left( p^{k}\right) $) and
similarly for the classical solution $\phi ^{i}$.

The key assumption\footnote{%
Let us stress that this assumption is a hypothesis which
should be questioned and tested.} behind the resummed approach to $%
\chi PT$ is that the functional $Z[v,a,p,s]$ and the safe observables obtained from it by linear operations are the only basic objects for which the chiral expansion can be, in a restricted sense, trusted
(by linear operations we mean performing functional differentiation with respect to the sources
with subsequent Fourier transform, taking the residue at the poles and
the expansion coefficients at points of analyticity, far away
from the thresholds). We do not assume a strict hierarchy of orders,  but require a global convergence only. This  notion can be quantified by assuming that for such a safe observable, denoted
generically as $G$ in what follows, the $O(p^{6})$ remainder $\Delta_{G}$, defined by
\begin{equation}
G=G^{(2)}+G^{(4)}+\Delta _{G},  \label{globally_convergent_expansion}
\end{equation}%
is reasonably small, typically 
\begin{equation}
\delta _{G}\equiv \frac{\Delta _{G}}{G}\sim 0.1.
\end{equation}%

As can be seen, nothing is assumed about the relative value of $G$ and
the leading order (LO) and next-to-leading order (NLO) terms of the chiral expansion, $%
G^{(2)}$ and $G^{(4)}$, respectively. In particular, the cases when $G^{(2)}\ll G$
or $G^{(2)}\sim G^{(4)}$ are not a priori excluded. This is in contrast with the
standard $\chi PT$ assumption 
\begin{eqnarray}
X_{G} &\equiv &\frac{G^{(2)}}{G}\sim 1,  \label{X_G_definition} \\
\overline{X}_{G} &\equiv &\frac{G^{(4)}}{G^{(2)}}\sim 0.3.
\label{barX_G_definition}
\end{eqnarray}%

Accepting the possibility of such irregularities of the chiral expansion and
combining it with the key assumption discussed above, we are forced to put
some constraints on the manipulations with the chiral series. Namely, the
physical observables which are related to any safe observable $G$
non-linearly are \emph{dangerous} in the sense
that their global convergence is not granted anymore - the formal expansion
of such a dangerous observable, respecting the chiral orders strictly, might
generate unusually large $O(p^{6})$ remainders. An example of such a dangerous quantity is the inverse of
a safe observable $1/G$, the formal expansion of which can be written as
\begin{equation}
\frac{1}{G}=\frac{1}{G^{(2)}}-\frac{G^{(4)}}{\left( G^{(2)}\right) ^{2}}%
+\Delta _{1/G}.
\end{equation}%
The remainder 
\begin{equation}
\Delta _{1/G}=\frac{1}{G}\left( -\frac{\delta _{G}}{X_{G}^{2}}+\left( \frac{%
1-X_{G}}{X_{G}}\right) ^{2}\right)
\end{equation}%
might be large when $X_G\ll 1$, even for
a small original remainder $\Delta _{G}=G\delta _{G}$. Therefore, in order to get
numerically reliable values for such dangerous observables, the formal
chiral expansion can not be performed and the original algebraic form has to
be preserved by holding the remainders explicit. These parameters then
estimate the theoretical uncertainty of the result. For our toy example this
means to use, instead of the chiral expansion, an algebraic identity 
\begin{equation}
\frac{1}{G}=\frac{1}{G^{(2)}+G^{(4)}+\Delta _{G}}.
\end{equation}%
Note that if we dropped the remainder $\Delta _{G}$, we could interpret the result
of such an approach as a partial resummation of (some of the) higher orders
of the chiral expansion%
\begin{equation}
\frac{1}{G}\rightarrow \frac{1}{G^{(2)}}\sum_{n=0}^{\infty }\left( -1\right)
^{n}\left( \frac{G^{(4)}}{G^{(2)}}\right) ^{n}.
\end{equation}%
On the other hand, by not neglecting the remainder, we get an exact algebraic identity
valid to all orders\footnote{%
Let us note that this approach can be straightforwardly extended to the
next-to-next-to-leading order (NNLO). We can write 
\begin{equation}
G=G^{(2)}+G^{(4)}+G^{(6)}+\overline{\Delta }_{G},
\end{equation}%
where $\overline{\Delta }_{G}$ is now assumed to be a reasonably small $%
O(p^{8})$ remainder. For the toy example of the chiral expansion
of the dangerous observable $1/G$, we get 
\begin{equation}
\frac{1}{G}=\frac{1}{G^{(2)}}-\frac{G^{(4)}}{\left( G^{(2)}\right) ^{2}}-%
\frac{1}{G^{(2)}}\left[ \left( \frac{G^{(4)}}{G^{(2)}}\right) ^{2}-\frac{%
G^{(6)}}{G^{(2)}}\right] +\overline{\Delta }_{1/G}.
\end{equation}%
The $O(p^{8})$ remainder is 
\begin{equation}
\overline{\Delta }_{1/G}=\frac{1}{G}\left( -\frac{\delta _{G}}{X_{G}^{2}}%
+\left( \frac{1-X_{G}}{X_{G}}\right) ^{2}-\frac{\overline{X}_{G}^{2}}{X_{G}}%
\right),
\end{equation}%
which can be large when $X_{G}\ll 1$ and/or $\overline{X}_{G}\sim 1$.
Therefore, the irregularities in the first two terms of the chiral expansion
can produce large higher order $O(p^{8})$ remainders when
dangerous operations are carried out, even when performed on globally convergent safe observables. Throwing away the remainders, as usually done within the standard NNLO, is therefore not safe in such a case.} .

\bigskip

\section{Amplitudes in terms of safe observables in the presence of mixing 
\label{mixing_subsection}}

As we have mentioned above, a necessary ingredient of the resummed
approach is the identification of the globally convergent safe observables.
In this section, we will briefly recapitulate the connection between the
safe observables and the amplitudes, and also introduce a
generalization for the case of the $\pi ^{0}$-$\eta$ mixing.

The key assumption of resummed $\chi PT$ is that the safe observables
are derived by means of linear operations from the generating functional $%
Z[v,a,p,s]$. Therefore, the functional derivatives of $Z[v,a,p,s]$ with
respect to the axial vector sources, their Fourier transforms and residues
at the one particle poles belong to the set of safe observables. Such safe
observables are directly connected with the physical amplitudes of the
processes with PGB. Indeed, for $p^{2}\rightarrow
M_{P}^{2}$, we symbolically have 
\begin{eqnarray}
\int \mathrm{d}^{4}xe^{ip\cdot x}\frac{\delta Z[v,a,p,s]}{\mathrm{i}\delta
a^{i}(x)} &\equiv&\langle 0|T\widetilde{j_{\mu 5}^{i}}(p)\ldots |0\rangle  \notag
\\
&=&\frac{\mathrm{i}}{p^{2}-M_{P}^{2}}\langle 0|j_{\mu 5}^{i}(0)|p,P\rangle
\langle p,P|\cdots |0\rangle +reg,
\end{eqnarray}%
where $j_{\mu 5}^{i}$ is the axial vector current, $|p,P\rangle $ is the
one-particle PGB state with mass $M_{P}$ and $reg=O\left( 1\right) $ denotes a
regular contribution. The PGB states couple to the operators $j_{\mu 5}^{i}$
and we have the following general relation 
\begin{equation}
\langle 0|j_{\mu 5}^{i}(0)|p,P\rangle =\mathrm{i}p_{\mu }F^{Pi}.
\label{F_nondiag}
\end{equation}%
Therefore, the residue of $\langle 0|\widetilde{j_{\mu _{1}5}^{i_{1}}}%
(p_{1})\ldots \widetilde{j_{\mu _{n}5}^{i_{n}}}(p_{n})|0\rangle $ at the
simultaneous poles at $p_{i}^{2}\rightarrow M_{P_{i}}^{2}$ correspond to a
safe observable $G_{i_{1}i_{2}\ldots i_{n}}^{P_{1}P_{2}\ldots
P_{n}}(p_{1},\ldots p_{n})$, for which we can write (no summation over $P_{k}$)
\begin{equation}
G_{i_{1}i_{2}\ldots i_{n}}^{P_{1}P_{2}\ldots P_{n}}(p_{1},\ldots
p_{n})=F^{P_{1}i_{1}}\ldots F^{P_{n}i_{n}}A_{P_{1}P_{2}\ldots
P_{n}}(p_{1},\ldots p_{n}).
\end{equation}%
$A_{P_{1}P_{2}\ldots P_{n}}(p_{1},\ldots
p_{n})$ are the $S-$matrix elements with PGB $P_{1}P_{2}\ldots P_{n}$ in the
in and out states.

In the absence of mixing, when the mass states have definite isospin, we have 
\begin{equation}
\langle 0|j_{\mu 5}^{i}(0)|p,P\rangle =\mathrm{i}p_{\mu }F_{i}\delta ^{iP},
\label{F_diag}
\end{equation}%
where $F_{i}$ are the corresponding pseudoscalar decay constants. We obtain $F_{P}^{2}$ and $F_{P}^{2}M_{P}^{2}$ as the simplest examples of safe observables, related to the residue of $\langle 0|T\widetilde{%
j_{\mu 5}^{i}}(p)j_{\nu 5}^{i}(0)|0\rangle $ at $p^2\rightarrow M_P^2$%
\begin{eqnarray}
\langle 0|T\widetilde{j_{\mu 5}^{P}}(p)j_{\nu 5}^{P}(0)|0\rangle &=&\frac{%
p_{\mu }p_{\nu }F_{P}^{2}}{p^{2}-M_{P}^{2}}+reg \\
\langle 0|T\widetilde{j_{\mu 5}^{P}}(p)j_{5}^{\mu P}(0)|0\rangle &=&\frac{%
M_{P}^{2}F_{P}^{2}}{p^{2}-M_{P}^{2}}+reg.
\end{eqnarray}%
On the other hand, the first powers of the decay constants $F_{P}$ and of the
masses $M_{P}$ cannot be considered as safe observables, as they are
linked to $F_{P}^{2}$ and $F_{P}^{2}M_{P}^{2}$ by nonlinear relations. In
the same spirit, the amplitudes $A_{i_{1}i_{2}\ldots i_{n}}(p_{1},\ldots
p_{n})$ do not represent safe observables either, being non-linearly related to $%
G_{i_{1}i_{2}\ldots i_{n}}(p_{1},\ldots p_{n})$ and $F_{i}^{2}$%
\begin{equation}
A_{P_{1}P_{2}\ldots P_{n}}(p_{1},\ldots p_{n})=F_{i_{1}}^{-1}\ldots
F_{i_{n}}^{-1}G_{i_{1}i_{2}\ldots i_{n}}^{P_{1}P_{2}\ldots
P_{n}}(p_{1},\ldots p_{n}).  \label{A_G_relation_diag}
\end{equation}

In the case of the $\pi ^{0}$-$\eta$ mixing, when the isospin symmetry is explicitly broken, the
matrix $F^{Pi}$ in the relations (\ref{F_nondiag}) is not diagonal. In the
first order of the isospin breaking, the non-diagonal terms of the
matrix $F^{Pi}$ directly correspond to the the $\pi ^{0}$-$\eta$ mixing sector. Hence we can
define 
\begin{equation}
{\bold{F}}=\left( 
\begin{array}{cc}
F^{\pi ^{0}3} & F^{\pi ^{0}8} \\ 
F^{\eta 3} & F^{\eta 8}%
\end{array}%
\right) =\left( 
\begin{array}{cc}
F_{\pi } & \varepsilon _{\pi }F_{\pi } \\ 
-\varepsilon _{\eta }F_{\eta } & F_{\eta }%
\end{array}%
\right).  \label{F_matrix}
\end{equation}%
$F_{\pi ,\eta }$ are the pion and eta decay constants and $\varepsilon
_{\pi ,\eta }$ are the mixing angles at the leading order in the isospin
breaking 
\begin{equation}
\varepsilon _{\pi ,\eta }=O\left( \frac{1}{R}\right),
\end{equation}%
where 
\begin{equation}
R=\frac{m_{s}-\widehat{m}}{m_{d}-m_{u}}.
\end{equation}

As was shown in \cite{Gasser:1984gg} (see also appendix \ref{Op4_mixing} for
details), the chiral expansion (up to and including the order $O(p^{4})$) of
the safe observables $G_{i_{1}i_{2}\ldots i_{n}}^{P_{1}P_{2}\ldots P_{n}}$
is related to the chiral expansion of the Fourier transforms of the
coefficient functionals $Z_{i_{1}i_{2}\ldots i_{n}}[0,0,0,0]$, introduced in (%
\ref{Z_i_definition}). In the general case, we have the following relation 
\begin{equation}
Z_{i_{1}i_{2}\ldots i_{n}}(p_{1},\ldots p_{n})=\sum_{P_{1}\ldots
P_{n}}G_{i_{1}i_{2}\ldots i_{n}}^{P_{1}P_{2}\ldots P_{n}}(p_{1},\ldots p_{n}),
\end{equation}%
which can also be understood as a definition of an extension of the right hand
side off the mass shell.

The $Z_{i_{1}i_{2}\ldots i_{n}}$'s, being linear combinations of the safe
observables $G_{i_{1}i_{2}\ldots i_{n}}^{P_{1}P_{2}\ldots P_{n}}$, are
therefore safe observables too. In the absence of the mixing, we simply have $%
G_{i_{1}i_{2}\ldots i_{n}}=Z_{i_{1}i_{2}\ldots i_{n}}$ and the amplitude $%
A_{P_{1}P_{2}\ldots P_{n}}$ is given by (\ref{A_G_relation_diag}), where the
inverse powers of decay constants $F_{i}$ are assumed not to be expanded but
substituted by their physical values. However, for the non-diagonal matrix ${\bold{F}}$,
we need a nonperturbative inverse of ${\bold{F}}$ in order to obtain the amplitude $A_{P_{1}P_{2}\ldots
P_{n}}$ from the safe observables $Z_{i_{1}i_{2}\ldots i_{n}}$
\begin{equation}
A_{P_{1}P_{2}\ldots P_{n}}(p_{1},\ldots p_{n})=\sum_{j_{1}\ldots
j_{n}}({\bold{F}}^{-1})^{P_{1}j_{1}}\ldots ({\bold{F}}^{-1})^{P_{n}j_{n}}Z_{j_{1}j_{2}\ldots
j_{n}}(p_{1},\ldots p_{n}).  \label{A-Z_general}
\end{equation}%
In the $\pi ^{0}$-$\eta$ sector, as we work in the first order of the isospin breaking, we get  
\begin{equation}
{\bold{F}}^{-1}=\left( 
\begin{array}{ll}
F_{\pi }^{-1} & -\varepsilon _{\pi }F_{\eta }^{-1} \\ 
\varepsilon _{\eta }F_{\pi }^{-1} & F_{\eta }^{-1}%
\end{array}%
\right) +O\left( \frac{1}{R^{2}}\right).  \label{F_inversion}
\end{equation}

Now, let us apply the above general recipe to the case of the amplitude $%
A(s,t;u)$ of the decay $\eta (p)\rightarrow \pi ^{+}(p_{+})\pi
^{-}(p_{-})\pi ^{0}(p_{0})$, defined by (\ref{A_charged}). According to (\ref%
{A-Z_general}) and (\ref{F_inversion}), we then have the following resummed
representation of the amplitude $A(s,t;u)$ in terms of safe observables $%
Z_{83+-}(s,t;u)$, $Z_{33+-}(s,t;u)$, $Z_{88+-}(s,t;u)$ and the matrix $F$ of
physical decay constants\footnote{%
Since we calculate the amplitude at the first order in the isospin breaking, we have
put $F_{\pi ^{\pm }}=F_{\pi ^{0}}=F_{\pi }$ on the left hand side of the
formula (\ref{amplitude_Z}).} 
\begin{equation}
F_{\pi }^{3}F_{\eta }A(s,t;u)\equiv G(s,t;u)=Z_{83+-}(s,t;u)-\varepsilon
_{\pi }Z_{33+-}(s,t;u)+\varepsilon _{\eta }Z_{88+-}(s,t;u),
\label{amplitude_Z}
\end{equation}%
 Note that $F_{\pi }^{-4}Z_{33+-}(s,t;u)$
and $F_{\pi }^{-2}F_{\eta }^{-2}$ $Z_{88+-}(s,t;u)$ can be identified as the
off-shell extensions of the $\pi \pi $ and $\pi \eta $ scattering amplitudes,
respectively, calculated in the limit of conserved isospin.

\section{Mixing angles \label{mixangles_section}}

Provided we knew all the entries of the matrix $\bold{F}$ (\ref{F_matrix})
from experimental measurements with good enough precision, we could proceed further. 
However, with the exception of $F_{\pi }$, this is not the case.
Let us therefore calculate the remaining matrix elements $F_{ij}$, which can also be viewed as an
illustration of the machinery of resummed $\chi PT$ applied to dangerous observables.

As we have discussed above, the matrix $F$ is directly related to the part
of the generating functional $Z[a]$ which is at most quadratic in the fields 
$\phi ^{i}[a]$. Let
us write the generating functional in the form $\ $($Z[a]\equiv
Z[v,a,p,s]|_{v=p=s=0}$) 
\begin{eqnarray}
Z[a] &=&\frac{1}{2\,}\int \mathrm{d}^{4}x[Z_{33}(\partial \pi
_{3}[a]-a_{3}F_{0})^{2}+Z_{88}(\partial \eta _{8}[a]-a_{8}F_{0})^{2}  \notag
\\
&&+2Z_{38}(\partial \pi _{3}[a]-a_{3}F_{0})\cdot (\partial \eta
_{8}[a]-a_{8}F_{0})  \notag \\
&&-\mathcal{M}_{33}\pi _{3}[a]^{2}-\mathcal{M}_{88}\eta _{8}[a]^{2}-2%
\mathcal{M}_{38}\pi _{3}[a]\eta _{8}[a]]+\ldots +O(p^{6}),
\end{eqnarray}%
where $F_{0}$ is the PGB decay constant in the chiral limit and where $Z_{ij}$
and $M_{ij}$ accumulate the $O(p^{2})$ and $O(p^{4})$ contributions
according to (\ref{Z_expansion}) 
\begin{eqnarray}
Z_{ij} &=&\delta _{ij}+Z_{ij}^{(4)} \\
\mathcal{M}_{ij} &=&\delta _{ij}\overset{o}{m}_{j}^{2}+\mathcal{M}%
_{ij}^{(4)}.
\end{eqnarray}%
Here $\overset{o}{m}_{j}$ denotes the $O(p^{2})$ masses. The off-diagonal
terms 
\begin{eqnarray}
Z_{38} &=&Z_{38}^{(4)} \\
\mathcal{M}_{38} &=&\mathcal{M}_{38}^{(2)}+\mathcal{M}_{38}^{(4)}
\end{eqnarray}%
are taken at the first order in the isospin breaking, 
\begin{equation}
Z_{38},\mathcal{M}_{38}=O\left( \frac{1}{R}\right) .
\end{equation}%
According to our discussion in the previous section, $Z_{ij}$ and $%
\mathcal{M}_{ij}$ represent safe observables.

It can be shown (see appendix \ref{Op4_mixing}) that up to higher
order corrections 
\begin{equation}
{\bold{F}}^{T}\cdot{\bold{F}}=F_{0}^{2}{\bold{Z}}+O(p^{6}).
\end{equation}%
Here 
\begin{equation}
Z=\left( 
\begin{array}{cc}
Z_{33}, & Z_{38} \\ 
Z_{38}, & Z_{88}%
\end{array}%
\right)
\end{equation}%
and 
\begin{equation}
{\bold{F}}^{T}\cdot\mathrm{diag}(M_{\pi }^{2},M_{\eta }^{2})\cdot{\bold{F}}=F_{0}^{2}\mathcal{M}+O(p^{6}),
\end{equation}%
where 
\begin{equation}
{\mathcal{M}}=\left( 
\begin{array}{cc}
{\mathcal{M}}_{33}, & {\mathcal{M}}_{38} \\ 
{\mathcal{M}}_{38}, & {\mathcal{M}}_{88}%
\end{array}%
\right) .
\end{equation}%
The above matrix relations can be written in components 
\begin{eqnarray}
Z_{33}F_{0}^{2} &=&F_{\pi }^{2}(1-\delta _{F_{\pi }^{2}}) \\
Z_{88}F_{0}^{2} &=&F_{\eta }^{2}(1-\delta _{F_{\eta }^{2}}) \\
Z_{38}F_{0}^{2} &=&(\varepsilon _{\pi }F_{\pi }^{2}-\varepsilon _{\eta
}F_{\eta }^{2})(1-\delta _{\varepsilon F})
\end{eqnarray}%
and 
\begin{eqnarray}
F_{0}^{2}\mathcal{M}_{33} &=&F_{\pi }^{2}M_{\pi }^{2}(1-\delta _{F_{\pi
}^{2}M_{\pi }^{2}}) \\
F_{0}^{2}\mathcal{M}_{88} &=&F_{\eta }^{2}M_{\eta }^{2}(1-\delta _{F_{\eta
}^{2}M_{\eta }^{2}}) \\
F_{0}^{2}\mathcal{M}_{38} &=&(\varepsilon _{\pi }F_{\pi }^{2}M_{\pi
}^{2}-\varepsilon _{\eta }F_{\eta }^{2}M_{\eta }^{2})(1-\delta _{\varepsilon
FM}),
\end{eqnarray}%
where we have defined a unique remainders for each observable, e.g. 
\begin{equation}
F_{\pi }^{2}=Z_{33}F_{0}^{2}+F_{\pi }^{2}\delta _{F_{\pi }^{2}}
\end{equation}%
and similarly in the rest of the cases. As a last step, we algebraically invert these
relations in order to find the resummed expressions for $F_{\eta }^{2}$
and the mixing angles $\varepsilon _{\pi ,\eta }$ 
\begin{eqnarray}
F_{\eta }^{2} &=&Z_{88}F_{0}^{2}+F_{\eta }^{2}\delta _{F_{\eta }^{2}}
\label{F_eta_expansion} \\
\varepsilon _{\pi ,\eta } &=&-\frac{F_{0}^{2}}{F_{\pi ,\eta }^{2}}\frac{%
\mathcal{M}_{38}(1-\delta _{\varepsilon FM})^{-1}-Z_{38}M_{\eta ,\pi
}^{2}(1-\delta _{\varepsilon F})^{-1}}{M_{\eta }^{2}-M_{\pi }^{2}}.
\label{mixing_angles_expansion}
\end{eqnarray}%
The explicit form of the strict chiral expansion of the mixing parameters $M_{38}$ and $Z_{38}$
can be found in appendix \ref{strict_expansion_appendix}, while a detailed
analysis of $F_{\eta }$ within resummed $\chi PT$ has been done in \cite%
{Kolesar:2008fu}.

Because the remainders are not neglected, the formulae (\ref{F_eta_expansion}) and (\ref%
{mixing_angles_expansion}) are exact algebraic identities valid to all
orders in the chiral expansion. Let us note that the standard chiral
expansion of the denominators in (\ref{mixing_angles_expansion}) should not be
performed because of the possible generation of large $O(p^{6})$
remainders. In this sense, the mixing angles $\varepsilon _{\pi ,\eta }$ are
typical examples of dangerous observables.

\section{Strict chiral expansion for $G(s,t;u)$ \label{strictexp_section}}

In the context of resummed $\chi PT$, we understand \emph{strict expansion}
as the chiral expansion of a safe observable expressed in
terms of original parameters of the chiral Lagrangian without any
reparametrization in terms of physical observables and without any
potentially dangerous operation, like the expansion of the denominators. 
Also, loop graphs are constructed strictly from
propagators and vertices derived from the $O(p^{2})$ part of the
Lagrangian. In particular, the propagator masses are held at their LO values, which
we denote as $\overset{o}{m}_{P}$ ($P=\pi ,K,\eta $).

According to (\ref{amplitude_Z}), the result for the re-scaled
amplitude $G(s,t;u)$ can be expressed in terms of the safe observables $%
Z_{ab+-}$ (where $a,b=3,8$) and physical mixing angles $\varepsilon _{\pi ,\eta
}$ 
\begin{equation}
G(s,t;u)=Z_{83+-}(s,t;u)-\varepsilon _{\pi }Z_{33+-}(s,t;u)+\varepsilon
_{\eta }Z_{88+-}(s,t;u).
\end{equation}%
The expansion up to $O(p^{4})$ can be written in the form 
\begin{equation}
Z_{ab+-}=Z_{ab+-}^{(2)}+Z_{ab+-,\,\mathrm{ct}}^{(4)}+Z_{ab+-,\,\mathrm{tad}%
}^{(4)}+Z_{ab+-,\,\mathrm{unit}}^{(4)}+\Delta_{Z_{ab+-}},  \label{Z_abpm_strict_expansion}
\end{equation}%
where the individual terms on the right hand side represent the $O(p^{2})$
contribution, the $O(p^{4})$ counterterm, tadpole and unitarity corrections
and the $O(p^{6})$ remainder, respectively. Let us note that the splitting
of the loop correction into the tadpole and the unitarity part is not unique. Here
we follow the splitting of the generating functional introduced in \cite%
{Gasser:1984gg}.

The explicit form of the strict expansion (\ref{Z_abpm_strict_expansion}) is
rather long and is therefore postponed to appendix \ref%
{strict_expansion_appendix}. The schematic form of the
final result for the amplitude $G_{\mathrm{strict}}(s,t;u)$ is
\begin{eqnarray}
G_{\mathrm{strict}}(s,t;u)
&=&R(s,t;u)+\sum_{PQ}Q_{PQ}^{(s)}(s,t;u)J_{PQ}^{r}(s)  \notag \\
&&+\sum_{PQ}Q_{PQ}^{(t)}(s,t;u)J_{PQ}^{r}(t)+%
\sum_{PQ}Q_{PQ}^{(u)}(s,t;u)J_{PQ}^{r}(u) + \Delta_G,  \label{G_strict_general_form}
\end{eqnarray}%
where $R$, $Q_{PQ}^{(s)}$, $Q_{PQ}^{(t)}$ and $Q_{PQ}^{(u)}$
(where $P,Q=\pi ,K,\eta $) are second order polynomials in the
Mandelstam variables. $%
J_{PQ}^{r}(s)$ is the $\overline{MS}_{\chi PT}$ subtracted  scalar bubble 
\begin{equation}
J_{PQ}^{r}(s)=-\mathrm{i}\int \frac{\mathrm{d}^{4-2\varepsilon }k}{(2\pi
)^{4-2\varepsilon }}\frac{1}{\left( k^{2}-\overset{o}{m}_{P}^{2}+\mathrm{i}0\right) 
\left( \left( k-p\right) ^{2}-\overset{o}{m}_{Q}^{2}+\mathrm{i}0\right)}-\lambda,  \label{J_r}
\end{equation}%
where $\overset{o}{m}_{P,Q}$ are $O(p^{2})$ masses and
\begin{equation}
\lambda=\mu^{-2\varepsilon}\left(\frac{1}{\varepsilon}+{\rm{ln}}4\pi-\gamma+1\right).
\end{equation}
In this formula, $d=4-2\varepsilon $ is the dimension, $\mu $ is the
renormalization scale of the dimensional regularization scheme and $s=p^{2}$.

\section{Bare expansion: matching with a dispersive representation\label{matching section}}

The calculation of the strict expansion is only the first
step in the construction of the amplitude within resummed $\chi PT$. Because
it strictly respects the chiral orders, it suffers from some pathologies
which have to be cured. The most serious one is that the position of the
unitarity cuts in the complex $s$, $t$ and $u$ planes is determined by the
$O(p^{2})$ masses and not by the physical masses of the particles
inside the loops. Also, the presence of the $O(p^{2})$ masses in the
arguments of chiral logarithms can generate undesirable singularities in
the $B_{0}$ dependence of the amplitude\footnote{%
Let us remind that $B_{0}$ is typically taken as a free parameter within resummed $%
\chi PT$.}. These pathologies can be removed either by hand (by means of
some well defined ad hoc prescription, see \cite{Bernard:2010ex}) or, more
systematically, by means of a matching with a dispersive representation of the amplitude, as we have introduced in \cite{Kolesar:2008jr}. The latter procedure, which we term as the construction of the \emph{bare expansion},
is recalled in this section.

Let us briefly introduce the most general model independent form of
the amplitude $G(s,t;u)$ to the order $O(p^{6})$ respecting unitarity,
analyticity and crossing symmetry. Such $G(s,t;u)$ can be constructed using
the reconstruction theorem, which has been developed originally for the $\pi \pi 
$ scattering amplitude \cite{Knecht:1995tr} and which can be easily
generalized to other processes with PGB \cite{Zdrahal:2008bd} (see also \cite%
{Kolesar:2008jr} for a more in-depth discussion of the subtleties connected with
applications in resummed $\chi PT$). According to this theorem, we can
write the $O(p^{4})$ amplitude in the form of a dispersive representation 
\begin{equation}
G(s,t;u)=P(s,t;u)+F_{\pi }^{3}F_{\eta }U(s,t;u)+O(p^{6}),
\label{G_general_form}
\end{equation}%
where $P(s,t;u)$ is a second order polynomial in Mandelstam variables 
\begin{equation}
P(s,t;u)=A_{P}+B_{P}(s-s_{0})+C_{P}(s-s_{0})^{2}+D_{P}\left[
(t-s_{0})^{2}+(u-s_{0})^{2}\right]
\end{equation}%
and $U(s,t;u)$ represent the unitarity corrections, which can be written in
the form 
\begin{eqnarray}
U(s,t;u) &=&\frac{1}{3}\left( W_{0}(s)-W_{2}(s)\right)  \notag \\
&&+\frac{1}{2}\left( 3(s-u)W_{1}(t)+W_{2}(t)\right)  \notag \\
&&+\frac{1}{2}\left( 3(s-t)W_{1}(u)+W_{2}(u\right) .  \label{U_in_terms_W_i}
\end{eqnarray}%
As discussed in detail in \cite{Knecht:1995tr} and \cite{Zdrahal:2008bd}, the
reconstruction theorem, together with the two-particle unitarity relation for
the partial waves, can be used for the iterative construction of the
amplitude up to and including the order $O(p^{6})$ (for the most general
result of such a construction without invoking the Lagrangian formalism,
we refer to \cite{Kampf:2011wr}). Here we use it as a tool for an appropriate
modification of the results described in the
previous section, i.e. for the construction of the bare expansion.

For this purpose we reconstruct and fix the unitarity part $U(s,t;u)$ of the
amplitude from the $O(p^{2})$ amplitudes calculated within resummed $\chi PT$. 
Let us note that there is some
ambiguity in the choice of the form of the $O(p^{2})$ amplitudes, which we
take as an input for the reconstruction theorem. The reason is that there
are (at least) two possibilities how to connect the generic physical $%
O\left( p^{2}\right) $ amplitude $A^{\left( 2\right) }$ of the process $%
AB\rightarrow CD$ (which is a dangerous observable) and the corresponding
safe observable $G^{\left( 2\right) }$, namely either%
\begin{equation}
A^{\left( 2\right) }=\frac{G^{\left( 2\right) }}{F_{A}F_{B}F_{C}F_{D}},
\label{choice_1}
\end{equation}%
where $F_{A},\ldots ,F_{D}$ are the physical decay constants or%
\begin{equation}
A^{\left( 2\right) }=\frac{G^{\left( 2\right) }}{F_{0}^{4}}.  \label{choice_2}
\end{equation}%
The choice between this alternatives is in fact a part of the definition of
the direct remainders (cf. (\ref{G_bare_expansion}) below) . See also \cite%
{Kolesar:2008jr} and appendix \ref{reconstruction_appendix} for more detail.

After calculating the unitarity part, the remaining polynomial part of the
amplitude is then fixed by means of matching of the strict chiral expansion
obtained in the previous section with the general form (\ref{G_general_form}%
). The list of all relevant $O(p^{2})$ amplitudes and a reconstruction of $%
U(s,t;u)$ from them is given in detail in appendix \ref{reconstruction_appendix}. 
The corresponding $W_{i}(s)$ (see (\ref%
{U_in_terms_W_i})) have the following schematic form (cf. (\ref%
{G_strict_general_form})) 
\begin{equation}
W_{i}(s)=\sum_{PQ}w_{i}^{PQ}(s)\overline{J}_{PQ}(s),  \label{Wi's}
\end{equation}%
where $w_{i}^{PQ}(s)$ is a second order polynomial with
coefficients depending on the $O(p^{2})$ parameters of the chiral Lagrangian and 
\begin{equation}
\overline{J}_{PQ}(s)=\frac{s}{16\pi ^{2}}\int_{(M_{P}+M_{Q})^{2}}^{\infty }%
\frac{\mathrm{d}x}{x}\frac{\lambda ^{1/2}(x,M_{P}^{2},M_{Q}^{2})}{x}\frac{1}{%
x-s-\mathrm{i}0}  \label{J_bar}
\end{equation}%
is a once subtracted scalar bubble with physical masses $M_{P}$ and \ $M_{Q}$ inside
the loop. In (\ref{Wi's}), the sum is taken over all two-particle
intermediate states $PQ$ in the given channel. Let us note that the general
form of the reconstructed $U(s,t;u)$ is similar to the last three terms of
the strict expansion of $G(s,t;u)$ (\ref{G_strict_general_form}), with the
exception that $J_{PQ}^{r}(s)$ with unphysically situated cuts are replaced
with $\overline{J}_{PQ}(s)$ for which the cuts are placed at the physical
two particle thresholds $s=(M_{P}+M_{Q})^{2}$. This enables us to match both
representations as follows.

In (\ref{G_strict_general_form}), let us write $J_{PQ}^{r}(s)=\overline{%
J_{PQ}^{r}}(s)+J_{PQ}^{r}(0)$, where the renormalization scale independent
part $\overline{J_{PQ}^{r}}(s)\equiv J_{PQ}^{r}(s)-J_{PQ}^{r}(0)$ of $%
J_{PQ}^{r}(s)$ is nothing else but the scalar bubble (\ref{J_r}) subtracted
at $s=0$. \ As a result, we can write 
\begin{equation}
G_{\mathrm{strict}}(s,t;u)=G_{\mathrm{pol}}(s,t;u)+F_{\pi }^{3}F_{\eta
}U^{r}(s,t;u),
\end{equation}%
where 
\begin{equation}
G_{\mathrm{pol}}(s,t;u)=G(s,t;u)|_{\overline{J_{PQ}^{r}}\rightarrow 0}
\end{equation}%
is a second order polynomial. Let us remark that a polynomial part of the amplitude constructed in this way 
does not depend on the renormalization scale $\mu $.

The matching can be performed by means of an identification of the
polynomial $P(s,t;u)$ from the dispersive representation (\ref%
{G_general_form}) with $G_{\mathrm{pol}}(s,t;u).$ This means we write the
amplitude in the form 
\begin{equation}
G(s,t;u)=G_{\mathrm{pol}}(s,t;u)+F_{\pi }^{3}F_{\eta }U(s,t;u),
\end{equation}%
with the $U(s,t;u)$ from (\ref{G_general_form}) and (\ref{U_in_terms_W_i})
constructed according to the reconstruction theorem. Note that such a $%
G(s,t;u)$ satisfies the requirements of the perturbative unitarity exactly.
We then get for the polynomial part of the amplitude
\begin{equation}
G_{\mathrm{pol}}(s,t;u)=G^{(2)}(s,t;u)+G_{\mathrm{ct}}^{(4)}(s,t;u)+G_{%
\mathrm{tad}}^{(4)}(s,t;u)+G_{\mathrm{pol,u}}^{(4)}(s,t;u),
\end{equation}%
where $G_{\mathrm{ct}}^{(4)}$ and $G_{\mathrm{tad}}^{(4)}$ (the counterterm and tadpole contributions) can be found in
appendix \ref{strict_expansion_appendix}, while $G_{\mathrm{pol,u}}^{(4)}$ (the unitarity contribution)
in appendix \ref{disp_poly_appendix}.

As a last step, we replace, in the expressions for $G_{\mathrm{pol}}(s,t;u)$,
the $O(p^{2})$ masses inside the chiral logarithms and inside $J_{PQ}^{r}(0)$
with the physical masses (but we keep them in all other places they
appeared). This ad hoc prescription avoids the unwanted logarithmic
singularities in the limit $X\rightarrow 0$ in our final formula for the bare
expansion of the amplitude $G(s,t;u)$%
\begin{equation}
G_{\mathrm{bare}}(s,t;u)=G_{\mathrm{pol}}(s,t;u)+F_{\pi }^{3}F_{\eta
}U(s,t;u)+\Delta _{G}(s,t;u).  \label{G_bare_expansion}
\end{equation}%
The latter formula can be also understood as a definition of the $O(p^{6})$
remainder $\Delta _{G}$. However, because we do not know the detailed
analytic structure of $\Delta _{G}$, we parametrize it in the form of
a polynomial in the variables $s$, $t$ and $u$ 
\begin{equation}
\Delta _{G}=A\delta _{A}+B\delta _{B}(s-s_{0})+C\delta
_{C}(s-s_{0})^{2}+D\delta _{D}\left[ (t-s_{0})^{2}+(u-s_{0})^{2}\right],
\label{G_remainders}
\end{equation}%
where the  observables $A$, \ldots ,$D$ are the coefficients of the
expansion of the amplitude $G(s,t;u)$ at the center of the Dalitz plot
\begin{eqnarray}
G(s,t;u) &=&A+B(s-s_{0})+C(s-s_{0})^{2}+D\left[ (t-s_{0})^{2}+(u-s_{0})^{2}%
\right]  \notag \\
&&+\Delta _{G}+O\left( (s-s_{0})^{3},\ldots \right).
\end{eqnarray}
The bare expansion of $A,\dots ,D$, derived form (\ref{G_bare_expansion}), is
explicitly given in appendix \ref{A_D_bare_expansion_appendix}.

While it is natural to assume the coefficients $A$, $B$, $C$ and $D$ to be safe observables,
strictly speaking, this assumption goes beyond the general definition  given in section \ref{good_observables_section}. Note that the global convergence of the Green function $G(s,t;u)$ does not automatically imply a convergence of its derivatives at the center of the Dalitz plot.  Moreover, the parameters $C$ and $D$ start at  $O(p^4)$ and therefore the criterion of the global convergence merely implies that  their  $O(p^8)$ reminders are reasonably small. The assumption about the natural size of their  $O(p^6)$ remainders is thus  in fact an additional conjecture about the regularity of the chiral series.
Therefore, considering  $A$, $B$, $C$ and $D$ as safe observables
should rather be taken as a working hypothesis. Actually, one of the
issues of this work is probing this assumption. 

The conventional Dalitz plot parameters are related to these coefficients by
means of the nonlinear relations (\ref{ABCD_Dalitz}) and thus should be
regarded as dangerous observables and expressed nonperturbatively with all
the remainders kept explicitly 
\begin{eqnarray}
a &=&-2R_{\eta }\mathrm{Re}\left( \frac{B}{A}\right) \frac{1-\delta _{A}}{%
1-\delta _{B}}  \label{Dalitz_a} \\
b &=&R_{\eta }^{2}\left( \left\vert \frac{B}{A}\right\vert ^{2}\left( \frac{%
1-\delta _{A}}{1-\delta _{B}}\right) ^{2}+\mathrm{Re}\left( \frac{2C\left(
1-\delta _{C}\right) ^{-1}+D\left( 1-\delta _{D}\right) ^{-1}}{A\left(
1-\delta _{A}\right) ^{-1}}\right) \right)  \label{Dalitz_b} \\
d &=&3R_{\eta }^{2}\mathrm{Re}\left( \frac{D}{A}\right) \left( \frac{%
1-\delta _{A}}{1-\delta _{D}}\right)  \label{Dalitz_d} \\
\alpha &=&\frac{1}{4}\left( d+b-R_{\eta }^{2}\left\vert \frac{B}{A}%
\right\vert ^{2}\left( \frac{1-\delta _{A}}{1-\delta _{B}}\right)
^{2}\right) .  \label{Dalitz_alpha}
\end{eqnarray}

\section{Reparametrization of LECs in terms of physical observables\label{repar_section}}

The approach of resummed $\chi PT$ to the problem of reparametrization of
the chiral expansion, i.e. to the exclusion of some parameters of the
Lagrangian in terms of physical observables, differs substantially from
the one used in standard $\chi PT$. The reason is that the usual
reparametrization procedure, consisting of the expansion of the $O(p^{2})$
parameters $F_{0}$ and $B_{0}m_{q}$, where $m_{q}$ are the light quark
masses, in terms of $F_{\pi }$, $F_{K}$ and pseudoscalar masses $M_{\pi }$, $%
M_{K}$ and $M_{\eta }$ is in general an operation, which can
generate a large higher order remainder. Indeed, on one hand the above
mentioned quantities are not safe observables, as we have discussed in
the section \ref{mixing_subsection}, and thus their expansion
in terms of the original parameters of the Lagrangian might include
large remainders. On the other hand, even for the safe observables, like $%
F_{P}^{2}$ and $F_{P}^{2}M_{P}^{2}$, an inversion of the expansion might
generate large remainder as well. 

As an illustration, let us assume the
following toy example. We can simplify things and assume there is only
one leading order parameter $G_{0}$. Suppose there exists a safe observable $%
G $, for which the globally convergent expansion has the form 
\begin{equation}
G=G_{0}+G^{(4)}(G_{0})+G^{(6)}(G_{0})+G\delta _{G}.
\end{equation}%
The $O(p^{4})$ and $O(p^{6})$ terms of this expansion generally depend on $%
G_{0}$. The usual reparametrization procedure up to the $O(p^{6})$ order
needs an inversion of this expansion and expressing $G_{0}$ in terms of $G$
\begin{equation}
G_{0}=G-G^{(4)}(G)-\left[ G^{(6)}(G)-G^{(4)^{\prime }}(G)G^{(4)}(G)\right]
+G_{0}\delta _{G_{0}},  \label{inverted_expansion}
\end{equation}%
where we have explicitly grouped the various chiral orders together. The
remainder $G_{0}\delta _{G_{0}}$ generated by such a procedure is then
dropped. However, we get an identity 
\begin{equation}
\delta _{G_{0}}=-\frac{1-X_{G}}{X_{G}}-\frac{1}{X_{G}}\overline{X}_{G_{0}}-%
\frac{1}{X_{G}}\overline{X}_{G_{0}}\overline{\overline{X}}_{G_{0}},
\end{equation}%
where the ratios 
\begin{eqnarray}
\overline{X}_{G_{0}} &=&\frac{G^{(4)}(G)}{G} \\
\overline{\overline{X}}_{G_{0}} &=&\frac{G^{(6)}(G)-G^{(4)^{\prime
}}(G)G^{(4)}(G)}{G^{(4)}(G)}
\end{eqnarray}%
are analogues of (\ref{X_G_definition}) and (\ref{barX_G_definition}) for the
expansion (\ref{inverted_expansion}) and probes the apparent convergence of
the inverted expansion. Even if \ the inverted expansion (\ref%
{inverted_expansion}) seems to converge well in the sense that $\overline{X}%
_{G_{0}}$ and $\overline{\overline{X}}_{G_{0}}$ are reasonably small, the
neglected remainder might be large for $X_{G}\ll 1$. 

Therefore, we keep the dependence on $F_{0}$ and $B_{0}m_{q}$ in terms of the following parameters 
\begin{eqnarray}
X &=&\frac{2B_{0}F_{0}^{2}\widehat{m}}{F_{\pi }^{2}M_{\pi }^{2}}  \label{X}
\\
Z &=&\frac{F_{0}^{2}}{F_{\pi }^{2}},  \label{Z}
\end{eqnarray}%
which probe the regularity of the bare chiral expansion of the safe
observables $F_{\pi }^{2}M_{\pi }^{2}$ and $F_{\pi }^{2}$ in the sense of
definition (\ref{X_G_definition}). 

The quark masses are left as free parameters as well, in terms of the ratios\footnote{%
In numerical calculations, we take $r$ from the lattice, which fixes the Kaplan-Manohar ambiguity} 
\begin{equation}
r=\frac{m_{s}}{\widehat{m}},
\end{equation}%
and\footnote{At the first order in the isospin breaking, the factor $1/R$ appears
only as an overall normalization factor.} 
\begin{equation}
R=\frac{m_{s}-\widehat{m}}{m_{d}-m_{u}}.
\end{equation}%
For further convenience, we also introduce the following frequently appearing
combination 
\begin{equation}
Y=\frac{X}{Z},
\end{equation}%
which is a measure of the regularity of the expansion of the dangerous observable $%
M_{\pi }^{2}$. Let us note that within the standard approach the parameters 
$X$, $Z$ and $r$ are fixed by means of the potentially dangerous inverted
expansions of the type (\ref{inverted_expansion}), namely
\begin{eqnarray}
X_{std} &=&1+O(p^{2})  \notag \\
Z_{std} &=&1+O(p^{2})  \notag \\
r_{std} &=&2\frac{M_{K}^{2}}{M_{\pi }^{2}}-1+O(p^{2}).
\label{dangerous_reparametrization}
\end{eqnarray}

Nevertheless, the bare expansion of the safe observables $F_{P}^{2}$ and $%
F_{P}^{2}M_{P}^{2}$ can be used for an alternative reparametrization, which is,
in contrast to the standard one, safe and do not suffer from dangerous
manipulations, which might generate uncontrollably large remainders. 
Because of the linear dependence of the $O(p^{4})$ order on the $L_{4}$-$L_{8}$ LECs, 
it is possible to express these constants by exact algebraic
identities in terms of these safe observables and their $O(p^{6})$
remainders $\delta _{F_{P}}$ and $\delta _{F_{P}M_{P}}$, schematically 
\begin{equation}
L_{i}=\sum_{P}a_{P}F_{P}^{2}(1-\delta
_{F_{P}})+\sum_{P}b_{P}F_{P}^{2}M_{P}^{2}(1-\delta _{F_{P}M_{P}}).
\label{L_i_reparametrization}
\end{equation}%
Due to the linearity of these relations, the corresponding remainders $\Delta
_{L_{l}}$
\begin{equation}
\Delta _{L_{l}}=-\sum_{P}\left( a_{P}F_{P}^{2}\delta
_{F_{P}}+b_{P}F_{P}^{2}M_{P}^{2}\delta _{F_{P}M_{P}}\right),
\label{L_i_remainders}
\end{equation}%
are well under control. The explicit form of these relations have been
published in \cite{DescotesGenon:2000di}, \cite{DescotesGenon:2003cg} (and
also in \cite{Kolesar:2008jr}, which is closest to the notation used here).

Concerning the LECs $L_{1}$-$L_{3}$, we don't have a similar procedure ready at
this point. We therefore treat $L_{1}$-$L_{3}$ as
independent parameters in our approach. Fortunately, as will be demonstrated in section \ref%
{Numerical_analysis}, the results depend on these parameters in a very
weak way only. This is not surprising in the cases of $L_1$ and $L_2$ which occur only in the combination  $(\varepsilon_{\pi}-\varepsilon_{\eta})L_i$. Such terms effectively correspond to a NNLO effect.

To summarize, when concerning the approach to reparametrization of a safe observable $G$, we proceed along the following points:

\begin{itemize}
\item dangerous observables are not used

\item the potentially dangerous inverted expansions are not performed

\item an explicit dependence on the parameters of the $O(p^{2})$ Lagrangian is kept ($F_{0}$ and $B_{0}m_{q}$)

\item some of the $O(p^{4})$ LECs, namely $L_{4},\ldots ,L_{8}$, are
expressed in terms of $F_{P}^{2}$ and $F_{P}^{2}M_{P}^{2}$, using safe
nonperturbative algebraical identities

\item the result of the reparametrization is then expressed in terms of $%
O(p^{2})$ parameters $X$, $Y$, $r$ and $R$, the remaining $O(p^{4})$ LECs $\
L_{1}$, $L_{2}$ and $L_{3}$ and the direct ($\Delta _{G}$) and indirect ($%
\delta _{F_{P}}$, $\delta _{F_{P}M_{P}}$, $\delta _{\varepsilon F}$, $\delta _{\varepsilon FM}$) remainders
\end{itemize}

In the case of the observables $A,\dots ,D$, connected with the $\eta
\rightarrow \pi ^{+}\pi ^{-}\pi ^{0}$, only the polynomial part of $A$ and $B$ 
depend nontrivially on $L_{4}$-$L_{8}$,
therefore only these have to be reparametrized. Explicit results are
given in appendix \ref{reparametrization_appendix}.

This step completes
the construction of the $\eta \rightarrow \pi ^{+}\pi ^{-}\pi ^{0}$
amplitude in terms of safe observables within resummed $\chi PT$.

\section{Treatment of free parameters and remainders \label{Remainders}}

As we have discussed above, the chiral symmetry breaking parameters $%
X$ and $Z$ are treated as free. This approach is the opposite of the usual
treatment of these parameters within standard $\chi PT$, where they are
predicted by means of the chiral expansion order by order, in terms of the
physical masses, decay constants and LECs. Because the dependence of the observables on $X$ and $Z$ is held explicitly, it can also serve as a source of
information on the mechanism of the chiral symmetry breaking in QCD.

Similarly, the ratio of the quark masses $r=m_{s}/\widehat{m}$  is
left as a free parameter too. In what follows, we fix this
parameter to a recent averaged lattice QCD value, obtained  by FLAG \cite{Aoki:2013ldr}
\begin{equation}
r=27.5\pm 0.4.  \label{r_lattice}
\end{equation}
Analogously, in the case of the quark mass ratio  $R=(m_s -\hat{m})/(m_d-m_u)$, we take an averaged lattice value \cite{Aoki:2013ldr} as well
\begin{equation}
R=35.8\pm 2.6.
 \label{R_lattice1}
\end{equation}

A second kind of free parameters present in our calculation are the remainders. We have the direct remainders (\ref{G_remainders}), namely 
\begin{equation}
\delta _{A},~\delta _{B},~\delta _{C},~\delta _{D},
\end{equation}%
which parametrize the unknown higher order contributions to the amplitude $%
G(s,t;u)$. Then there are the indirect remainders
\begin{equation}
\delta _{F_{\pi }^{2}M_{\pi }^{2}},~\delta _{F_{\pi }^{2}},~\delta
_{F_{K}^{2}M_{K}^{2}},~\delta _{F_{K}^{2}},\delta _{F_{\eta }^{2}M_{\eta
}^{2}},~\delta _{F_{\eta }^{2}},~\delta _{\varepsilon FM},~\delta
_{\varepsilon F},
\end{equation}%
which stem from the higher order contributions to the safe observables $%
F_{P}^{2}M_{P}^{2}$, $F_{P}^{2}\,$, $(\varepsilon _{\pi
}F_{\pi }^{2}-\varepsilon _{\eta }F_{\eta }^{2})$,~$(\varepsilon _{\pi
}F_{\pi }^{2}M_{\pi }^{2}-\varepsilon _{\eta }F_{\eta }^{2}M_{\eta }^{2})$,
i.e. emerge in the procedure of the reparametrization of the strict
chiral expansion in terms of the masses and decay constants. 

As we have discussed in section \ref{matching section}, the definition of the direct remainders depends on the
fixing of the unitarity part of the amplitude, i.e. on the choice between (%
\ref{choice_1}) and (\ref{choice_2}). 
In the numerical analysis below, we use the alternative (\ref{choice_2}). The rationale for such a choice
is that it is closer to the strict expansion, for which we assume
the remainders should respect the global convergence criteria. The possibility (\ref{choice_1}), on the other hand, leads to a significant suppression of the unitarity corrections due to the factor $F_0^4/(\prod_i F_{P_i}^2)$ appearing in the loop functions, which appears quite unnatural.

In the presented paper, we treat the higher order remainders as a source of
uncertainty of the theoretical prediction.  Let us remind that contributions of the  NNLO LECs $C_{i}$ are implicitly included in the remainders. Because the $C_i$'s are not well known, they represent an important source of theoretical uncertainty of the standard NNLO $\chi PT$ calculations, which are hard to quantify. In the resummed $\chi PT$ approach this uncertainty is supposed to be under better control as a consequence of the assumption of the global convergence.

Of course, any additional
information on the actual values of the remainders both from inside the theory (higher order
calculations) or from outside $\chi PT$ (estimates based on resonance
contributions or a unitarization of the amplitude) can be used to reduce such an uncertainty. 

As a first step, we do not assume any supplemental information. In what follows, we take the remainders as independent,
uncorrelated, normally distributed random variables with zero mean value and
a standard deviation attributed to them according to a rule based
on the general expectation about the convergence of the chiral expansion of
the safe observables. In accord with \cite{DescotesGenon:2003cg}, we take
\begin{eqnarray}
\delta ^{NLO} &=&0.0\pm 0.3  \label{deltaNLO} \\
\delta ^{NNLO} &=&0.0\pm 0.1  \label{deltaNNLO}
\end{eqnarray}%
for NLO ($\delta _{C},~\delta _{D}$) and NNLO
(the rest) remainders, respectively.

Besides the higher order remainders and the parameters $X$, $Z$, $r$ and $R$, the
resulting formulae for the amplitude and Dalitz plot parameters depend on
the constants $L_{1}$, $L_{2}$ and $L_{3}$. For these a similar
reparameterization procedure as that described above for $L_{4}$-$L_{8}$ is
not available. We therefore collect standard $\chi PT$ fits 
\cite{Bijnens:1994ie,Amoros:2001cp,Bijnens:2011tb, Bijnens:2014lea} and by taking the mean
and spread of such a set, we obtain an estimate of the influence of these
LECs: 
\begin{eqnarray}
L_{1}^{r}(M_{\rho }) &=&(0.57\pm 0.18)\cdot 10^{-3}  \notag \\
L_{2}^{r}(M_{\rho }) &=&(0.82\pm 0.27)\cdot 10^{-3}  \notag \\
L_{3}^{r}(M_{\rho }) &=&(-2.95\pm 0.37)\cdot 10^{-3}.  \label{L123}
\end{eqnarray}%
We ignore the reported error bars of the fits, as they are relatively small and
in some cases there is quite a substantial disagreement among them. The variance we
obtain is a source of theoretical uncertainty and we treat it on the same
footing as the uncertainty stemming from the higher order remainders.

\section{Numerical analysis \label{Numerical_analysis}}

\begin{figure}[t]
\begin{center}
\includegraphics[scale=0.5]{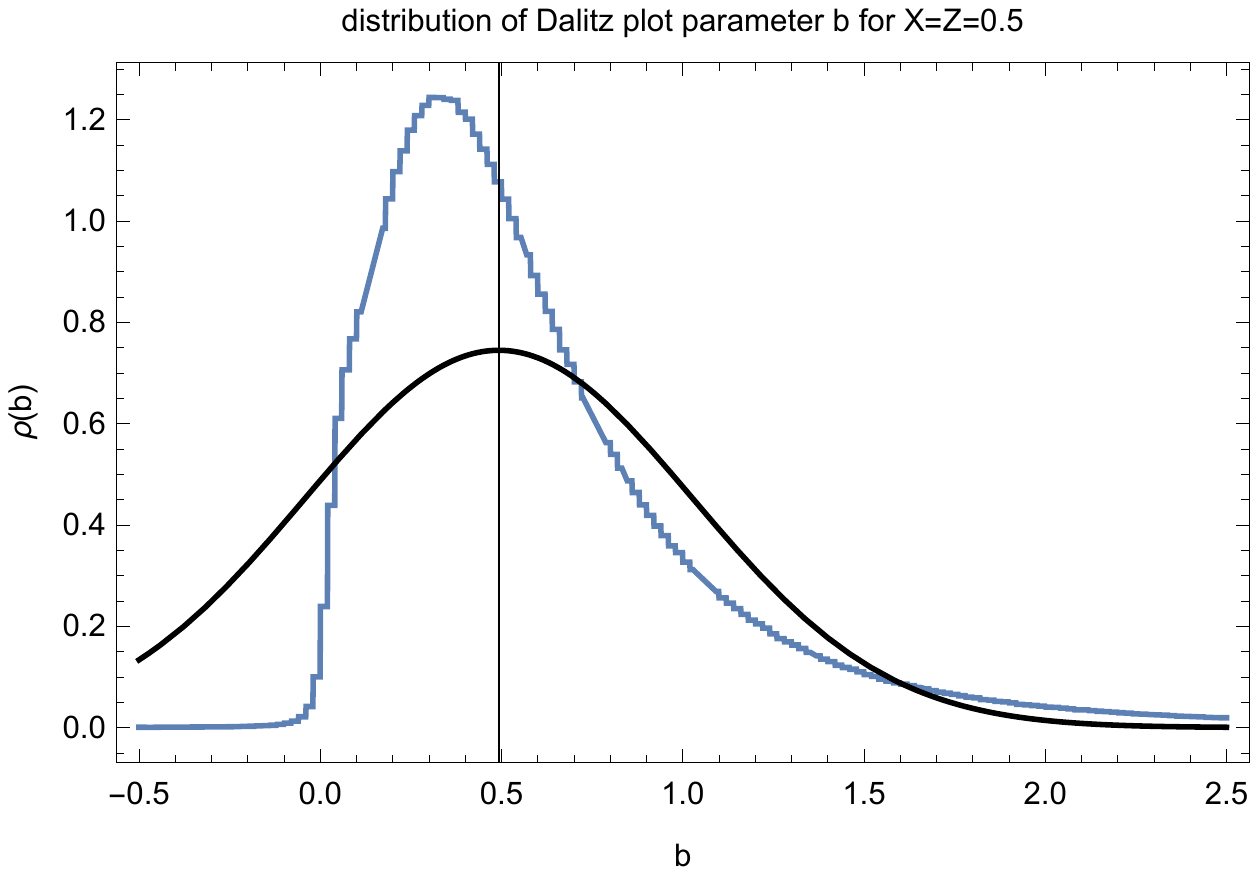} %
\includegraphics[scale=0.5]{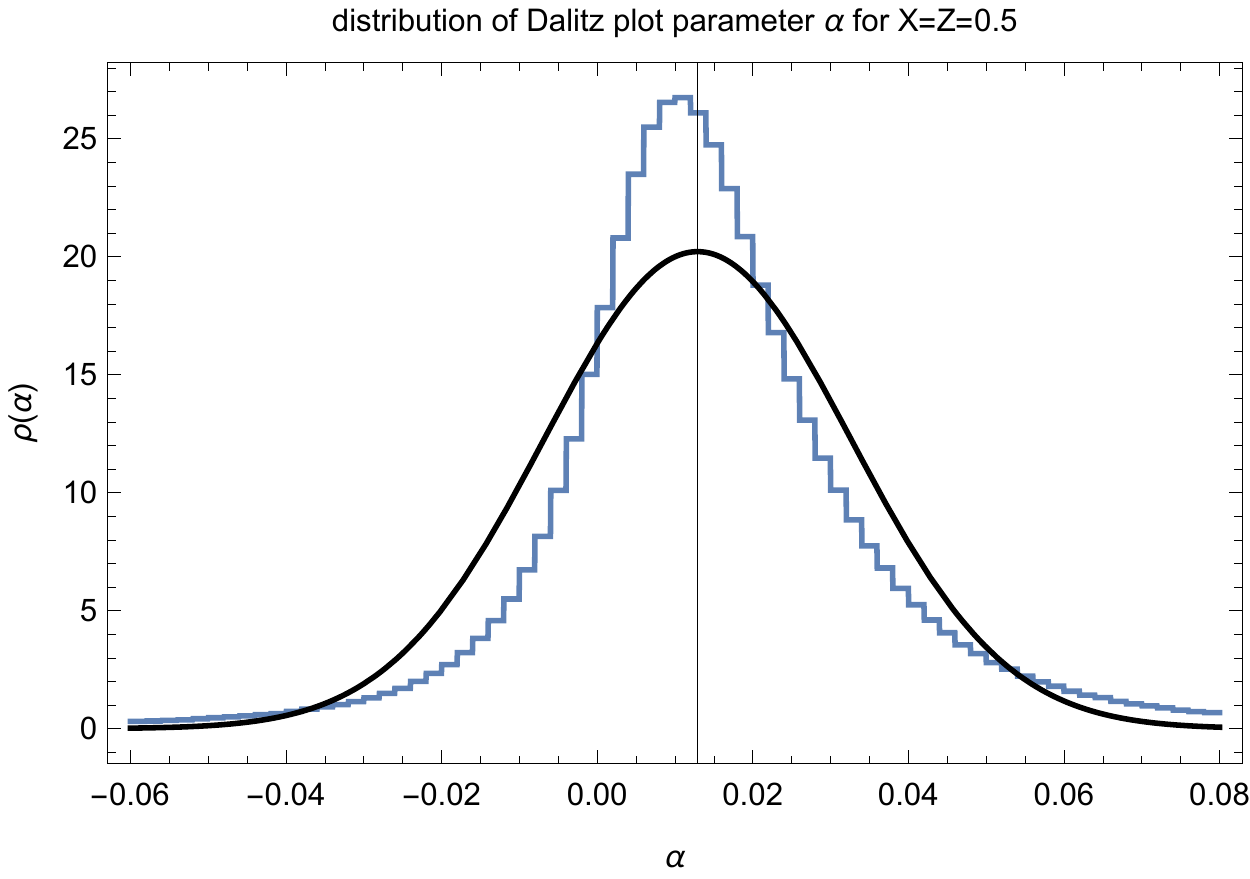}
\end{center}
\caption{Distribution of the predictions for the Dalitz parameters $b$
and $\protect\alpha$ at $X=Z=0.5$ (histograms). For comparison, the solid line depicts a normal
distribution with the same median and standard deviation around the median.}
\label{f0}
\end{figure}

As explained above, we treat the remainders, the LECs $L_{1}$-$L_{3}$ and the quark mass parameters $r$ and $R$ as normally distributed random variables. This implies
that at this stage our predictions are of
stochastic nature. In what follows, we therefore numerically generate an ensemble of $4\times 10^{5}$ normally distributed random sets of these parameters according to (\ref{r_lattice}), (\ref{R_lattice1}), (\ref{deltaNLO}), (\ref{deltaNNLO}), (\ref{L123}) and compute distributions for the observables under interest.\footnote{We use the current PDG values \cite{PDG:2014kda} for quantities not explicitly discussed above} 
Because the observables depend on these random variables in a complex nonlinear
manner, the obtained range of theoretical predictions is in general distributed according to
non-gaussian distributions (see fig. \ref{f0} for an
illustration). In particular, the mean value of such a distribution can often be different than the median. 
The median of the distributions, however, in most of our cases coincide with the value
obtained by setting the free parameters to their means very well.
We therefore quote the median rather than
the mean value in the following and the reported error bars correspond to (a generally
non-symmetric) one-sigma contour around it\footnote{This approach  is different from the one we used in  the preliminary  report \cite{Kolesar:2011wn}, where a gaussian distribution was implicitly assumed.}

\begin{table}[t]
\begin{tabular}{|l|l|l|l|l|l|l|l|}
\hline
& $X_{\mathrm{std}}$ & $Z_{\mathrm{std}}$ & $r_{\mathrm{std}}$ & $a$ & $b$ & 
$d$ & $\alpha $ \\ \hline\hline
$R\chi PT$+BE14 & $0.628$ & $0.593$ & $27.5$ & $-1.38_{-0.56}^{+0.47}$ & $%
0.48_{-0.27}^{+0.48}$ & $0.081_{-0.042}^{+0.060}$ & $0.012_{-0.015}^{+0.018}$
\\ \hline
$R\chi PT$+free fit & $0.452$ & $0.482$ & $27.5$ & $-1.43_{-0.65}^{+0.53}$ & $%
0.51_{-0.30}^{+0.58}$ & $0.084_{-0.047}^{+0.068}$ & $0.013_{-0.016}^{+0.020}$
\\ \hline\hline
KLOE '16, '10 &  &  &  & $-1.095(4)$ & $%
0.145(6)$ & $0.081(7)$ &  $%
-0.0301(50)$ \\ \hline
NNLO $\chi PT$ &  &  &  & $-1.271(75)$ & $0.394(102)$ & $0.055(57)$ & $%
0.013(32)$ \\ \hline
\end{tabular}%
\caption{Illustration of the predictions of  resummed $\chi PT$ for the Dalitz plot parameters in comparison with the experimental data. The input values of the parameters   $X$, $Z$ and $r$ are taken from the most recent fit \cite{Bijnens:2014lea}. The upper and lower bounds
correspond to a one-sigma interval around the median.}
\label{XZr}
\end{table}

To get a flavour of the values of the Dalitz plot parameters  and  of the uncertainty generated by the
unknown remainders, as the first step, we provide resummed $\chi PT$ predictions of $a$, $b$, 
$d$ and $\alpha $ for a set of fixed values of $X$ and $Z$. In table \ref{XZr}, we set $X$ and $Z$ to values obtained by the  most recent standard $\chi$PT fits \cite{Bijnens:2014lea}. In these fits, the parameters $X$ and $Z$ are obtained from the results for the NLO LECs, while $r$ is fixed to the lattice value (\ref{r_lattice}).

\begin{figure}[p]
\begin{center}
\includegraphics[scale=0.6]{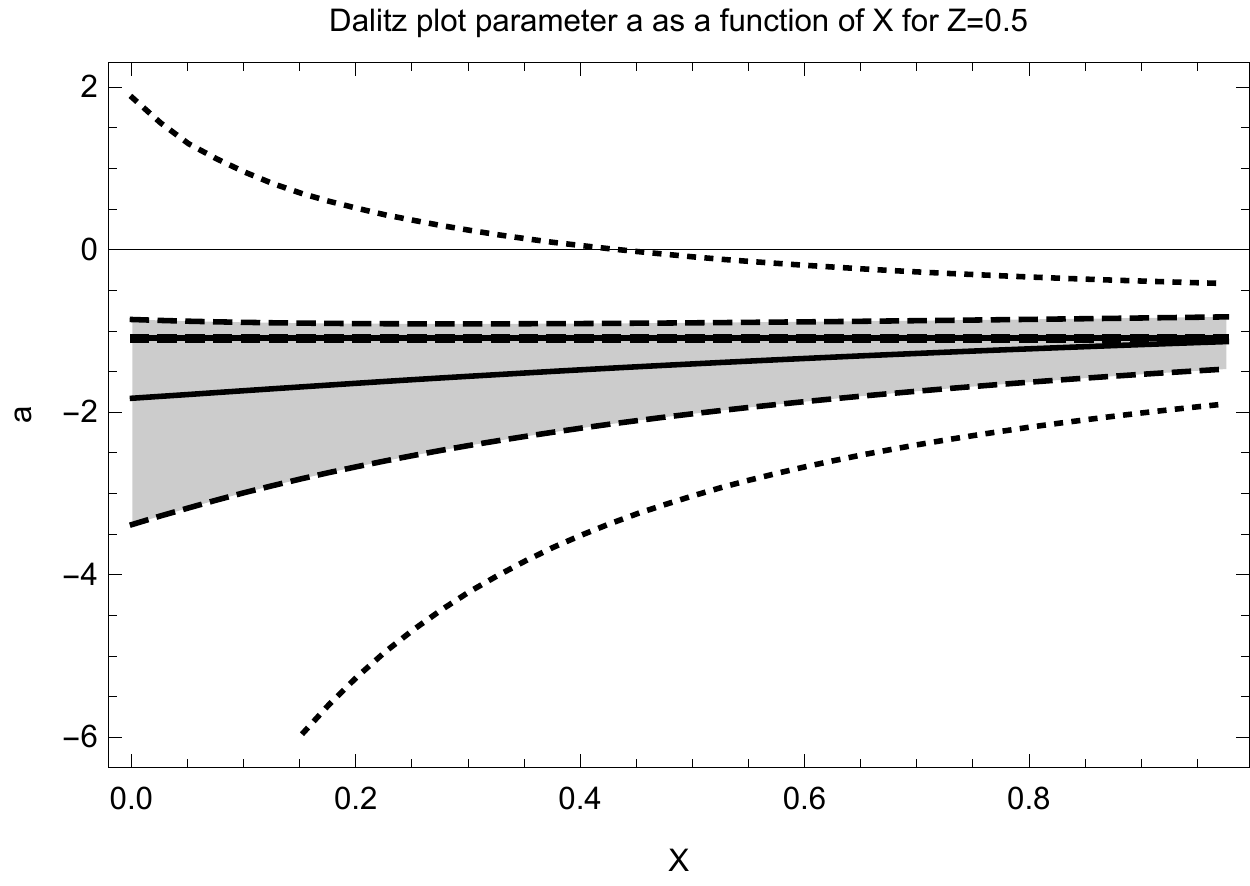}
\includegraphics[scale=0.6]{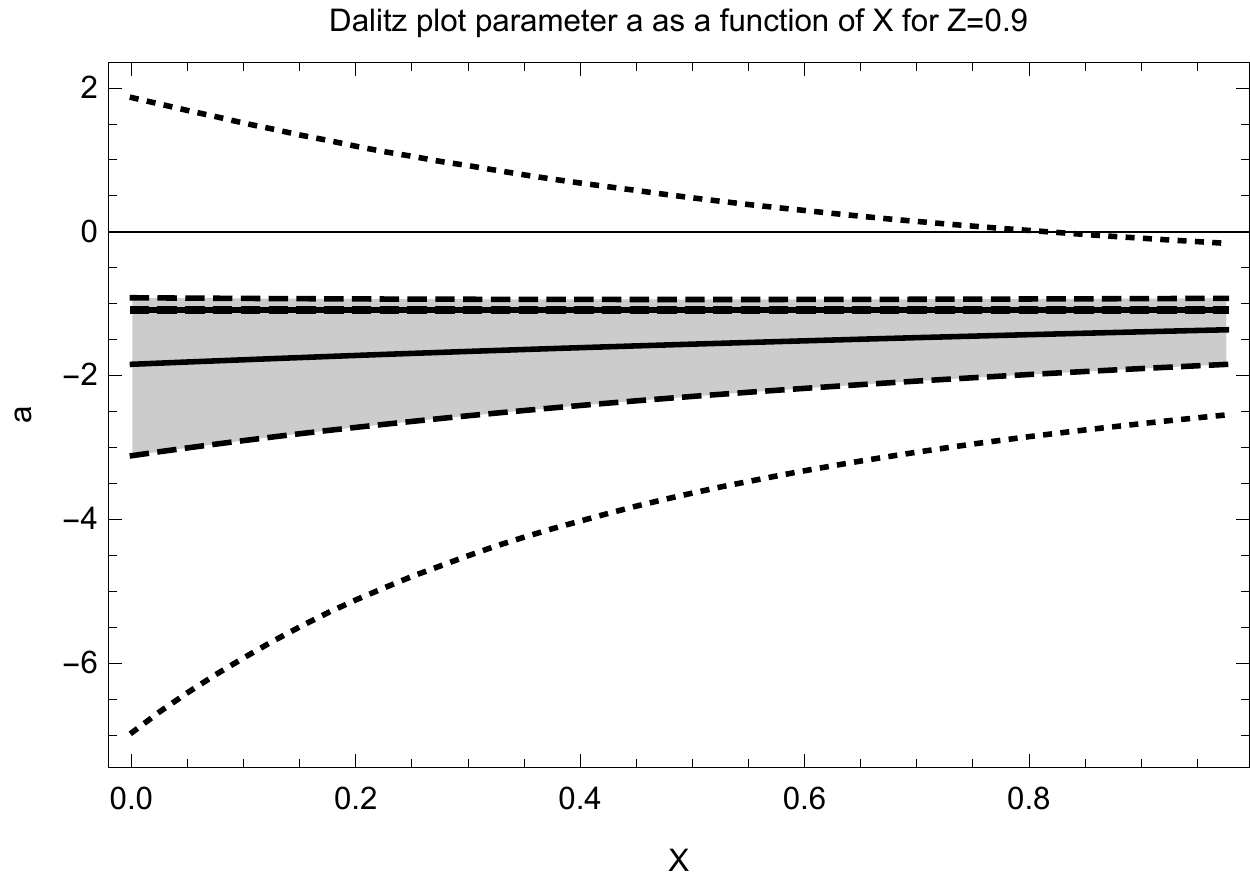}
\includegraphics[scale=0.6]{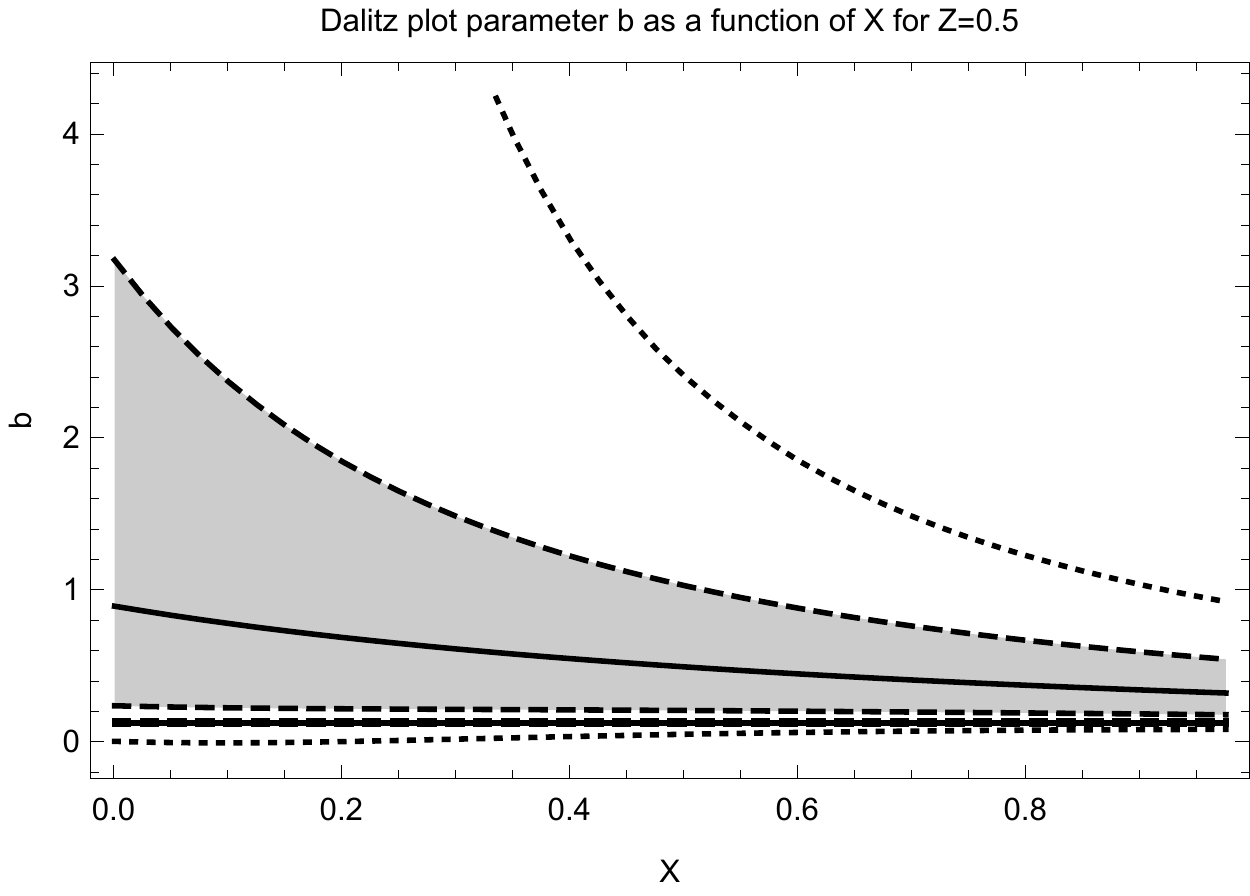} 
\includegraphics[scale=0.6]{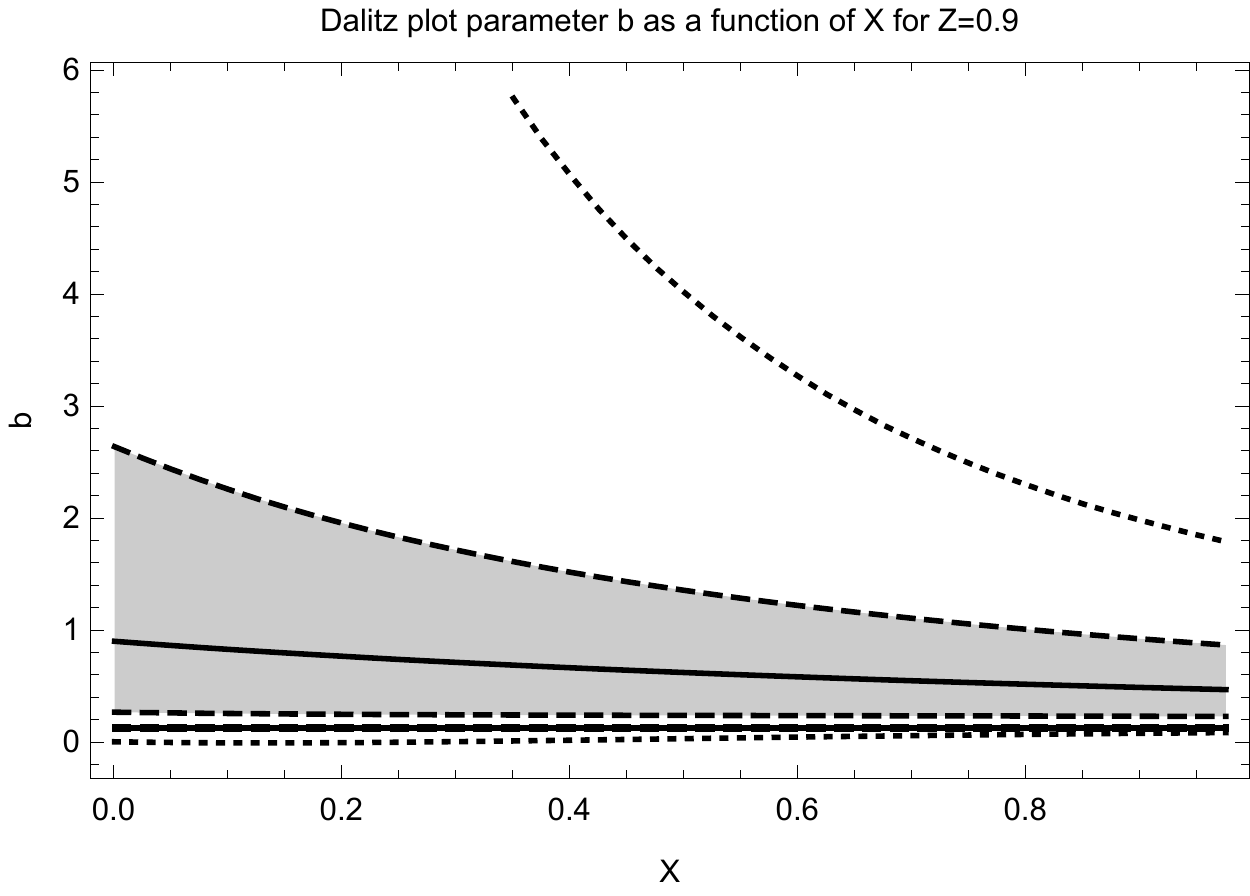}
\includegraphics[scale=0.6]{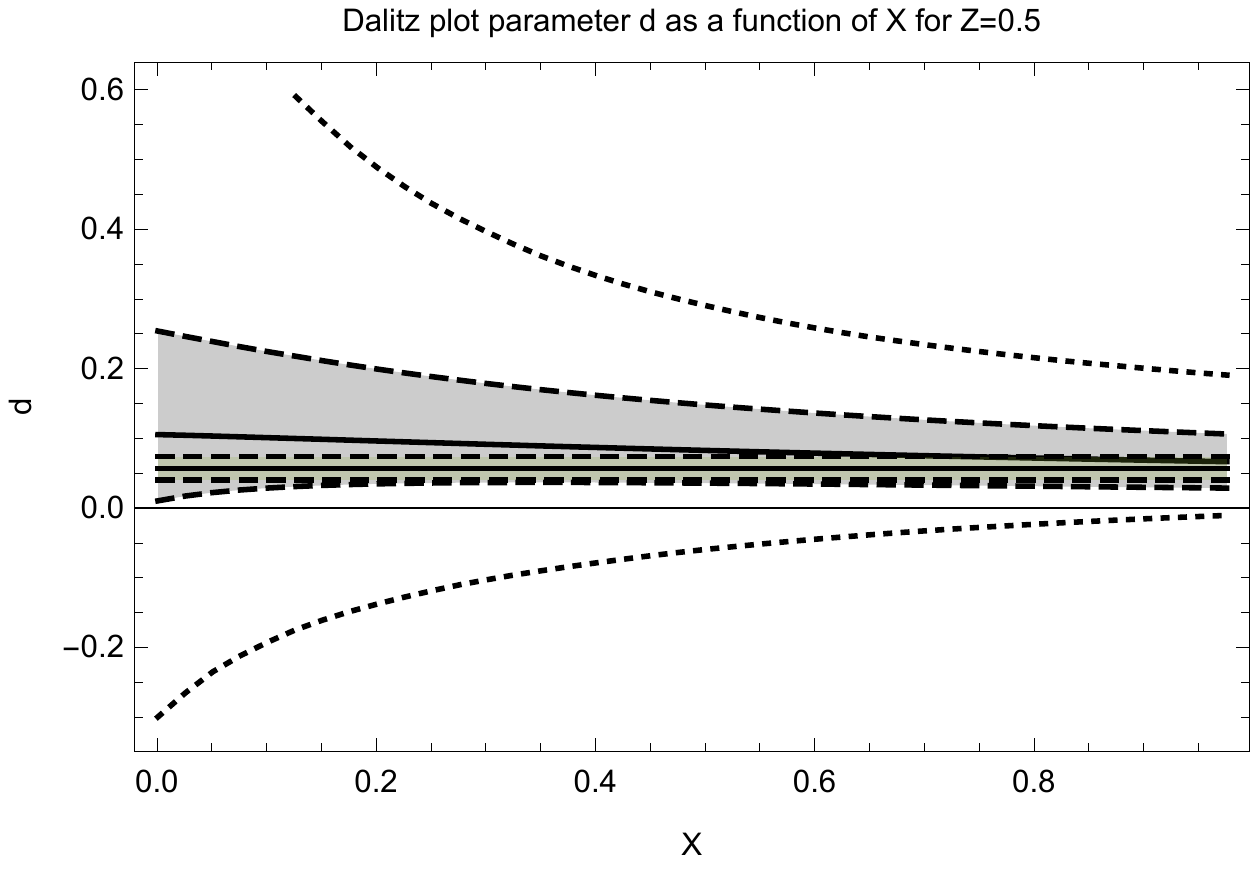}
\includegraphics[scale=0.6]{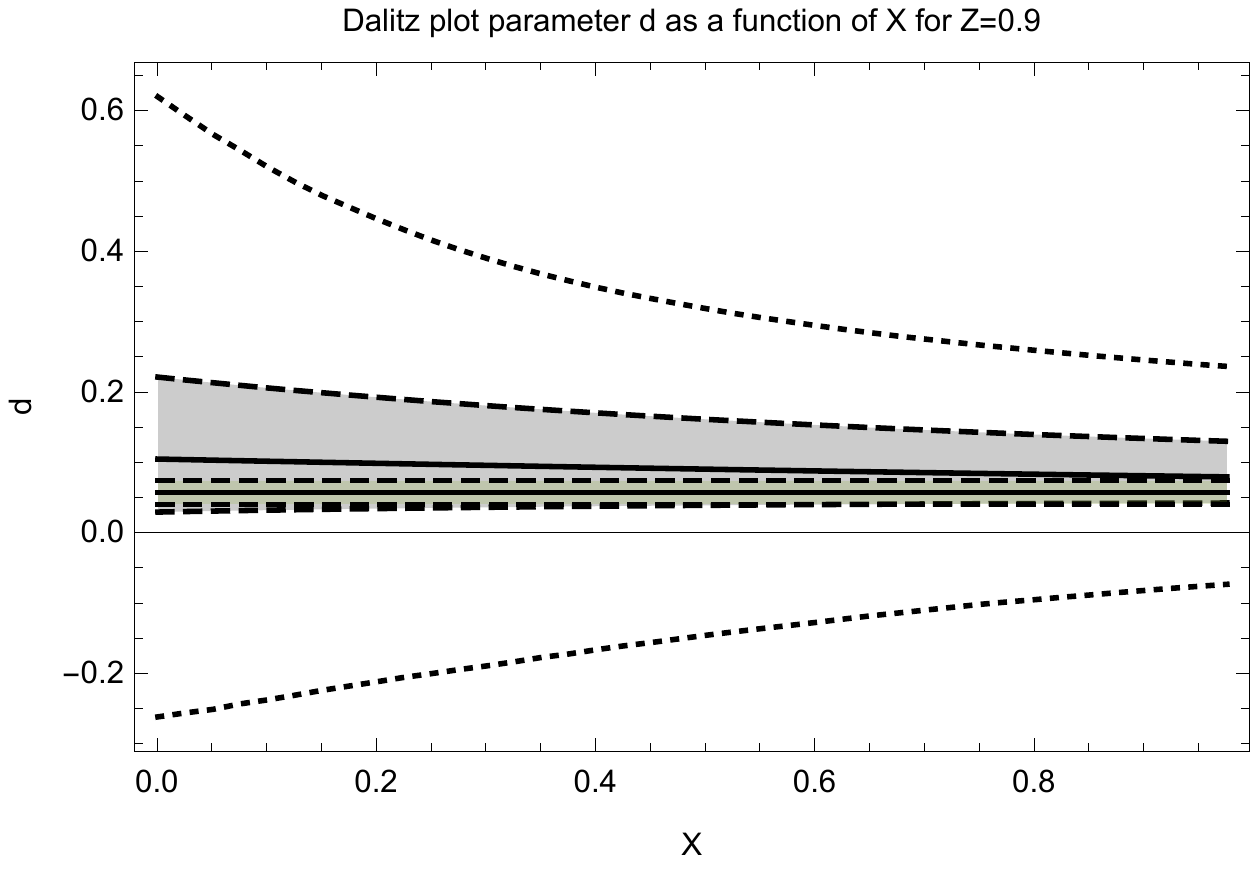}
\end{center}
\caption{The parameters $a$, $b$ and $d$ as a function of $X$ for $%
Z=0.5 $ and $0.9$. The median (solid line), the one-sigma band (dashed,
shadowed) and two-sigma band (dotted) are depicted together with the
experimental value (solid horizontal line with dashed error band).}
\label{f1}
\end{figure}

The results collected in table \ref{XZr} show values which are consistent with the NNLO $\chi PT$ predictions \cite{Bijnens:2007pr} and with each other as well.  However, while the parameters $a$ and $d$ are also compatible  with the experimental data, the predictions for the parameters $b$ and $\alpha $ do not encompass the experimental values  within the  one-sigma uncertainty band. For the parameter $\alpha$, we reproduce the positive sign of the standard $\chi PT$ prediction \cite{Bijnens:2007pr} and the apparent disagreement with the experimental value.  We might thus ask a question, whether the suggested tension really implies an incompatibility of the prediction with the experimental data at the indicated confidence level, and if yes, what is the reason for it. One possible explanation could then be that the assumed values of $X$ and 
$Z$ are not compatible with experiment, another that the assumption about
the distributions of the remainders is not adequate and the bare chiral
expansion of the apparently safe observables does not satisfy the
criteria of global convergence. 

\begin{figure}[t]
\begin{center}
\includegraphics[scale=0.6]{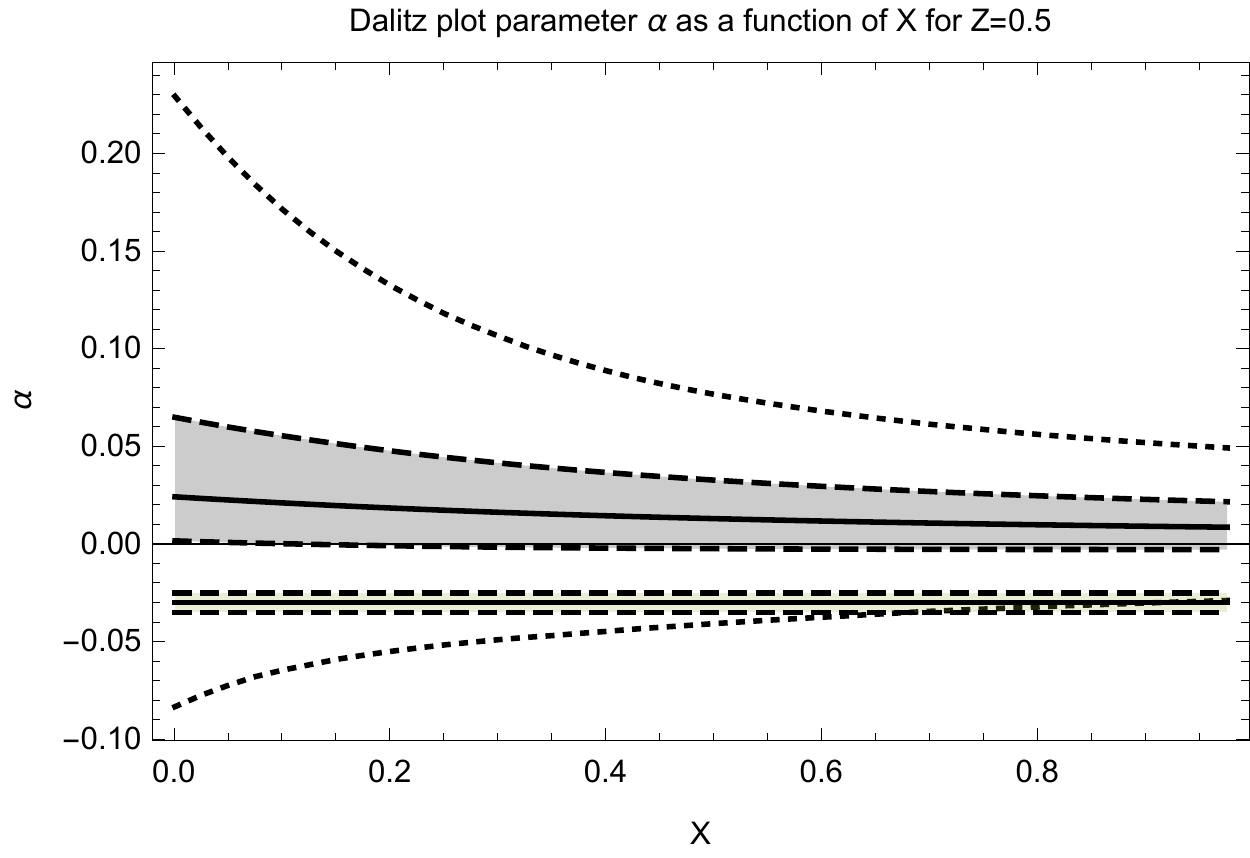}
\includegraphics[scale=0.6]{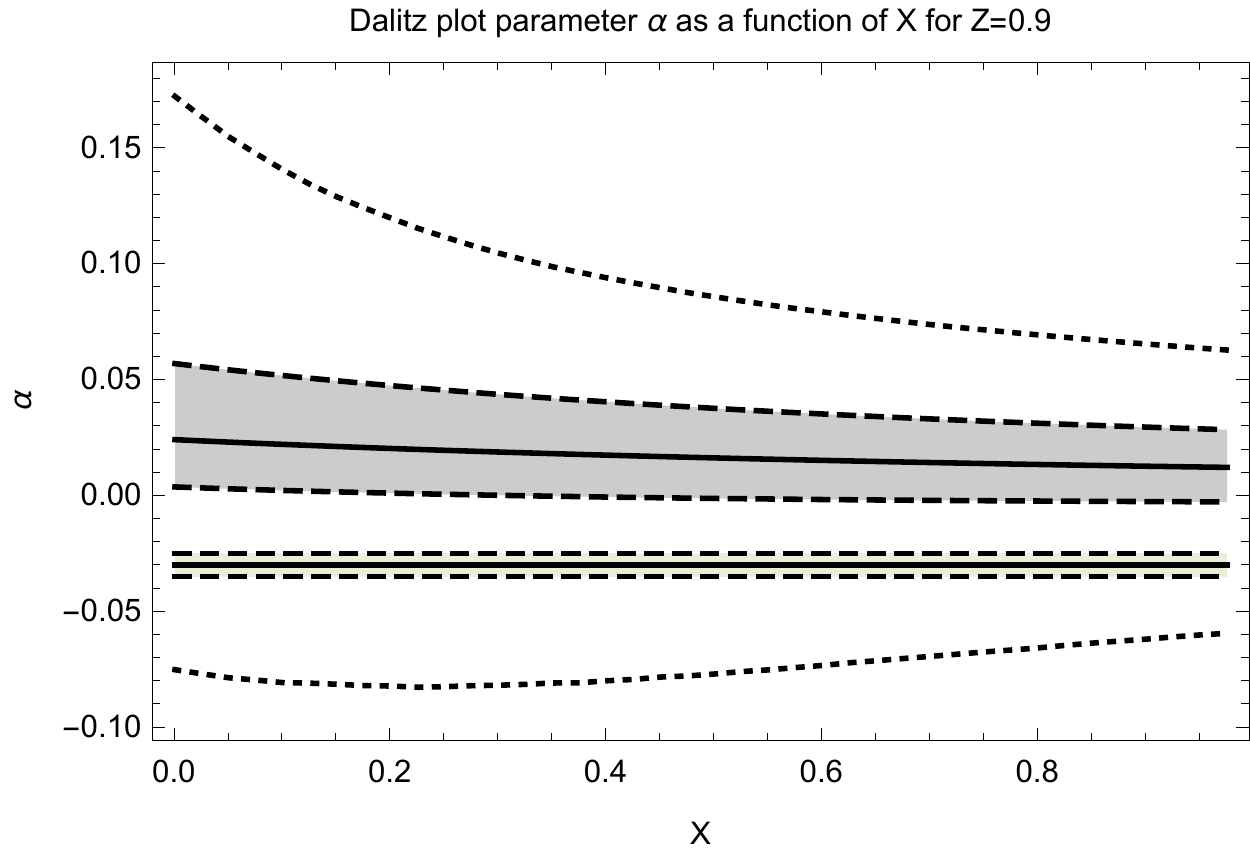}
\end{center}
\caption{The parameter $\protect\alpha$ as a function of $X$ for $%
Z=0.5 $ and $0.9$. The median (solid line), the one-sigma band (dashed,
shadowed) and two-sigma band (dotted) are depicted together with the
experimental value (solid horizontal line with dashed error band).}
\label{f2}
\end{figure}

\begin{figure}[p]
\begin{center}
\includegraphics[scale=0.6]{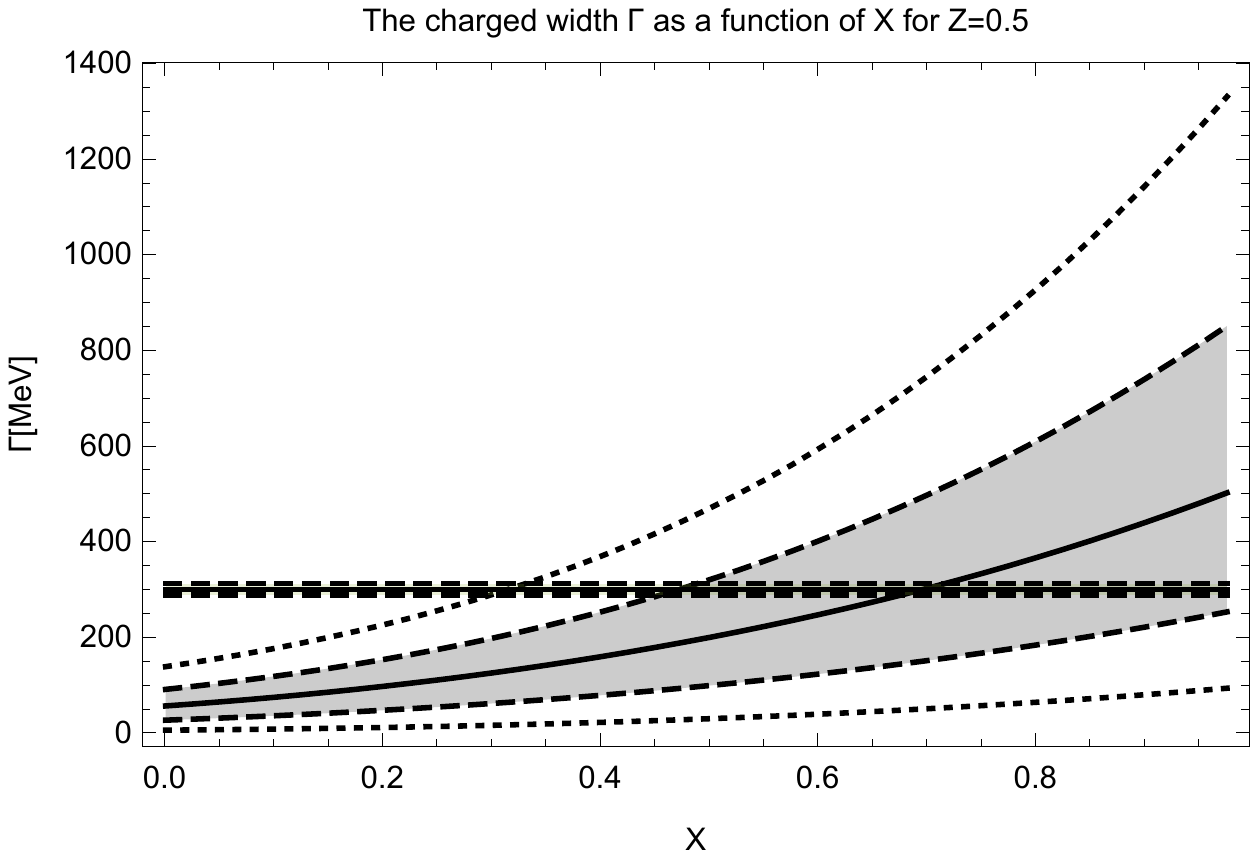} %
\includegraphics[scale=0.6]{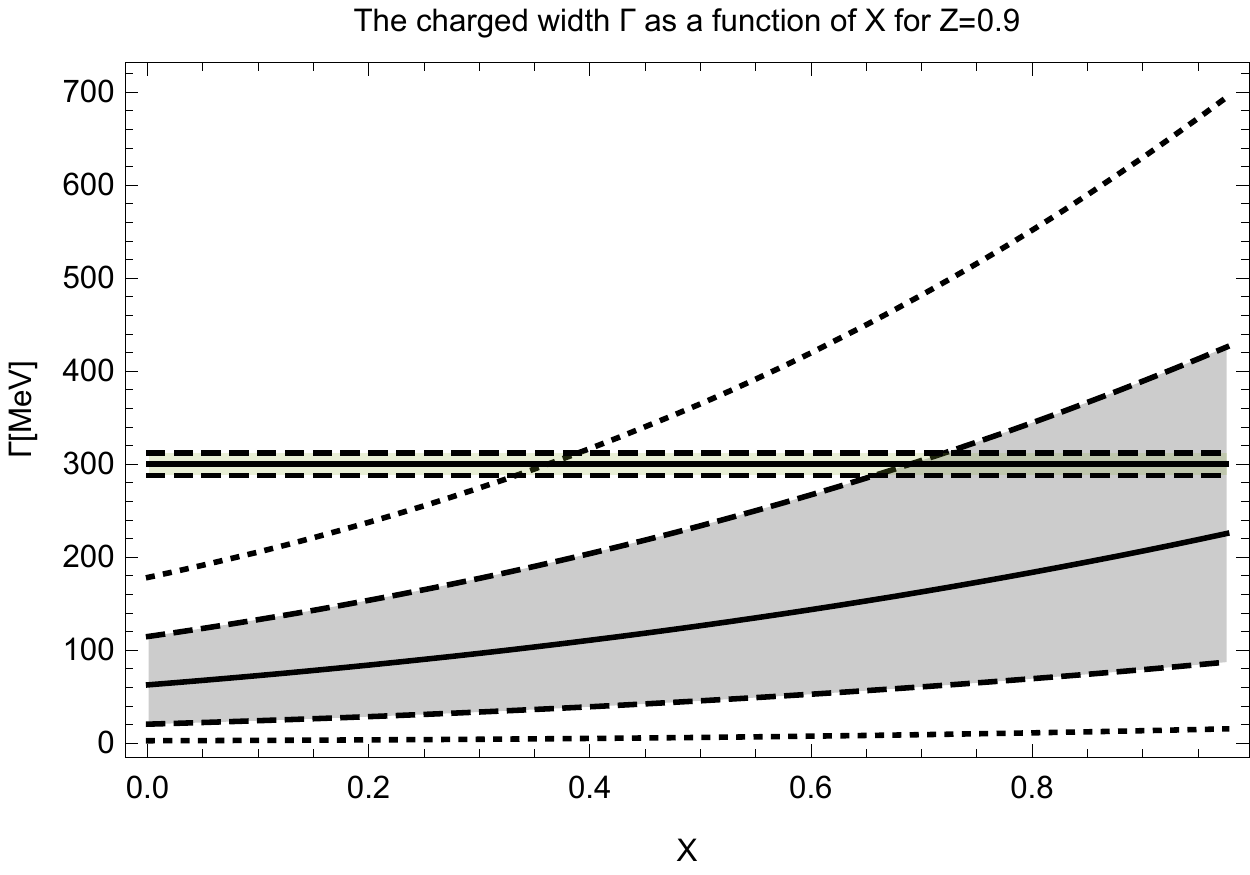} %
\includegraphics[scale=0.6]{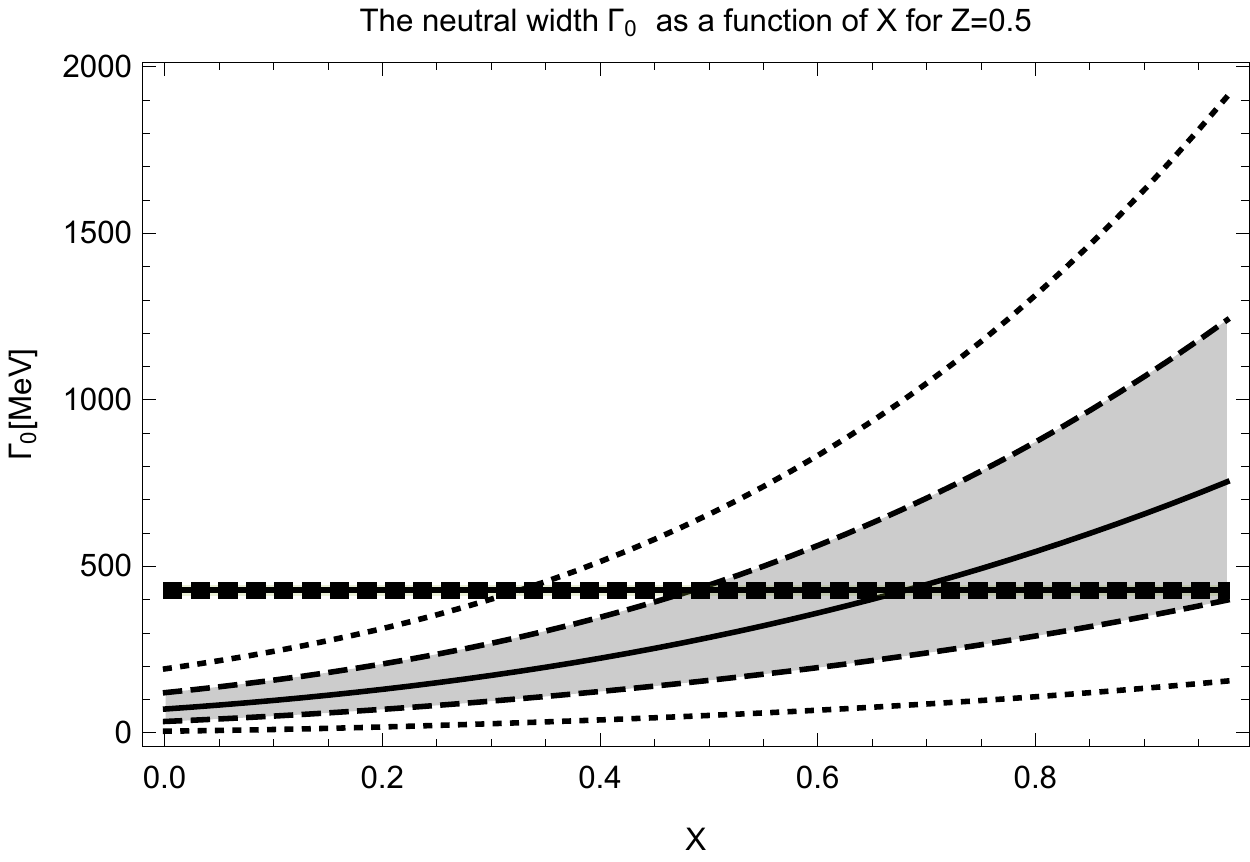} %
\includegraphics[scale=0.6]{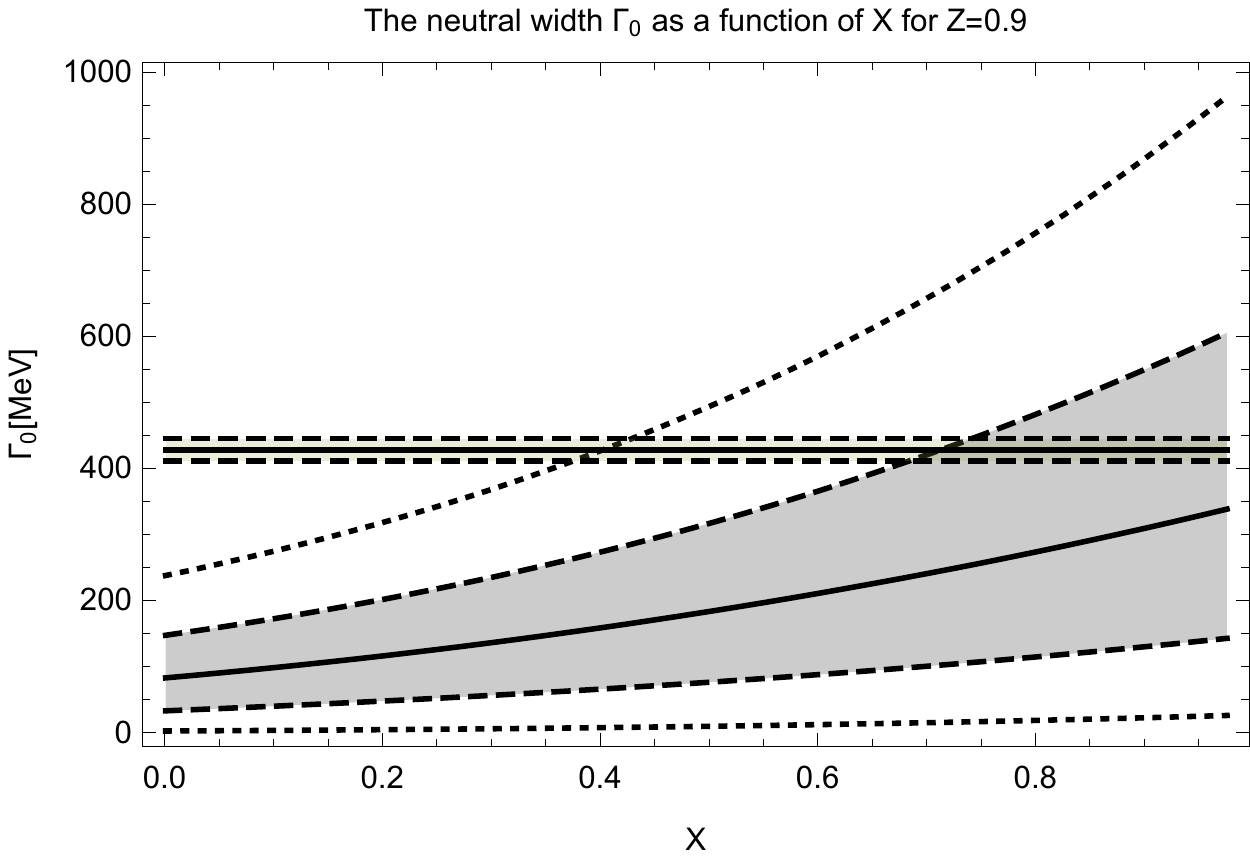} %
\includegraphics[scale=0.6]{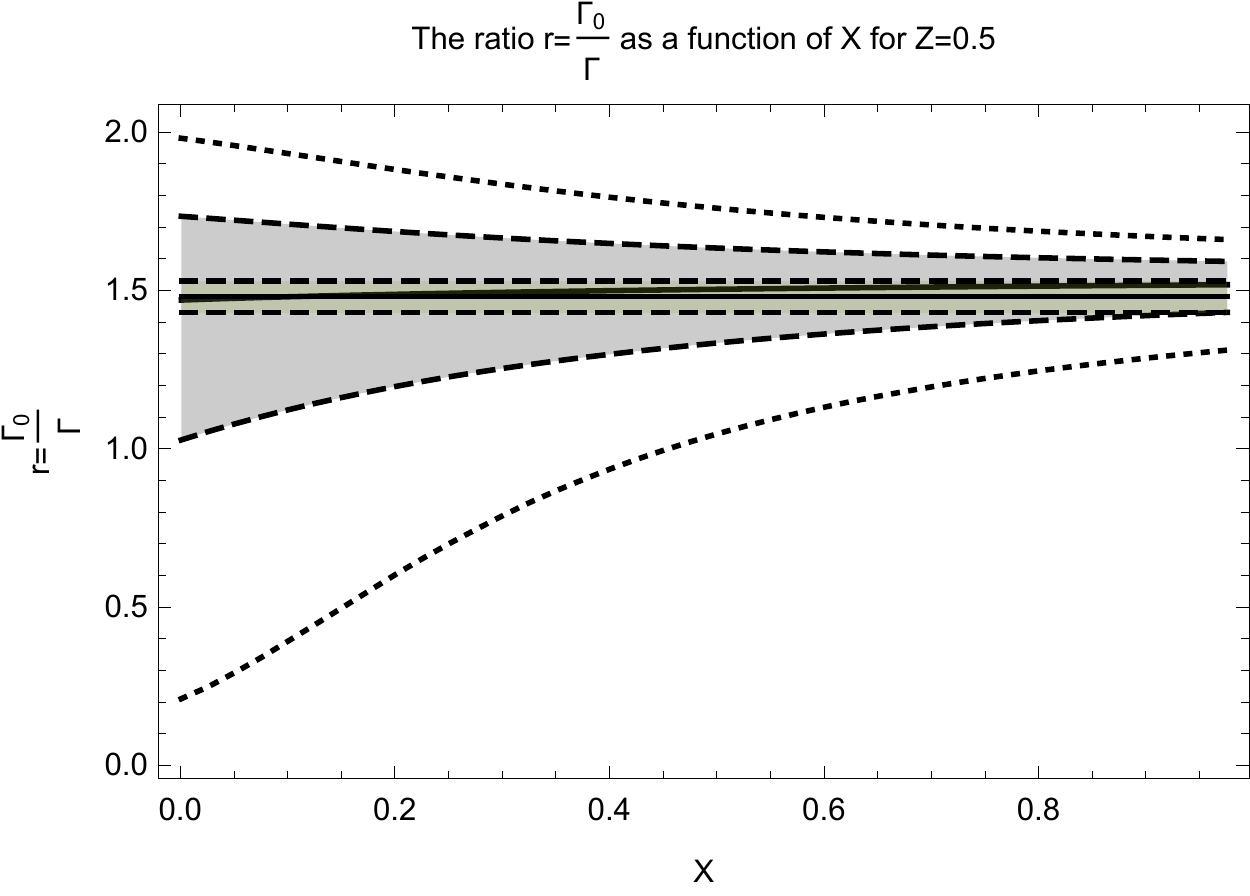} %
\includegraphics[scale=0.6]{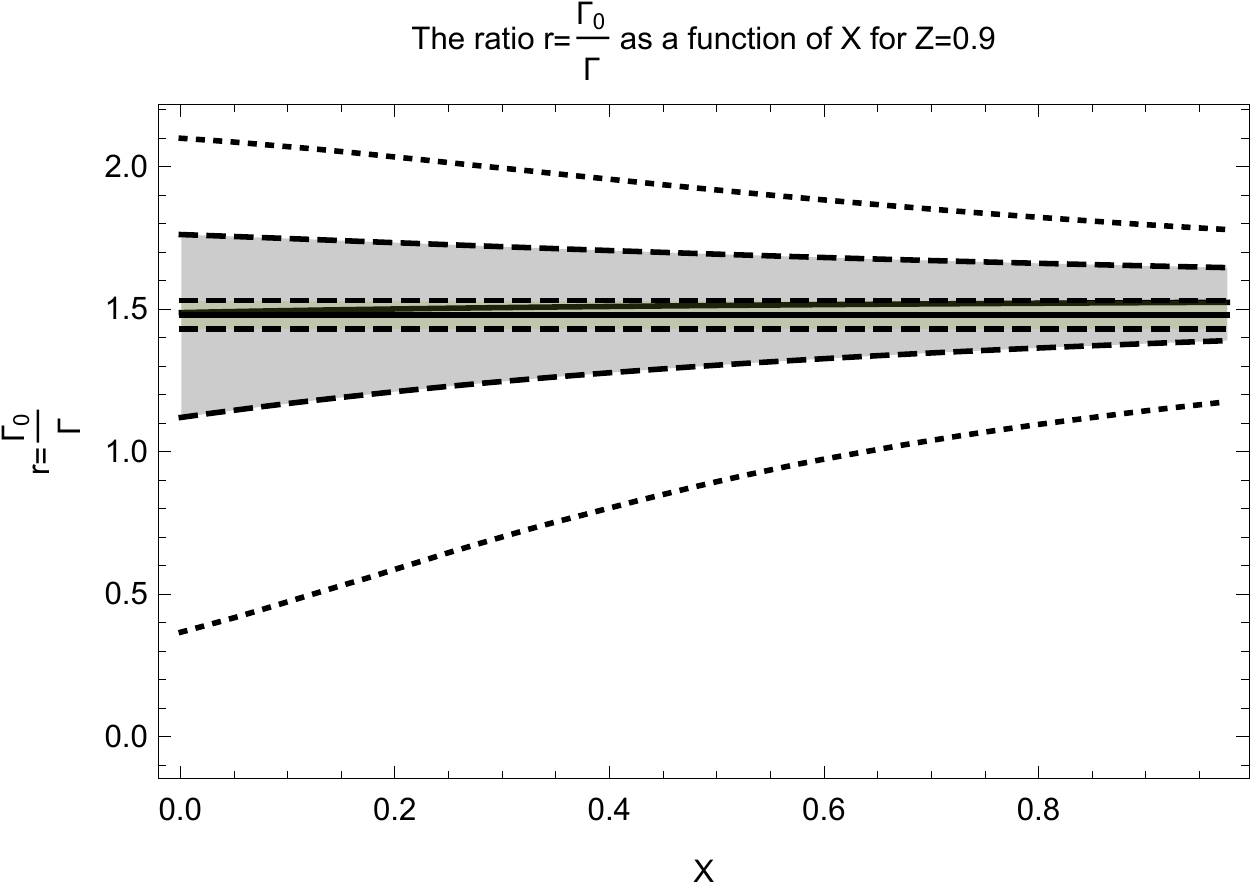}
\end{center}
\caption{The charged and neutral decay widths and their ratio as a function
of $X$ for $Z=0.2,\,0.5 $ and $0.9$. The median (solid line), the one-sigma
band (dashed, shadowed) and two-sigma band (dotted) are depicted together
with the experimental value (solid horizontal line with dashed error band).}
\label{f3}
\end{figure}

Let us therefore take a closer look at the predicted distributions, while assuming a global
convergence of the bare expansions and treating the remainders as above, and allow a variation of the parameters $X$ and $Z$ in a wider range.
Namely, we will set $Z$ according to two scenarios ($Z$=0.5 and $Z$=0.9)
and vary $X$ in the full range $0<X<1$. The results are depicted in figures \ref{f1} and \ref{f2},
where we have shown the median (solid line)
one-(dashed) and two-(dotted) sigma contours, as well as the experimental
value (solid horizontal line with dashed error band). From the figures it is
visible that the experimental value of the observable $a$ is compatible
within the one-sigma contour for almost all the range of values of $X$ and $%
Z $, the same is true for the parameter $d$ as well.

As for the Dalitz parameter $b$, its experimental value is located close but inside the two-sigma contour (fig.\ref{f1}). We could thus  conclude that we have a marginal 
compatibility. Note, however,  that the theoretical distribution is non-gaussian and  strongly constrained from below, see fig.\ref{f0}. Hence the one- and two-sigma contours are very close to each other and it is therefore difficult to make a definite statements on the compatibility of the theory and experiment.

Concerning the neutral
decay parameter $\alpha $, the dependence of the median on the parameters of $X$ and $Z$ is relatively mild (fig.\ref{f2}). The theoretical  distribution is non-gaussian again, with a long tail, as can be seen in fig.\ref{f0} as well. The experimental value lies inside the two-sigma contour in most of the range of $X$ and $Z$ (with an exception of very low values of $Z$, not shown here), but always very far from the one-sigma one. Note that by assuming a gaussian distribution with the same one sigma contour, one would be tempted to conclude  that the two-sigma contour was much more narrow and that the experimental value were clearly incompatible, as was our preliminary result in \cite{Kolesar:2011wn}.

Let us now investigate the qualitative predictions of resummed $\chi PT$ for
the charged and neutral decay widths, $\Gamma^+$ and $\Gamma^{0}$,
respectively. These are $R$ dependent observables, in contrast to the 
Dalitz plot parameters discussed above. We use the lattice average value for $R$ (\ref{R_lattice1}),
with the corresponding error bar as a further source of the uncertainty of
the prediction. The results are depicted in fig. \ref{f3}. As can be seen, the obtained
distributions of both widths are strongly $X$ and $Z$
dependent. For a relatively large range of the parameters, we observe
good compatibility with the experimental values, while other regions can be excluded at 2$\sigma$ C.L.
Of course, changing the value of $R$, which is in
principle a free parameter in the framework of resummed $\chi PT$, might
modify the details of this picture. Qualitatively, however, we expect a similar
behavior, as $R$ is present only through an overall normalization factor $1/R$ in the amplitude. 
The sensitivity of the observables $\Gamma^+$ and $\Gamma^{0}$ on
the chiral symmetry breaking parameters $X$ and $Z$ and the existence of
both compatibility and incompatibility regions seems to be promising for
a more in-depth analysis of the parametric space of resummed $\chi
PT$ with the aim of extracting the values of $X$ and $Z$. This issue we be discussed in a separate paper \cite{Kolesar:prep}, preliminary results are already available \cite{Kolesar:2014zra}. 

Overall, we can conclude that there is no indication that the apparently terrible convergence of the decay widths, as discussed in the Introduction, imply a violation of the assumption of the global convergence of the chiral series and a large value of some of the higher order remainders.

For completeness, we have also depicted the $R$ independent
ratio $r_\Gamma=\Gamma^{0}/\Gamma^+$, see fig \ref{f3}. The prediction is compatible with the experimental value $r_\Gamma=1.48\pm 0.05$ (PDG average \cite{PDG:2014kda}) in the whole region of $X$ and $Z$.

\begin{figure}[t]
\begin{center}
\includegraphics[scale=0.6]{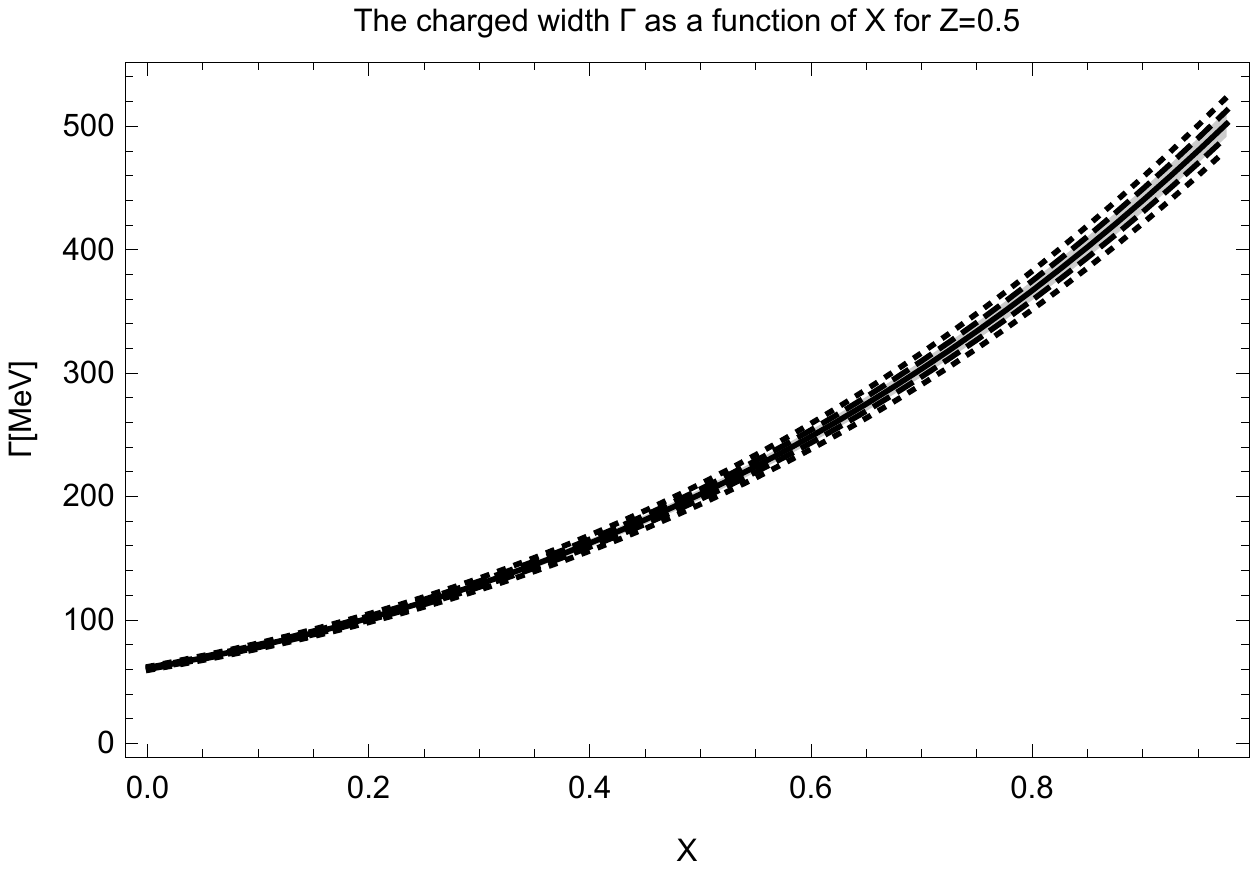}
\includegraphics[scale=0.6]{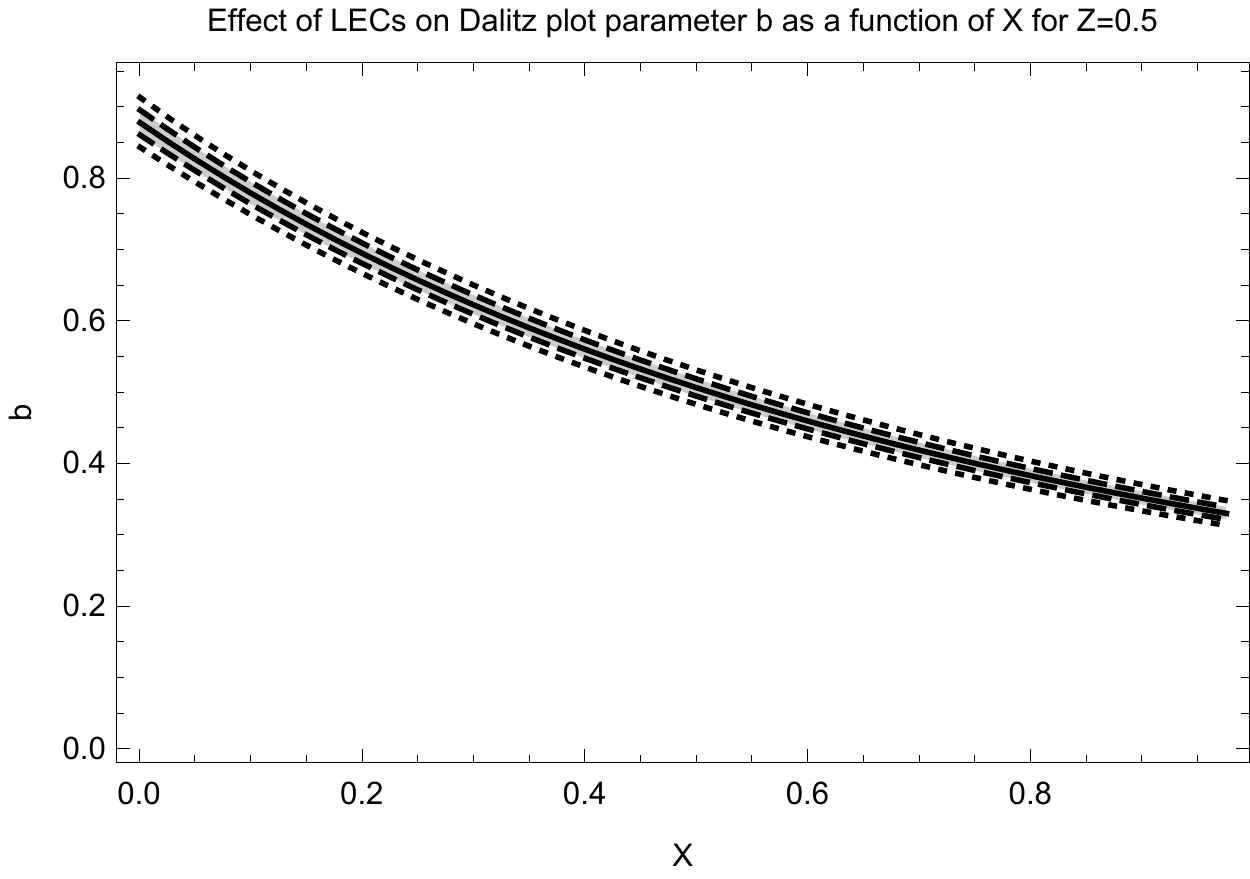} 
\end{center}
\caption{The effect of the uncertainty of the LECs on the charged width $\Gamma^+$ and on the parameter $b$ and and  as a function of $X$ for $Z=0.5$. The median (solid line), the one-sigma band (dashed,
shadowed) and two-sigma band (dotted) are depicted.}
\label{f4}
\end{figure}

Finally, let's have a look on the issue of the dependence of the results on the uncertainty stemming from the weak knowledge of the LECs $L_1$-$L_3$. We can put all the other free parameters to their mean values and leave 
only the estimated uncertainty (\ref{L123}) at play. We have found the resulting errors to be negligible in all cases, as illustrated on two examples in fig.\ref{f4}.

\section{Summary and outlook \label{discussion_section}}

The main purpose of this paper was an application of
the formalism of resummed $\chi PT$ to the $\eta$$\,\to$$3\pi$ decays 
and addressing questions concerning convergence properties of various observables related to these
decays. 

As we have explained in detail, the standard assumption on the convergence
of the chiral expansion has to be taken with some care and not all of the
observables can be trusted to be automatically well convergent. The working
hypothesis of the resummed approach is that only a limited set of safe observables has the property of global convergence, i.e. that the NNLO remainders are of a natural order of magnitude. Observables derived from the safe ones by means of nonlinear relations do not in general satisfy the criteria for global convergence due
to the possible irregularities of the chiral series. Therefore, it is
necessary to express such dangerous observables in terms of the safe ones in a non-perturbative way. This can be understood as a general procedure in cases when one encounters an expansion with significant irregularities.
Also, one has to keep the higher order remainders explicit and not neglect them. In this paper, we have
treated them as a source of theoretical uncertainty of the predictions.

As for the observables, we have concentrated on the Dalitz plot parameters $a$, $b$, $d$ of the charged channel, the parameter $\alpha$ of the neutral mode and on the
decay widths of both channels. All these observables are dangerous in the above
sense. Our results depend, besides the higher order remainders, on several free parameters - the chiral condensate, the chiral decay constant, the strange quark mass and the difference of light quark masses. These are expressed in terms of convenient parameters: $X$, $Z$, $r$ and $R$, respectively. The quark mass parameters have been fixed from lattice QCD averages \cite{Aoki:2013ldr}. There is also a residual dependence on NLO LECs $L_1$-$L_3$, which we have shown to be very mild.

We have treated the uncertainties in the higher order remainders and other parameters statistically and numerically generated a large range of theoretical predictions, which have been then confronted with experimental information. Let us stress that at this point our goal is not to provide sharp predictions, as the theoretical uncertainties are large. Nevertheless, in this form, the approach is suitable for addressing questions which might be difficult to ask within the standard framework.

In the case of the decay widths, the experimental values can be reconstructed for a reasonable range of the free parameters and thus no tension is observed, in spite of what some of the traditional calculations suggest \cite{Osborn:1970nn,Gasser:1984pr}. We have found a strong dependence of the widths on $X$ and $Z$ and an appearance of both compatibility ($<1\sigma$ C.L.) and incompatibility ($>2\sigma$ C.L.) regions. Such a behavior is not necessarily in contradiction with the global convergence assumption and,
moreover, it might be promising for constraining the parameter space and an investigation of possible scenarios of the chiral symmetry breaking \cite{Kolesar:prep, Kolesar:2014zra}.

As for the Dalitz plot parameters, $a$ and $d$ can be described very well too, within $1\sigma$ C.L. However, when $b$ and $\alpha$ are concerned, we find a mild tension for the whole range of the free parameters, at less than 2$\sigma$ C.L. This marginal compatibility is not entirely unexpected. In the case of derivative parameters, obtained by expanding the amplitude in a specific kinematic point, in our case the center of the Dalitz plot, and depending on NLO quantities, the global convergence assumption is questionable, as discussed in section \ref{strictexp_section}. Also, the distribution of the theoretical uncertainties is found to be significantly non-gaussian, so the consistency cannot be simply judged by the 1$\sigma$ error bars.

This paper constitutes the first stage of our effort to gain information from the $\eta$$\,\to$$3\pi$ decays. One application is the extraction of the parameters $X$ and $Z$ - the chiral condensate and the chiral decay constant. The theory seems to work well for the decay widths and the Dalitz plot parameter $a$ and thus it seems to be safe to use them for further analysis, which is under preparation \cite{Kolesar:prep,Kolesar:2014zra}. Due to theoretical considerations mentioned above, one should be a bit more careful with regard to the parameter $d$, although it has been reconstructed just fine in this work.

The  marginal compatibility in the case of the parameters  $b$ and $\alpha$ can be interpreted in two ways - either some of the higher order corrections are indeed unexpectedly large or there is a specific configuration of the remainders, which is, however, not completely improbable.
 This warrants a further investigation of the higher order remainders by including additional information. Work is under way in analyzing $\pi\pi$ rescattering effects and resonance contributions, some preliminary results can be found in \cite{Kolesar:2011wn}.
\\ \\ \\
{\bf{Acknowledgments:}}\\ \\
 We would like to thank Sebastien Descotes-Genon and Marc Knecht for valuable discussions at early stages of this project. 
 This work was supported by the Czech Science Foundation (grant no. GACR 15-18080S).

\appendix

\section{Explicit form of the strict expansion of $G(s,t;u)$ \label{strict_expansion_appendix}}

In this appendix, we give a summary of formulae for various contributions to
the strict chiral expansion of the amplitude $G(s,t;u)$ and to the mixing
parameters $Z_{38}$ and $\mathcal{M}_{38}$. We write the amplitude $G(s,t;u)$
in the form 
\begin{equation}
G(s,t;u)=Z_{83+-}(s,t;u)-\varepsilon _{\pi }Z_{33+-}(s,t;u)+\varepsilon
_{\eta }Z_{88+-}(s,t;u)
\end{equation}%
and split the expansion of $Z_{ab+-}$ up to $O(p^{4})$ according to 
\begin{equation}
Z_{ab+-}=Z_{ab+-}^{(2)}+Z_{ab+-,\,\mathrm{ct}}^{(4)}+Z_{ab+-,\,\mathrm{tad}%
}^{(4)}+Z_{ab+-,\,\mathrm{unit}}^{(4)}+\Delta_{Z_{ab+-}},
\end{equation}%
where the individual terms denote the $O(p^{2})$, $O(p^{4})$ counterterms,
the tadpoles, the unitary contributions and the $O(p^{6})$ remainder, respectively.

\subsection{$O(p^{2})$ contribution $Z_{ab+-}^{(2)}$}

\begin{eqnarray}
Z_{83+-}^{(2)}(s,t;u) &=&-\frac{F_{0}^{2}\overset{o}{m}_{\pi }^{2}(r-1)}{6%
\sqrt{3}R} \\
Z_{33+-}^{(2)}(s,t;u) &=&\frac{F_{0}^{2}}{3}\left( 3(s-s_{0})+\overset{o}{m}%
_{\pi }^{2}\right) \\
Z_{88+-}^{(2)}(s,t;u) &=&\frac{1}{3}F_{0}^{2}\overset{o}{m}_{\pi }^{2}
\end{eqnarray}

\subsection{$O(p^{4})$ counterterm contributions $Z_{ab+-,\,\mathrm{ct}%
}^{(4)}$}

{\normalsize 
\begin{eqnarray}
Z_{83+-,\mathrm{ct}}^{(4)}(s,t;u) &=&-\frac{2\overset{o}{m}_{\pi }^{2}(r-1)}{%
3\sqrt{3}R}[6L_{4}(s-2M_{\pi }^{2})+3L_{5}(s-2s_{0})  \notag \\
&&+4\overset{o}{m}_{\pi }^{2}(L_{6}(r+8)-2L_{7}(r-4)+8L_{8})] \\
&&  \notag \\
Z_{33+-,\mathrm{ct}}^{(4)}(s,t;u) &=&4(2L_{1}+L_{3})(s-M_{\pi }^{2}-M_{\eta
}^{2})(s-2M_{\pi }^{2})  \notag \\
&&+4L_{2}\left[ (t-M_{\pi }^{2}-M_{\eta }^{2})(t-2M_{\pi }^{2})\right. 
\notag \\
&&\left. +(u-M_{\pi }^{2}-M_{\eta }^{2})(u-2M_{\pi }^{2})\right]  \notag \\
&&+8L_{4}\overset{o}{m}_{\pi }^{2}\left[ s(r+4)-s_{0}(r+5)\right]  \notag \\
&&+4L_{5}\overset{o}{m}_{\pi }^{2}(4s-5s_{0})  \notag \\
&&+\frac{16}{3}\overset{o}{m}_{\pi }^{4}\left[ L_{6}(r+8)+4L_{8}\right] \\
&&  \notag \\
Z_{88+-,\mathrm{ct}}^{(4)}(s,t;u) &=&\frac{4}{3}(6L_{1}+L_{3})(s-M_{\pi
}^{2}-M_{\eta }^{2})(s-2M_{\pi }^{2})  \notag \\
&&+\frac{4}{3}(3L_{2}+L_{3})\left[ (t-M_{\pi }^{2}-M_{\eta }^{2})(t-2M_{\pi
}^{2})\right.  \notag \\
&&\left. +(u-M_{\pi }^{2}-M_{\eta }^{2})(u-2M_{\pi }^{2})\right]  \notag \\
&&+\frac{8}{3}L_{4}\overset{o}{m}_{\pi }^{2}\left[ (s-2M_{\pi
}^{2})(2r+4)-3\Delta _{\pi \eta }\right]  \notag \\
&&-4L_{5}\overset{o}{m}_{\pi }^{2}s_{0}  \notag \\
&&+\frac{16}{3}\overset{o}{m}_{\pi }^{4}\left[ L_{6}(5r+4)-4L_{7}(r-1)+4L_{8}%
\right]
\end{eqnarray}%
}

\subsection{$O(p^{4})$ tadpole contributions $Z_{ab+-,\,\mathrm{tad}}^{(4)}$}

\begin{eqnarray}
Z_{83+-,\mathrm{tad}}^{(4)} &=&\frac{F_{0}^{2}\overset{o}{m}_{\pi }^{2}}{6%
\sqrt{3}R}\left[ \mu _{\pi }(3r-4)+2\mu _{K}(r-1)\right.  \notag \\
&&\left. +\frac{1}{3}\mu _{\eta }(r+2)+2R\Delta \mu _{K}\right]  \notag \\
&&+\frac{\sqrt{3}}{4}F_{0}^{2}(s-s_{0})\Delta \mu _{K} \\
&&  \notag \\
Z_{33+-,\mathrm{tad}}^{(4)} &=&-\frac{F_{0}^{2}\overset{o}{m}_{\pi }^{2}}{3}%
\left( 3\mu _{\pi }+2\mu _{K}+\frac{1}{3}\mu _{\eta }\right)  \notag \\
&&-7F_{0}^{2}(s-s_{0})\left( \mu _{\pi }+\frac{1}{2}\mu _{K}\right) \\
&&  \notag \\
Z_{88+-,\mathrm{tad}}^{(4)} &=&-\frac{F_{0}^{2}\overset{o}{m}_{\pi }^{2}}{3}%
\left[ 3\mu _{\pi }+2\mu _{K}+\frac{1}{3}\mu _{\eta }\right]
\end{eqnarray}%
We denote 
\begin{eqnarray}
\mu _{P} &=&\frac{\overset{o}{m}_{P}^{2}}{32\pi ^{2}F_{0}^{2}}\ln \frac{%
\overset{o}{m}_{P}^{2}}{\mu ^{2}}=-\frac{\overset{o}{m}_{P}^{2}}{2F_{0}^{2}}%
\left( J_{PP}^{r}(0)+\frac{1}{16\pi ^{2}}\right) \\
\Delta \mu _{K} &=&\mu _{K^{0}}-\mu _{K^{+}}=\frac{1}{32\pi ^{2}F_{0}^{2}}%
\Delta \overset{o}{m}_{K}^{2}\left( \ln \frac{\overset{o}{m}_{K}^{2}}{\mu
^{2}}+1\right) =-\frac{1}{2F_{0}^{2}}\Delta \overset{o}{m}%
_{K}^{2}J_{KK}^{r}(0) \\
\Delta \overset{o}{m}_{K}^{2} &=&\Delta mB_{0}=\frac{B_{0}\widehat{m}(r-1)}{R%
} \\
\Delta M_{K}^{2} &=&M_{K^{0}}^{2}-M_{K^{+}}^{2} \\
\Delta _{\pi \eta } &=&M_{\eta }^{2}-M_{\pi }^{2} \\
\Sigma _{\pi \eta } &=&M_{\eta }^{2}+M_{\pi }^{2}
\end{eqnarray}

\subsection{$O(p^{4})$ unitarity contributions $Z_{ab+-,\,\mathrm{unit}%
}^{(4)}$}

\begin{eqnarray}
Z_{83+-,\mathrm{unit}}^{(4)}(s,t;u) &=&-\frac{1}{4\sqrt{3}R}\overset{o}{m}%
_{\pi }^{2}\left[ (s-2M_{\pi }^{2})(2r-3)+\overset{o}{m}_{\pi }^{2}(3r-4)%
\right] J_{\pi \pi }^{r}(s)  \notag \\
&&-\frac{1}{8\sqrt{3}R}\overset{o}{m}_{\pi }^{2}(r-1)\left[ s-2M_{\pi }^{2}+2%
\overset{o}{m}_{\pi }^{2}\right] J_{KK}^{r}(s)  \notag \\
&&-\frac{\sqrt{3}}{16}\left[ s-2M_{\pi }^{2}+2\overset{o}{m}_{\pi }^{2}%
\right] \left[ s-\Sigma _{\pi \eta }+\frac{2}{3}\overset{o}{m}_{\pi }^{2}%
\right] \Delta J_{KK}^{r}(s)  \notag \\
&&-\frac{(r+2)}{36\sqrt{3}R}\overset{o}{m}_{\pi }^{4}J_{\eta \eta }^{r}(s) 
\notag \\
&&-\frac{1}{4\sqrt{3}R}\overset{o}{m}_{\pi }^{2}\left[ s-2M_{\pi }^{2}+\frac{%
2}{3}\overset{o}{m}_{\pi }^{2}\right] J_{\pi \eta }^{r}(s)  \notag \\
&&-\frac{1}{8\sqrt{3}R}\overset{o}{m}_{\pi }^{2}\left[ t-2M_{\pi }^{2}\right]
J_{\pi \pi }^{r}(t)  \notag \\
&&-\frac{1}{8\sqrt{3}R}\overset{o}{m}_{\pi }^{2}\left[ \frac{2}{3}\overset{o}%
{m}_{\pi }^{2}(r-1)-(t-2M_{\pi }^{2})\right] J_{\pi \eta }^{r}(t)  \notag \\
&&-\frac{\sqrt{3}}{16R}\overset{o}{m}_{\pi }^{2}(r-1)\left[ \frac{2}{3}%
\overset{o}{m}_{\pi }^{2}+t-M_{\pi }^{2}-M_{\eta }^{2}\right] J_{KK}^{r}(t) 
\notag \\
&&-\frac{1}{8\sqrt{3}R}\overset{o}{m}_{\pi }^{2}\left[ u-2M_{\pi }^{2}\right]
J_{\pi \pi }^{r}(u)  \notag \\
&&-\frac{1}{8\sqrt{3}R}\overset{o}{m}_{\pi }^{2}\left[ \frac{2}{3}\overset{o}%
{m}_{\pi }^{2}(r-1)-(u-2M_{\pi }^{2})\right] J_{\pi \eta }^{r}(u)  \notag \\
&&-\frac{\sqrt{3}}{16R}\overset{o}{m}_{\pi }^{2}(r-1)\left[ \frac{2}{3}%
\overset{o}{m}_{\pi }^{2}+u-M_{\pi }^{2}-M_{\eta }^{2}\right] J_{KK}^{r}(u)
\end{eqnarray}%
where 
\begin{eqnarray}
\Delta J_{KK}^{r}(s) &=&J_{K^{0}}^{r}(s)-J_{K^{+}}^{r}(s)  \notag \\
&=&-2\Delta \overset{o}{m}_{K}^{2}\frac{J_{K}^{r}(s)-J_{K}^{r}(4\overset{o}{m%
}_{K}^{2})}{s-4\overset{o}{m}_{K}^{2}}+O\left( \frac{1}{R^{2}}\right)
\end{eqnarray}%
\begin{eqnarray}
Z_{33+-,\mathrm{unit}}^{(4)}(s,t;u) &=&\frac{1}{2}\left[ s-2M_{\pi }^{2}+%
\overset{o}{m}_{\pi }^{2}\right] \left[ s-2M_{\pi }^{2}+3\overset{o}{m}_{\pi
}^{2}\right] J_{\pi \pi }^{r}(s)  \notag \\
&&-\frac{1}{2}\Delta _{\pi \eta }\left[ s-2M_{\pi }^{2}+2\overset{o}{m}_{\pi
}^{2}\right] J_{\pi \pi }^{r}(s)  \notag \\
&&+\frac{1}{8}\left[ s-2M_{\pi }^{2}+2\overset{o}{m}_{\pi }^{2}\right]
^{2}J_{KK}^{r}(s)  \notag \\
&&-\frac{1}{8}\Delta _{\pi \eta }\left[ s-2M_{\pi }^{2}+2\overset{o}{m}_{\pi
}^{2}\right] J_{KK}^{r}(s)  \notag \\
&&+\frac{1}{18}\overset{o}{m}_{\pi }^{4}J_{\eta \eta }^{r}(s)  \notag \\
&&+\frac{1}{12}\left[ (s-u)(t-4\overset{o}{m}_{\pi }^{2})+3(t-2M_{\pi
}^{2})^{2}\right] J_{\pi \pi }^{r}(t)  \notag \\
&&-\frac{1}{4}\Delta _{\pi \eta }\left[ t-2M_{\pi }^{2}\right] J_{\pi \pi
}^{r}(t)  \notag \\
&&+\frac{1}{288\pi ^{2}}(s-u)(t-6\overset{o}{m}_{\pi }^{2})+\frac{1}{3}%
(s-u)F_{0}^{2}\mu _{\pi }  \notag \\
&&+\frac{1}{24}(s-u)\left[ t-4\overset{o}{m}_{K}^{2}\right] J_{KK}^{r}(t) 
\notag \\
&&+\frac{1}{576\pi ^{2}}(s-u)\left[ t-6\overset{o}{m}_{K}^{2}\right] +\frac{1%
}{6}(s-u)F_{0}^{2}\mu _{K}  \notag \\
&&+\frac{1}{12}\left[ (s-t)(u-4\overset{o}{m}_{\pi }^{2})+3(u-2M_{\pi
}^{2})^{2}\right] J_{\pi \pi }^{r}(u)  \notag \\
&&-\frac{1}{4}\Delta _{\pi \eta }\left[ u-2M_{\pi }^{2}\right] J_{\pi \pi
}^{r}(u)  \notag \\
&&+\frac{1}{288\pi ^{2}}(s-t)(u-6\overset{o}{m}_{\pi }^{2})+\frac{1}{3}%
(s-t)F_{0}^{2}\mu _{\pi }  \notag \\
&&+\frac{1}{24}(s-t)\left[ u-4\overset{o}{m}_{K}^{2}\right] J_{KK}^{r}(u) 
\notag \\
&&+\frac{1}{576\pi ^{2}}(s-t)\left[ u-6\overset{o}{m}_{K}^{2}\right] +\frac{1%
}{6}(s-t)F_{0}^{2}\mu _{K}
\end{eqnarray}%
\begin{eqnarray}
Z_{88+-,\mathrm{unit}}^{(4)}(s,t;u) &=&\frac{1}{3}\overset{o}{m}_{\pi }^{2}%
\left[ s-2M_{\pi }^{2}+\frac{3}{2}\overset{o}{m}_{\pi }^{2}\right] J_{\pi
\pi }^{r}(s)  \notag \\
&&+\frac{1}{8}\left[ s-2M_{\pi }^{2}+2\overset{o}{m}_{\pi }^{2}\right] 
\notag \\
&&\times \left[ 3s-6M_{\eta }^{2}+4\overset{o}{m}_{\eta }^{2}-\frac{2}{3}%
\overset{o}{m}_{\pi }^{2}\right] J_{KK}^{r}(s)  \notag \\
&&+\frac{3}{8}\Delta _{\pi \eta }\left[ s-2M_{\pi }^{2}+2\overset{o}{m}_{\pi
}^{2}\right] J_{KK}^{r}(s)  \notag \\
&&+\frac{2}{9}\overset{o}{m}_{\pi }^{2}\left[ \overset{o}{m}_{\eta }^{2}-%
\frac{1}{4}\overset{o}{m}_{\pi }^{2}\right] J_{\eta \eta }^{r}(s)  \notag \\
&&+\frac{1}{9}\overset{o}{m}_{\pi }^{4}J_{\pi \eta }^{r}(t)  \notag \\
&&+\frac{3}{8}\left[ t-M_{\pi }^{2}-M_{\eta }^{2}+\frac{2}{3}\overset{o}{m}%
_{\pi }^{2}\right] ^{2}J_{KK}^{r}(t)  \notag \\
&&+\frac{3}{8}\Delta _{\pi \eta }\left[ t-M_{\pi }^{2}-M_{\eta }^{2}+\frac{2%
}{3}\overset{o}{m}_{\pi }^{2}\right] J_{KK}^{r}(t)  \notag \\
&&+\frac{1}{9}\overset{o}{m}_{\pi }^{4}J_{\pi \eta }^{r}(u)  \notag \\
&&+\frac{3}{8}\left[ u-M_{\pi }^{2}-M_{\eta }^{2}+\frac{2}{3}\overset{o}{m}%
_{\pi }^{2}\right] ^{2}J_{KK}^{r}(u)  \notag \\
&&+\frac{3}{8}\Delta _{\pi \eta }\left[ u-M_{\pi }^{2}-M_{\eta }^{2}+\frac{2%
}{3}\overset{o}{m}_{\pi }^{2}\right] J_{KK}^{r}(u)
\end{eqnarray}%
Up to now, we have kept the masses at their $O(p^{2})$ values in all the loop functions $%
J_{PQ}^{r}(s)$.

\subsection{Mixing parameters $Z_{38}$ and $\mathcal{M}_{38}$}

The strict chiral expansion of the parameters $Z_{38}$ and $M_{38}$ to 
$O(p^{4})$ reads
\begin{eqnarray}
Z_{38} &=&\frac{8(1-r)B_{0}\widehat{m}}{\sqrt{3}F_{0}^{2}R}L_{5}+\sqrt{3}%
\Delta \mu _{K} \\
\mathcal{M}_{38} &=&\frac{(1-r)B_{0}\widehat{m}}{\sqrt{3}R}+\frac{32B_{0}^{2}%
\widehat{m}^{2}(1-r)}{\sqrt{3}F_{0}^{2}R}\left[ (r+2)L_{6}-2(r-1)L_{7}+2L_{8}%
\right]  \notag \\
&&+\frac{B_{0}\widehat{m}}{\sqrt{3}R}\left[ 2\Delta \mu _{K}+\frac{1}{3}\mu
_{\eta }(r+2)+2\mu _{K}(r-1)+\mu _{\pi }(3r-4)\right]
\end{eqnarray}

\bigskip

\section{Reconstruction of the unitarity part $U(s,t;u)$\label{reconstruction_appendix}}

According to the reconstruction theorem (for more details on the general
method see \cite{Knecht:1995tr}, \cite{Zdrahal:2008bd} and for the
application to resummed $\chi PT$, see \cite{Kolesar:2008jr}), we get the following general formula
for the unitarity part $U(s,t;u)$ of the amplitude 
\begin{eqnarray}
U(s,t;u) &=&\frac{1}{3}\left( W_{0}(s)-W_{2}(s)\right)  \notag \\
&&+\frac{1}{2}\left( 3(s-u)W_{1}(t)+W_{2}(t)\right)  \notag \\
&&+\frac{1}{2}\left( 3(s-t)W_{1}(u)+W_{2}(u\right) .
\end{eqnarray}%
Here, $W_{i}(s)$ are uniquely defined up to a subtraction polynomial by
appropriately subtracted dispersion integrals with discontinuities 
\begin{eqnarray}
\mathrm{disc}W_{0}(s) &=&-32\pi \sqrt{3}\theta (s-4M_{\pi }^{2})\mathrm{disc}%
A_{0}^{0,0}(s)  \notag \\
\mathrm{disc}W_{1}(s) &=&\mp 32\pi \sqrt{3}\theta (s-4M_{\pi }^{2})\mathrm{%
disc}\frac{A_{1}^{1\pm 1}(s)}{\lambda ^{1/2}(s,M_{\pi }^{2},M_{\eta
}^{2})\sigma (s)}  \notag \\
\mathrm{disc}W_{2}(s) &=&32\pi \sqrt{\frac{3}{2}}\theta (s-4M_{\pi }^{2})%
\mathrm{disc}A_{0}^{2,0}(s).  \label{disc_W_i}
\end{eqnarray}%
In the above expressions, $A_{l}^{I,I_{3}}(s)$ corresponds to an $l-$th
partial wave amplitude in the channel $\eta \pi ^{I_{3}}\rightarrow (\pi \pi
)^{I,I_{3}}$, with fixed isospin and its third component in the final state.
For the isospin decomposition, we use the Condon-Shortley phase convention 
\begin{eqnarray}
A^{0,0}(s,t;u) &=&-\frac{1}{\sqrt{3}}[3A(s,t;u)+A(t,s;u)+A(u,t;s)] \\
A^{2,0}(s,t;u) &=&\sqrt{\frac{2}{3}}[A(t,s;u)+A(u,t;s)] \\
A^{1,\pm 1}(s,t;u) &=&\mp \frac{1}{\sqrt{2}}[A(t,s;u)-A(u,t;s)].
\end{eqnarray}%
The discontinuities of $W_{i}(s)$ (where $\ I=0,1,2$) are fixed by
unitarity. Up to kinematic factors, they correspond to two-particle
intermediate state contributions to the right hand cut discontinuities of
$A_{l}^{I,I_{3}}$ 
\begin{eqnarray}
\mathrm{disc}A_{l}^{I,I_{3}}(s) &=&\sum_{PQ}\theta (s-(M_{P}+M_{Q})^{2}) 
\notag \\
&&\times 2\left( \frac{N^{^{\prime }}N^{^{\prime \prime }}}{NS}\right) \frac{%
\lambda ^{1/2}(s,M_{P}^{2},M_{Q}^{2})}{s}A_{l}^{\eta \pi ^{I_{3}}\rightarrow
PQ}(s)A_{l}^{PQ\rightarrow (\pi \pi )^{I,I_{3}}}(s)^{\ast }.
\label{two_particle_unitarity}
\end{eqnarray}%
Here $N^{^{\prime }}$, $N^{^{\prime \prime }}$ and $N$ are normalization
factors of the expansion of the amplitudes $A^{\eta \pi ^{I_{3}}\rightarrow
PQ}$, $A^{PQ\rightarrow (\pi \pi )^{I,I_{3}}}$ and $A^{I,I_{3}}(s)$ to
the partial waves $A_{l}^{\eta \pi ^{I_{3}}\rightarrow PQ}(s)$, $%
A_{l}^{PQ\rightarrow (\pi \pi )^{I,I_{3}}}(s)$ and $A_{l}^{I,I_{3}}(s)$,
respectively. Schematically%
\begin{equation}
A(s,t;u)=32\pi N\sum_{l=0}^{\infty }\left( 2l+1\right) A_{l}\left( s\right)
P_{l}\left( \cos \theta \right) .
\end{equation}%
$S$ \ is a symmetry factor of the intermediate state $PQ$. As a result
of the reconstruction, we get $W_{i}(s)$ as a sum of the contributions of
the two-particle intermediate states in each channel 
\begin{eqnarray}
W_{0}(s) &=&W_{0}^{(4)\pi \pi }(s)+W_{0}^{(4)\eta \pi }(s)+W_{0}^{(4)\eta
\eta }(s)+W_{0}^{(4)K\overline{K}}(s) \\
W_{1}(s) &=&W_{1}^{(4)\pi \pi }(s)+W_{1}^{(4)K\overline{K}}(s) \\
W_{2}(s) &=&W_{2}^{(4)\pi \pi }(s)+W_{2}^{(4)\eta \pi }(s)+W_{0}^{(4)K%
\overline{K}}(s).
\end{eqnarray}%
For the reconstruction of the $O(p^{4})$ functions $W_{I}^{(4)PQ}(s)$, with
a help of (\ref{disc_W_i}) and (\ref{two_particle_unitarity}), we need a
complete set of coupled $O\left( p^{2}\right) $ amplitudes. The relevant $%
O\left( p^{2}\right) $ amplitudes in the $\pi \pi $, $\pi \eta $, $\eta \eta $
and $K\overline{K}$ channels, as well as the explicit form for $%
W_{I}^{(4)PQ}(s)$, are given in the following subsections.

As explained in detail in \cite{Kolesar:2008jr}, we use two
possible ways how to treat the $O\left( p^{2}\right) $ amplitudes entering
the reconstruction theorem. The reason is that there are two possibilities
how to connect the generic physical $O\left( p^{2}\right) $ amplitude $%
A^{\left( 2\right) }$ of the process $AB\rightarrow CD$ (which is a
dangerous observable) and the corresponding safe observable $G^{\left(
2\right) }$. In what follows, we give the formulae in accord with the choice%
\begin{equation}
A^{\left( 2\right) }=\frac{G^{\left( 2\right) }}{F_{A}F_{B}F_{C}F_{D}},
\end{equation}
where $F_{P}$ is the physical decay constant of the PGB $P$. The second
possibility corresponds to a replacement of $F_{P}\rightarrow F_{0}$\ in the
above formula. For this second possibility, the $W_{I}^{(4)PQ}(s)$ are easily
obtained form the results presented below by means of a substitution of $%
F_{P}\rightarrow F_{0}$ on the right hand side of the expressions for $%
F_{\pi }^{3}F_{\eta }W_{I}^{(4)PQ}(s)$.

\subsection{$\protect\pi \protect\pi $ intermediate state}

The $\eta \pi ^{I_{3}}\rightarrow (\pi \pi )^{I,I_{3}}$ partial wave
amplitudes $A_{l}^{I,I_{3}}(s)$ are 
\begin{eqnarray}
F_{\pi }^{3}F_{\eta }A_{0}^{0,0}(s) &=&\frac{F_{0}^{2}}{16\pi \sqrt{3}}\left[
\varepsilon _{\pi }(s-s_{0})+\frac{5}{18}B_{0}\widehat{m}\left( \sqrt{3}%
\frac{(r-1)}{R}+6(\varepsilon _{\pi }-\varepsilon _{\eta })\right) \right]
\label{etapi000} \\
F_{\pi }^{3}F_{\eta }A_{1}^{1\pm 1}(s) &=&\pm \frac{F_{0}^{2}}{96\pi \sqrt{2}%
}\varepsilon _{\pi }\lambda ^{1/2}(s,M_{\pi }^{2},M_{\eta }^{2})\sigma (s) \\
F_{\pi }^{3}F_{\eta }A_{0}^{2,0}(s) &=&\frac{F_{0}^{2}}{32\pi }\sqrt{\frac{2%
}{3}}\left[ \varepsilon _{\pi }(s-s_{0})-\frac{2}{9}B_{0}\widehat{m}\left( 
\sqrt{3}\frac{(r-1)}{R}+6(\varepsilon _{\pi }-\varepsilon _{\eta })\right) %
\right],
\end{eqnarray}%
while the $\left( \pi \pi \right) ^{I}\rightarrow \left( \pi \pi \right) ^{I}$
partial wave amplitudes $A_{l}^{I}(s)$ can be written as
\begin{eqnarray}
F_{\pi }^{4}A_{0}^{0}(s) &=&\frac{F_{0}^{2}}{16\pi }\left[ (s-\frac{4}{3}%
M_{\pi }^{2})+\frac{5}{6}\overset{o}{m}_{\pi }^{2}\right] \\
F_{\pi }^{4}A_{1}^{1}(s) &=&\frac{F_{0}^{2}}{96\pi }(s-\frac{4}{3}M_{\pi
}^{2}) \\
F_{\pi }^{4}A_{0}^{0}(s) &=&-\frac{F_{0}^{2}}{32\pi }\left[ (s-\frac{4}{3}%
M_{\pi }^{2})-\frac{2}{3}\overset{o}{m}_{\pi }^{2}\right].
\end{eqnarray}%
We then get by using (\ref{disc_W_i}) and (\ref{two_particle_unitarity}) 
\begin{eqnarray}
F_{\pi }^{3}F_{\eta }W_{0}^{(4)\pi \pi }(s) &=&-\frac{F_{0}^{4}}{F_{\pi }^{4}%
}2\left[ \varepsilon _{\pi }(s-s_{0})+\frac{5}{18}B_{0}\widehat{m}\left( 
\sqrt{3}\frac{(r-1)}{R}+6(\varepsilon _{\pi }-\varepsilon _{\eta })\right) %
\right]  \notag \\
&&\times \left[ (s-\frac{4}{3}M_{\pi }^{2})+\frac{5}{6}\overset{o}{m}_{\pi
}^{2}\right] \overline{J}_{\pi \pi }(s) \\
F_{\pi }^{3}F_{\eta }W_{1}^{(4)\pi \pi }(s) &=&-\frac{F_{0}^{4}}{F_{\pi }^{4}%
}\frac{1}{18}\varepsilon _{\pi }(s-4M_{\pi }^{2})\overline{J}_{\pi \pi }(s)
\\
F_{\pi }^{3}F_{\eta }W_{2}^{(4)\pi \pi }(s) &=&-\frac{F_{0}^{4}}{F_{\pi }^{4}%
}\frac{1}{2}\left[ \varepsilon _{\pi }(s-s_{0})-\frac{2}{9}B_{0}\widehat{m}%
\left( \sqrt{3}\frac{(r-1)}{R}+6(\varepsilon _{\pi }-\varepsilon _{\eta
})\right) \right]  \notag \\
&&\times \left[ (s-\frac{4}{3}M_{\pi }^{2})-\frac{2}{3}\overset{o}{m}_{\pi
}^{2}\right] \overline{J}_{\pi \pi }(s).
\end{eqnarray}

\subsection{$\protect\eta \protect\pi $ intermediate state}

For the $\eta \pi \rightarrow \eta \pi $ amplitude $A_{0}^{\eta \pi }(s)$ ($%
S-$ wave only), we get at $O(p^{2})$%
\begin{equation}
F_{\pi }^{2}F_{\eta }^{2}A_{0}^{\eta \pi }(s)=\frac{F_{0}^{2}}{32\pi }\frac{%
\overset{o}{m}_{\pi }^{2}}{3}
\end{equation}%
and therefore, according to (\ref{disc_W_i}) and (\ref{two_particle_unitarity}%
) and \ with a help of (\ref{etapi000}) 
\begin{eqnarray}
F_{\pi }^{3}F_{\eta }W_{0}^{(4)\eta \pi }(s) &=&-\frac{F_{0}^{4}}{F_{\eta
}^{2}F_{\pi }^{2}}\frac{2}{3}\overset{o}{m}_{\pi }^{2}\left[ \varepsilon
_{\pi }(s-s_{0})+\frac{5}{18}B_{0}\widehat{m}\left( \sqrt{3}\frac{(r-1)}{R}%
+6(\varepsilon _{\pi }-\varepsilon _{\eta })\right) \right] \overline{J}%
_{\eta \pi }(s) \notag \\ \\
F_{\pi }^{3}F_{\eta }W_{2}^{(4)\eta \pi }(s) &=&\frac{F_{0}^{4}}{F_{\eta
}^{2}F_{\pi }^{2}}\frac{1}{3}\overset{o}{m}_{\pi }^{2}\left[ \varepsilon
_{\pi }(s-s_{0})-\frac{2}{9}B_{0}\widehat{m}\left( \sqrt{3}\frac{(r-1)}{R}%
+6(\varepsilon _{\pi }-\varepsilon _{\eta })\right) \right] \overline{J}%
_{\eta \pi }(s). \notag \\
\end{eqnarray}

\subsection{$\protect\eta \protect\eta $ intermediate state}

For the $\eta \pi ^{0}\rightarrow \eta \eta $ amplitude $A_{0}^{\eta \pi
^{0}\rightarrow \eta \eta }(s)$ ($S-$ wave only), we have
\begin{equation}
F_{\pi }F_{\eta }^{3}A_{0}^{\eta \pi ^{0}\rightarrow \eta \eta }(s)=\frac{%
F_{0}^{2}}{32\pi }\frac{\overset{o}{m}_{\pi }^{2}}{6}\left[ -\frac{(r-1)}{%
\sqrt{3}R}+\frac{2}{3}\varepsilon _{\eta }(1+8r)-6\varepsilon _{\pi }\right]
\end{equation}%
and the $\eta \eta \rightarrow (\pi \pi )^{I,I_{3}}$amplitude $A_{0}^{\eta \eta
00}(s)$ ($S-$ wave, $(I,I_{3})=(0,0)$ only) is
\begin{equation}
F_{\pi }^{2}F_{\eta }^{2}A_{0}^{\eta \eta 00}(s)=-\frac{F_{0}^{2}}{32\pi }%
\frac{\overset{o}{m}_{\pi }^{2}}{\sqrt{3}}.
\end{equation}%
We then get 
\begin{equation}
F_{\pi }^{3}F_{\eta }W_{0}^{(4)\eta \eta }(s)=\frac{F_{0}^{4}}{F_{\eta }^{4}}%
\frac{1}{12}\overset{o}{m}_{\pi }^{4}\left[ -\frac{(r-1)}{\sqrt{3}R}+\frac{2%
}{3}\varepsilon _{\eta }(1+8r)-6\varepsilon _{\pi }\right] \overline{J}%
_{\eta \eta }(s).
\end{equation}

\subsection{$K\overline{K}$ intermediate states}

The contribution of the $K\overline{K}$ intermediate states, where 
\begin{equation}
K=\left( 
\begin{array}{l}
K^{+} \\ 
K^{0}%
\end{array}%
\right) ,\,\,\overline{K}=\left( 
\begin{array}{l}
-K^{-} \\ 
\overline{K}^{0}%
\end{array}%
\right) ,
\end{equation}%
is a little bit less transparent. The reason is that to the first order of
the isospin breaking, both amplitudes $\eta \pi \rightarrow K\overline{K}$
and $K\overline{K}\rightarrow \pi \pi $ have both $\Delta I=0$ as well as $%
\Delta I=1$ parts and also the mass difference $\Delta
M_{K}^{2}=M_{K^{0}}^{2}-M_{K^{\pm }}^{2}$, which is of the first order in
the isospin breaking, must be taken into account. Let 
\begin{equation}
\langle \overline{K}K_{out}\mid \eta \pi (I_{3})_{in}\rangle =\mathrm{i}%
(2\pi )^{4}\delta ^{(4)}(P_{f}-P_{a})A_{\overline{K}K}^{I_{3}}(s,t;u)
\end{equation}%
and 
\begin{equation}
\langle \pi \pi (I,I_{3})_{out}\mid \overline{K}K_{in}\rangle =\mathrm{i}%
(2\pi )^{4}\delta ^{(4)}(P_{f}-P_{a})A_{\overline{K}K}^{I,I_{3}}(s,t;u).
\end{equation}%
Then it follows from the isospin decomposition of the amplitudes 
\begin{eqnarray}
A_{\overline{K}^{0}K(K^{-}K^{+})}^{0}(s,t;u) &=&C^{0}(s,t;u)\pm B^{0}(s,t;u)
\notag \\
&=&16\pi \sum_{l}(2l+1)(C_{l}^{0}(s)\pm B_{l}^{0}(s))P_{l}(\cos \theta _{0}),
\\
A_{\overline{K}^{0}K^{+}(K^{-}K^{0})}^{\pm 1}(s,t;u) &=&C^{\pm 1}(s,t;u)\pm
B^{\pm 1}(s,t;u)  \notag \\
&=&16\pi \sum_{l}(2l+1)(C_{l}^{\pm 1}(s)\pm B_{l}^{\pm 1}(s))P_{l}(\cos
\theta _{1}).
\end{eqnarray}%
Here $C^{I_{3}}$ is the isospin conserving $\Delta I=0$ and $B^{I_{3}}$ is
the isospin breaking $\Delta I=1$ part of the amplitudes, and, to the first
order in the isospin breaking (i.e. up to the corrections $O((\Delta
M_{K}^{2})^{2})$),
\begin{eqnarray}
\cos \theta _{0} &=&\frac{t-u}{\lambda ^{1/2}(s,M_{\eta }^{2},M_{\pi }^{2})}%
\left( 1-\frac{4M_{K}^{2}}{s}\right) ^{-1/2}\left( 1\pm \frac{\Delta
M_{K}^{2}}{s-4M_{K}^{2}}\right) \\
\cos \theta _{1} &=&\frac{t-u+\frac{1}{s}\Delta M_{K}^{2}\Delta }{\lambda
^{1/2}(s,M_{\eta }^{2},M_{\pi }^{2})}\left( 1-\frac{4M_{K}^{2}}{s}\right)
^{-1/2}.
\end{eqnarray}%
In particular, because $C^{\pm 1}(s,t;u)=C^{\pm 1}(s,u;t)$ as a consequence of
the $C-$symmetry, we have $C_{1}^{\pm 1}(s)=0$. In the same way 
\begin{eqnarray}
A_{\overline{K}^{0}K(K^{-}K^{+})}^{I,0}(s,t;u) &=&\pm
C^{I,0}(s,t;u)+B^{I,0}(s,t;u)  \notag \\
&=&16\pi \sqrt{2}\sum_{l}(2l+1)(\pm C_{l}^{I,0}(s)+B_{l}^{I,0}(s))P_{l}(\cos
\theta _{0}^{^{\prime }}) \\
A_{\overline{K}^{0}K^{+}(K^{-}K^{0})}^{1,\pm 1}(s,t;u) &=&C^{1,\pm
1}(s,t;u)\pm B^{1,\pm 1}(s,t;u)  \notag \\
&=&16\pi \sqrt{2}\sum_{l}(2l+1)(C_{l}^{1,\pm 1}(s)\pm B_{l}^{1,\pm
1}(s))P_{l}(\cos \theta _{1}^{^{\prime }}),
\end{eqnarray}%
where once again, $C^{I,I_{3}}$ and $B^{I,I_{3}}$ mean isospin conserving and
breaking parts, respectively, and to the first order in the isospin breaking 
\begin{eqnarray}
\cos \theta _{0}^{^{\prime }} &=&\frac{t-u}{\sqrt{s(s-4M_{\pi }^{2})}}\left(
1-\frac{4M_{K}^{2}}{s}\right) ^{-1/2}\left( 1\pm \frac{\Delta M_{K}^{2}}{%
s-4M_{K}^{2}}\right) \\
\cos \theta _{1}^{^{\prime }} &=&\frac{t-u}{\lambda ^{1/2}(s,M_{\eta
}^{2},M_{\pi }^{2})}\left( 1-\frac{4M_{K}^{2}}{s}\right) ^{-1/2}.
\end{eqnarray}%
Once again, due to the $C-$invariance, $B^{1,\pm 1}(s,t;u)=B^{1,\pm 1}(s,u;t)$,
so that $B_{1}^{1,\pm 1}(s)=0$. Using the following formulae, valid
up to the $O((\Delta M_{K}^{2})^{2})$ corrections, 
\begin{eqnarray}
\sqrt{1-\frac{4M_{K^{0},K^{\pm }}^{2}}{s}} &=&\sqrt{1-\frac{4M_{K}^{2}}{s}}%
\left( 1\mp \frac{\Delta M_{K}^{2}}{s-4M_{K}^{2}}\right) \\
\frac{\lambda ^{1/2}(s,M_{K^{0}}^{2},M_{K^{\pm }}^{2})}{s} &=&\sqrt{1-\frac{%
4M_{K}^{2}}{s}},
\end{eqnarray}%
we can write for the contribution of the $K\overline{K}$ intermediate states
to the discontinuities of the isospin partial waves along the right hand cut
up to the first order in the isospin breaking 
\begin{eqnarray}
\mathrm{disc}A_{0}^{I,0}(s) &=&\frac{1}{\sqrt{2}}\sqrt{1-\frac{4M_{K}^{2}}{s}%
}\theta (s-4M_{K}^{2})[-\frac{2\Delta M_{K}^{2}}{s-4M_{K}^{2}}%
C_{0}^{0}(s)C_{0}^{I,0}(s)  \notag \\
&&+2C_{0}^{0}(s)B_{0}^{I,0}(s)+2B_{0}^{0}(s)C_{0}^{I,0}(s)] \\
\mathrm{disc}A_{1}^{1,\pm 1}(s) &=&\mp \frac{1}{\sqrt{2}}\sqrt{1-\frac{%
4M_{K}^{2}}{s}}\theta (s-4M_{K}^{2})B_{1}^{\pm 1}(s)C_{1}^{1,\pm 1}(s).
\end{eqnarray}%
We further need the $\eta \pi \rightarrow \overline{K}K$ amplitudes, for
which we get 
\begin{eqnarray}
F_{\pi }F_{\eta }F_{K}^{2}C^{0}(s,t;u) &=&-\frac{\sqrt{3}F_{0}^{2}}{4}[(s-%
\frac{1}{3}M_{\eta }^{2}-\frac{1}{3}M_{\pi }^{2}-\frac{2}{3}M_{K}^{2}) 
\notag \\
&&-\frac{1}{3}(2\overset{o}{m}_{K}^{2}-\overset{o}{m}_{\pi }^{2}-\overset{o}{%
m}_{\eta }^{2})] \\
F_{\pi }F_{\eta }F_{K}^{2}B^{0}(s,t;u) &=&\frac{3F_{0}^{2}}{4}\left(
\varepsilon _{\eta }-\frac{1}{3}\varepsilon _{\pi }\right) (s-\frac{1}{3}%
M_{\eta }^{2}-\frac{1}{3}M_{\pi }^{2}-\frac{2}{3}M_{K}^{2})  \notag \\
&&-\frac{F_{0}^{2}}{8\sqrt{3}R}(2\overset{o}{m}_{K}^{2}-\overset{o}{m}_{\pi
}^{2}-\overset{o}{m}_{\eta }^{2}-2R\Delta M_{K}^{2})  \notag \\
&&-\frac{F_{0}^{2}}{6}\varepsilon _{\pi }(\overset{o}{m}_{K}^{2}+\overset{o}{%
m}_{\pi }^{2})+\frac{F_{0}^{2}}{6}\varepsilon _{\eta }(3\overset{o}{m}%
_{K}^{2}-\overset{o}{m}_{\pi }^{2}) \\
F_{\pi }F_{\eta }F_{K}^{2}C^{\pm 1}(s,t;u) &=&-\frac{\sqrt{6}F_{0}^{2}}{4}(s-%
\frac{1}{3}M_{\eta }^{2}-\frac{1}{3}M_{\pi }^{2}-\frac{2}{3}M_{K}^{2}) 
\notag \\
&&+\frac{F_{0}^{2}}{2\sqrt{6}}(2\overset{o}{m}_{K}^{2}-\overset{o}{m}_{\pi
}^{2}-\overset{o}{m}_{\eta }^{2}) \\
F_{\pi }F_{\eta }F_{K}^{2}B^{\pm 1}(s,t;u) &=&\frac{F_{0}^{2}}{2\sqrt{2}}%
\varepsilon _{\pi }(t-u)
\end{eqnarray}%
and also the $\overline{K}K\rightarrow (\pi \pi )^{I,I_{3}}$ amplitudes,
which read 
\begin{eqnarray}
F_{\pi }^{2}F_{K}^{2}C^{0,0}(s,t;u) &=&-\frac{\sqrt{3}F_{0}^{2}}{4}(s-\frac{2%
}{3}M_{\pi }^{2}-\frac{2}{3}M_{K}^{2})  \notag \\
&&-\frac{F_{0}^{2}}{2\sqrt{3}}(\overset{o}{m}_{K}^{2}+\overset{o}{m}_{\pi
}^{2}) \\
F_{\pi }^{2}F_{K}^{2}B^{0,0}(s,t;u) &=&\frac{F_{0}^{2}}{2}\varepsilon _{\eta
}(s-\frac{2}{3}M_{\pi }^{2}-\frac{2}{3}M_{K}^{2})-\frac{F_{0}^{2}}{9}%
\varepsilon _{\eta }(\overset{o}{m}_{K}^{2}-\overset{o}{m}_{\pi }^{2}) 
\notag \\
&&-\frac{5F_{0}^{2}}{12\sqrt{3}R}(\overset{o}{m}_{K}^{2}-\overset{o}{m}_{\pi
}^{2})+\frac{F_{0}^{2}}{4\sqrt{3}}\Delta M_{K}^{2} \\
F_{\pi }^{2}F_{K}^{2}C^{2,0}(s,t;u) &=&0 \\
F_{\pi }^{2}F_{K}^{2}B^{2,0}(s,t;u) &=&-\frac{F_{0}^{2}}{\sqrt{2}}%
\varepsilon _{\eta }(s-\frac{2}{3}M_{\pi }^{2}-\frac{2}{3}M_{K}^{2})+\frac{%
\sqrt{2}F_{0}^{2}}{9}\varepsilon _{\eta }(\overset{o}{m}_{K}^{2}-\overset{o}{%
m}_{\pi }^{2})  \notag \\
&&+\frac{F_{0}^{2}}{3\sqrt{6}R}(\overset{o}{m}_{K}^{2}-\overset{o}{m}_{\pi
}^{2}) \\
F_{\pi }^{2}F_{K}^{2}C^{1,1}(s,t;u) &=&-\frac{1}{2F_{\pi }^{2}}(t-u) \\
F_{\pi }^{2}F_{K}^{2}B^{1,1}(s,t;u) &=&0.
\end{eqnarray}%
Putting all these ingredients together, with the help of (\ref{disc_W_i})
and (\ref{two_particle_unitarity}), we get the final result 
\begin{eqnarray}
F_{\pi }^{3}F_{\eta }W_{0}^{(4)K\overline{K}}(s) &=&\frac{3\sqrt{3}}{8}\frac{%
F_{0}^{4}}{F_{K}^{4}}\Delta M_{K}^{2}[(s-\frac{1}{3}M_{\eta }^{2}-\frac{1}{3}%
M_{\pi }^{2}-\frac{2}{3}M_{K}^{2})  \notag \\
&&-\frac{1}{3}(2\overset{o}{m}_{K}^{2}-\overset{o}{m}_{\pi }^{2}-\overset{o}{%
m}_{\eta }^{2})]  \notag \\
&&\times \lbrack (s-\frac{2}{3}M_{\pi }^{2}-\frac{2}{3}M_{K}^{2})+\frac{2}{3}%
(\overset{o}{m}_{K}^{2}+\overset{o}{m}_{\pi }^{2})]  \notag \\
&&\times \frac{\overline{J}_{KK}(s)-\overline{J}_{KK}(4M_{K}^{2})}{%
s-4M_{K}^{2}}  \notag \\
&&+\frac{F_{0}^{4}}{F_{K}^{4}}\{\frac{3}{2}[\frac{3}{4}\left( \varepsilon
_{\eta }-\frac{1}{3}\varepsilon _{\pi }\right) (s-\frac{1}{3}M_{\eta }^{2}-%
\frac{1}{3}M_{\pi }^{2}-\frac{2}{3}M_{K}^{2})  \notag \\
&&-\frac{1}{8\sqrt{3}R}(2\overset{o}{m}_{K}^{2}-\overset{o}{m}_{\pi }^{2}-%
\overset{o}{m}_{\eta }^{2}-2R\Delta M_{K}^{2})  \notag \\
&&-\frac{1}{6}\varepsilon _{\pi }(\overset{o}{m}_{K}^{2}+\overset{o}{m}_{\pi
}^{2})+\frac{1}{6}\varepsilon _{\eta }(3\overset{o}{m}_{K}^{2}-\overset{o}{m}%
_{\pi }^{2})]  \notag \\
&&\times \lbrack (s-\frac{2}{3}M_{\pi }^{2}-\frac{2}{3}M_{K}^{2})+\frac{2}{3}%
(\overset{o}{m}_{K}^{2}+\overset{o}{m}_{\pi }^{2})]\overline{J}_{KK}(s) 
\notag \\
&&+\frac{3}{2}[\frac{1}{2}\varepsilon _{\eta }(s-\frac{2}{3}M_{\pi }^{2}-%
\frac{2}{3}M_{K}^{2})-\frac{1}{9}\varepsilon _{\eta }(\overset{o}{m}_{K}^{2}-%
\overset{o}{m}_{\pi }^{2})  \notag \\
&&-\frac{5}{12\sqrt{3}R}(\overset{o}{m}_{K}^{2}-\overset{o}{m}_{\pi }^{2})+%
\frac{1}{4\sqrt{3}}\Delta M_{K}^{2}]  \notag \\
&&\times \lbrack (s-\frac{1}{3}M_{\eta }^{2}-\frac{1}{3}M_{\pi }^{2}-\frac{2%
}{3}M_{K}^{2})-\frac{1}{3}(2\overset{o}{m}_{K}^{2}-\overset{o}{m}_{\pi }^{2}-%
\overset{o}{m}_{\eta }^{2})]\overline{J}_{KK}(s)\} \notag \\ \\
F_{\pi }^{3}F_{\eta }W_{1}^{(4)K\overline{K}}(s) &=&-\frac{F_{0}^{4}}{%
F_{K}^{4}}\frac{1}{36}\varepsilon _{\pi }(s-4M_{K}^{2})\overline{J}_{KK}(s)
\\
F_{\pi }^{3}F_{\eta }W_{2}^{(4)K\overline{K}}(s) &=&\frac{F_{0}^{4}}{%
F_{K}^{4}}\frac{3}{4}[\varepsilon _{\eta }(s-\frac{2}{3}M_{\pi }^{2}-\frac{2%
}{3}M_{K}^{2})-\frac{2}{9}\varepsilon _{\eta }(\overset{o}{m}_{K}^{2}-%
\overset{o}{m}_{\pi }^{2})  \notag \\
&&-\frac{1}{3\sqrt{3}R}(\overset{o}{m}_{K}^{2}-\overset{o}{m}_{\pi }^{2})] 
\notag \\
&&\times \lbrack (s-\frac{1}{3}M_{\eta }^{2}-\frac{1}{3}M_{\pi }^{2}-\frac{2%
}{3}M_{K}^{2})-\frac{1}{3}(2\overset{o}{m}_{K}^{2}-\overset{o}{m}_{\pi }^{2}-%
\overset{o}{m}_{\eta }^{2})]\overline{J}_{KK}(s). \notag \\
\end{eqnarray}

\section{Unitarity contribution to the polynomial part \label{disp_poly_appendix}}

In this appendix, we summarize the result of the matching of the strict
expansion with the dispersive reconstruction of the amplitude, as explained in\
section \ref{matching section}. Let us remind that the resulting polynomial
part of the amplitude can be written in the form 
\begin{equation}
G_{\mathrm{pol}}(s,t;u)=G^{(2)}(s,t;u)+G_{\mathrm{ct}}^{(4)}(s,t;u)+G_{%
\mathrm{tad}}^{(4)}(s,t;u)+G_{\mathrm{pol,u}}^{(4)}(s,t;u),
\end{equation}%
where the listed contributions correspond to the leading order,
countertems, tadpoles and unitarity part, respectively. The strict expansion
of the former three contributions can be found in appendix \ref%
{strict_expansion_appendix}. Here we will concentrate on the unitarity
contribution 
\begin{equation}
G_{\mathrm{pol,u}}^{(4)}(s,t;u)=G_{\mathrm{pol,u}83}^{(4)}(s,t;u)-%
\varepsilon _{\pi }G_{\mathrm{pol,u}33}^{(4)}(s,t;u)+\varepsilon _{\eta }G_{%
\mathrm{pol,u}88}^{(4)}(s,t;u),
\end{equation}%
defined as%
\begin{eqnarray}
G_{\mathrm{pol,u}83}^{(4)}(s,t;u) &=&Z_{83+-,\mathrm{unit}}(s,t;u)|_{%
\overline{J_{PQ}^{r}}\rightarrow 0} \\
G_{\mathrm{pol,u}33}^{(4)}(s,t;u) &=&Z_{33+-,\mathrm{unit}}(s,t;u)|_{%
\overline{J_{PQ}^{r}}\rightarrow 0} \\
G_{\mathrm{pol,u}88}^{(4)}(s,t;u) &=&Z_{88+-,\mathrm{unit}}(s,t;u)|_{%
\overline{J_{PQ}^{r}}\rightarrow 0}.
\end{eqnarray}%
As a result, we get 
\begin{eqnarray}
G_{\mathrm{pol,u}83}^{(4)}(s,t;u) &=&-\frac{1}{8\sqrt{3}R}\overset{o}{m}%
_{\pi }^{2}\left[ 2(s-2M_{\pi }^{2})(2r-3)+2\overset{o}{m}_{\pi
}^{2}(3r-4)+\Delta _{\pi \eta }-s\right] J_{\pi \pi }^{r}(0)  \notag \\
&&-\frac{1}{16\sqrt{3}R}\overset{o}{m}_{\pi }^{2}(r-1)\left[ 8\overset{o}{m}%
_{\pi }^{2}-3M_{\eta }^{2}-M_{\pi }^{2}-s\right] J_{KK}^{r}(0)  \notag \\
&&-\frac{(r+2)}{36\sqrt{3}R}\overset{o}{m}_{\pi }^{2}J_{\eta \eta }^{r}(0) 
\notag \\
&&-\frac{1}{8\sqrt{3}R}\overset{o}{m}_{\pi }^{2}\left[ \frac{4}{3}\overset{o}%
{m}_{\pi }^{2}r+3(s-s_{0})\right] J_{\pi \eta }^{r}(0)
\end{eqnarray}%
\begin{eqnarray}
G_{\mathrm{pol,u}33}^{(4)}(s,t;u) &=&\left\{ \frac{1}{2}\left[ s-2M_{\pi
}^{2}+\overset{o}{m}_{\pi }^{2}\right] \left[ s-2M_{\pi }^{2}+3\overset{o}{m}%
_{\pi }^{2}\right] \right.  \notag \\
&&\left. -\frac{1}{12}\left[ 12\overset{o}{m}_{\pi
}^{2}(s-s_{0})-(s-u)t-(s-t)u\right] \right.  \notag \\
&&\left. +\frac{1}{4}\left[ (t-2M_{\pi }^{2})^{2}+(u-2M_{\pi }^{2})^{2}%
\right] \right.  \notag \\
&&\left. -\frac{1}{4}\Delta _{\pi \eta }\left[ s-8M_{\pi }^{2}+4\overset{o}{m%
}_{\pi }^{2}+3s_{0}\right] \right\} J_{\pi \pi }^{r}(0)  \notag \\
&&+\left\{ \left[ s-2M_{\pi }^{2}+2\overset{o}{m}_{\pi }^{2}\right] \left[
s-M_{\eta }^{2}-M_{\pi }^{2}+2\overset{o}{m}_{\pi }^{2}\right] \right. 
\notag \\
&&\left. -\frac{1}{24}\left[ 12\overset{o}{m}_{K}^{2}(s-s_{0})-(s-u)t-(s-t)u%
\right] \right\} J_{KK}^{r}(0)  \notag \\
&&+\frac{1}{18}\overset{o}{m}_{\pi }^{4}J_{\eta \eta
}^{r}(0)+F_{0}^{2}(s-s_{0})\left( \mu _{\pi }+\frac{1}{2}\mu _{K}\right) 
\notag \\
&&+\frac{1}{192\pi ^{2}}(s-u)\left[ t-2\overset{o}{m}_{K}^{2}-4\overset{o}{m}%
_{\pi }^{2}\right]  \notag \\
&&+\frac{1}{192\pi ^{2}}(s-t)\left[ u-2\overset{o}{m}_{K}^{2}-4\overset{o}{m}%
_{\pi }^{2}\right]
\end{eqnarray}%
and 
\begin{eqnarray}
G_{\mathrm{pol,u}88}^{(4)}(s,t;u) &=&\frac{1}{3}\overset{o}{m}_{\pi }^{2}%
\left[ s-2M_{\pi }^{2}+\frac{3}{2}\overset{o}{m}_{\pi }^{2}\right] J_{\pi
\pi }^{r}(0)  \notag \\
&&+\frac{2}{9}\overset{o}{m}_{\pi }^{4}J_{\pi \eta }^{r}(0)+\frac{2}{9}%
\overset{o}{m}_{\pi }^{2}\left[ \overset{o}{m}_{\eta }^{2}-\frac{1}{4}%
\overset{o}{m}_{\pi }^{2}\right] J_{\eta \eta }^{r}(0)  \notag \\
&&+\left\{ \frac{1}{8}\left[ s-2M_{\pi }^{2}+2\overset{o}{m}_{\pi }^{2}%
\right] \right.  \notag \\
&&\left. \times \left[ 3s-6M_{\eta }^{2}+4\overset{o}{m}_{\eta }^{2}-\frac{2%
}{3}\overset{o}{m}_{\pi }^{2}\right] \right.  \notag \\
&&\left. +\frac{3}{8}\left[ t-M_{\pi }^{2}-M_{\eta }^{2}+\frac{2}{3}\overset{%
o}{m}_{\pi }^{2}\right] ^{2}\right.  \notag \\
&&\left. +\frac{3}{8}\left[ u-M_{\pi }^{2}-M_{\eta }^{2}+\frac{2}{3}\overset{%
o}{m}_{\pi }^{2}\right] ^{2}\right\} J_{KK}^{r}(0)  \notag \\
&&+\frac{3}{8}\Delta _{\pi \eta }\left[ -M_{\pi }^{2}-M_{\eta }^{2}+\frac{10%
}{3}\overset{o}{m}_{\pi }^{2}\right] J_{KK}^{r}(0).
\end{eqnarray}

\section{\protect\bigskip Bare expansion of the observables $A,\ldots ,D$ 
\label{A_D_bare_expansion_appendix}}

The observables $A,\ldots ,D$ correspond to an expansion of $G(s,t;u)$ in
the center of the Dalitz plot 
\begin{equation}
G(s,t;u)=A+B(s-s_{0})+C(s-s_{0})^{2}+D\left[ (t-s_{0})^{2}+(u-s_{0})^{2}%
\right] +\dots .
\end{equation}%
The splitting of the amplitude 
\begin{equation}
G(s,t;u)=Z_{83+-}(s,t;u)-\varepsilon _{\pi }Z_{33+-}(s,t;u)+\varepsilon
_{\eta }Z_{88+-}(s,t;u),
\end{equation}%
and the further splitting into the polynomial and the unitarity parts (which
corresponds to (\ref{G_bare_expansion})) 
\begin{equation}
Z_{ab+-}(s,t;u)=Z_{ab+-,\,\mathrm{pol}}(s,t;u)+Z_{ab+-,\,\mathrm{unitary}%
}(s,t;u),
\end{equation}%
induce analogous splitting for $A,\ldots ,D$. We, therefore, write 
\begin{equation}
\mathcal{C}=(\mathcal{C}_{83}^{p}+\mathcal{C}_{83}^{u})-\varepsilon _{\pi }(%
\mathcal{C}_{33}^{p}+\mathcal{C}_{33}^{u})+\varepsilon _{\eta }(\mathcal{C}%
_{88}^{p}+\mathcal{C}_{88}^{u}),
\end{equation}%
where $\mathcal{C}=A,\ldots ,D$. $\mathcal{C}_{ab}^{p}$ stems from
the polynomial part and $\mathcal{C}_{ab}^{u}$ from the unitarity
corrections. In the following formulae, we have abbreviated $L_{i}^{r}(\mu
)\equiv L_{i}$. The results are given in the following subsections (we write $\Delta _{0}=%
\overset{o}{m}_{\eta }^{2}-\overset{o}{m}_{\pi }^{2},\Delta
_{PQ}=M_{P}^{2}-M_{Q}^{2},~\Sigma _{PQ}=M_{P}^{2}+M_{Q}^{2},\,\,\Delta
M_{K}^{2}=(M_{K^{0}}^{2}-M_{K^{+}}^{2})_{QCD}$).

\subsection{Polynomial contributions}

{\normalsize 
\begin{eqnarray}
A_{83}^{p} &=&-\frac{F_{0}^{2}(\overset{o}{m}_{\eta }^{2}-\overset{o}{m}%
_{\pi }^{2})}{4\sqrt{3}R}  \notag \\
&&\frac{1}{16\sqrt{3}R}{\overset{o}{m}_{\pi }^{2}}M_{\eta }^{2}\left(
4\left( J_{\pi \pi }^{r}(0)+8L_{4}-4L_{5}\right) -5J_{KK}^{r}(0)\right) 
\notag \\
&&-\frac{\sqrt{3}}{16R}{\overset{o}{m}_{\pi }^{2}}M_{\pi }^{2}\left(
J_{KK}^{r}(0)+4\left( J_{\pi \pi }^{r}(0)+8L_{4}+4L_{5}\right) \right) 
\notag \\
&&-\frac{1}{12\sqrt{3}R}{\overset{o}{m}_{\pi }^{2}\overset{o}{m}_{\eta }^{2}}%
\left( 9J_{KK}^{r}(0)+3\left( J_{\pi \eta }^{r}(0)+6J_{\pi \pi
}^{r}(0)+96L_{6}+192L_{7}+128L_{8}\right) +J_{\eta \eta }^{r}(0)\right) 
\notag \\
&&+\frac{1}{16\sqrt{3}R}{\overset{o}{m}_{\eta }^{2}}M_{\eta }^{2}\left(
5J_{KK}^{r}(0)-4\left( J_{\pi \pi }^{r}(0)+8L_{4}-4L_{5}\right) \right) 
\notag \\
&&+\frac{\sqrt{3}}{16R}{\overset{o}{m}_{\eta }^{2}}M_{\pi }^{2}\left(
J_{KK}^{r}(0)+4\left( J_{\pi \pi }^{r}(0)+8L_{4}+4L_{5}\right) \right) 
\notag \\
&&+\frac{1}{48\sqrt{3}R}{\overset{o}{m}_{\pi }^{4}}\left(
45J_{KK}^{r}(0)+2\left( -J_{\eta \eta }^{r}(0)+2J_{\pi \eta
}^{r}(0)+44J_{\pi \pi }^{r}(0)+720L_{6}+864L_{7}+768L_{8}\right) \right) 
\notag \\
&&-\frac{1}{48\sqrt{3}R}\overset{o}{m}_{\eta }^{4}\left(
9J_{KK}^{r}(0)+2\left( J_{\eta \eta }^{r}(0)+144\left( L_{6}-2L_{7}\right)
\right) \right)  \notag \\
&&+\frac{25{\overset{o}{m}_{\pi }^{4}}}{768\sqrt{3}\pi ^{2}R}-\frac{7{%
\overset{o}{m}_{\pi }}^{2}\overset{o}{m}_{\eta }^{2}}{384\sqrt{3}\pi ^{2}R}-%
\frac{11\overset{o}{m}_{\eta }^{4}}{768\sqrt{3}\pi ^{2}R} \\
&&  \notag \\
-A_{33}^{p} &=&\frac{1}{3}F_{0}^{2}\overset{o}{m}_{\pi }^{2}+\frac{1}{12}{%
\overset{o}{m}_{\pi }^{2}}M_{\eta }^{2}\left( J_{KK}^{r}(0)+4\left( J_{\pi
\pi }^{r}(0)+8L_{4}+4L_{5}\right) \right)  \notag \\
&&+\frac{1}{4}{\overset{o}{m}_{\pi }^{2}}M_{\pi }^{2}\left(
J_{KK}^{r}(0)+4\left( J_{\pi \pi }^{r}(0)+8L_{4}+4L_{5}\right) \right) 
\notag \\
&&-\frac{1}{36}{\overset{o}{m}_{\pi }^{2}}\overset{o}{m}_{\eta }^{2}\left(
9J_{KK}^{r}(0)+2J_{\eta \eta }^{r}(0)+288L_{6}\right)  \notag \\
&&+\frac{1}{36}M_{\eta }^{4}\left( J_{KK}^{r}(0)+8\left( J_{\pi \pi
}^{r}(0)+8L_{1}+8L_{2}+4L_{3}\right) \right)  \notag \\
&&-\frac{1}{12}M_{\pi }^{2}M_{\eta }^{2}\left( J_{KK}^{r}(0)+8\left( J_{\pi
\pi }^{r}(0)+8L_{1}+8L_{2}+4L_{3}\right) \right)  \notag \\
&&-\frac{1}{36}\overset{o}{m}_{\pi }^{4}\left( 21J_{KK}^{r}(0)+2J_{\eta \eta
}^{r}(0)+72J_{\pi \pi }^{r}(0)+1440L_{6}+768L_{8}\right)  \notag \\
&&-\frac{7\overset{o}{m}_{\pi }^{4}}{192\pi ^{2}}-\frac{11{\overset{o}{m}%
_{\pi }^{2}}\overset{o}{m}_{\eta }^{2}}{576\pi ^{2}} \\
&&  \notag \\
A_{88}^{p} &=&\frac{1}{3}F_{0}^{2}\overset{o}{m}_{\pi }^{2}  \notag \\
&&-\frac{1}{36}{\overset{o}{m}_{\pi }^{2}}M_{\eta }^{2}\left(
25J_{KK}^{r}(0)-4J_{\pi \pi }^{r}(0)+192L_{4}+48L_{5}\right)  \notag \\
&&-\frac{1}{12}{\overset{o}{m}_{\pi }^{2}}M_{\pi }^{2}\left(
5J_{KK}^{r}(0)+4J_{\pi \pi }^{r}(0)+48L_{5}\right)  \notag \\
&&+\frac{1}{36}{\overset{o}{m}_{\pi }^{2}}\overset{o}{m}_{\eta }^{2}\left(
45J_{KK}^{r}(0)+10J_{\eta \eta }^{r}(0)+1440L_{6}-1152L_{7}\right)  \notag \\
&&+\frac{1}{6}\overset{o}{m}_{\eta }^{2}M_{\eta }^{2}\left(
J_{KK}^{r}(0)+16L_{4}\right) -\frac{1}{2R}\overset{o}{m}_{\eta }^{2}M_{\pi
}^{2}\left( J_{KK}^{r}(0)+16L_{4}\right)  \notag \\
&&-\frac{1}{36}M_{\eta }^{4}\left( 9J_{KK}^{r}(0)+32\left( 2\left(
L_{1}+L_{2}\right) +L_{3}\right) \right)  \notag \\
&&+\frac{1}{12}M_{\pi }^{2}M_{\eta }^{2}\left( 9J_{KK}^{r}(0)+32\left(
2\left( L_{1}+L_{2}\right) +L_{3}\right) \right)  \notag \\
&&+\frac{1}{36}\overset{o}{m}_{\pi }^{4}\left( 9J_{KK}^{r}(0)+4\left(
9\left( J_{\pi \pi }^{r}(0)+8L_{6}+32L_{7}\right) +2J_{\pi \eta
}^{r}(0)+192L_{8}\right) -2J_{\eta \eta }^{r}(0)\right)  \notag \\
&&+\frac{7\overset{o}{m}_{\pi }^{4}}{192\pi ^{2}}+\frac{11{\overset{o}{m}%
_{\pi }}^{2}\overset{o}{m}_{\eta }^{2}}{576\pi ^{2}} \\
&&  \notag \\
B_{83}^{p} &=&\frac{\sqrt{3}}{16R}\overset{o}{m}_{\pi }^{2}\left(
J_{KK}^{r}(0)-2J_{\pi \eta }^{r}(0)+6J_{\pi \pi
}^{r}(0)+32L_{4}+16L_{5}\right)  \notag \\
&&-\frac{\sqrt{3}}{16R}\overset{o}{m}_{\eta }^{2}\left(
J_{KK}^{r}(0)+4\left( J_{\pi \pi }^{r}(0)+8L_{4}+4L_{5}\right) \right) \\
B_{33}^{p} &=&F_{0}^{2}+\frac{1}{4}\overset{o}{m}_{\pi }^{2}\left(
3J_{KK}^{r}(0)+16\left( J_{\pi \pi }^{r}(0)+7L_{4}+4L_{5}\right) \right) 
\notag \\
&&\frac{3}{4}\overset{o}{m}_{\eta }^{2}\left( J_{KK}^{r}(0)+16L_{4}\right) -%
\frac{4}{3}\left( 2L_{1}-L_{2}+L_{3}\right) (M_{\eta }^{2}+3M_{\pi }^{2}) 
\notag \\
&&+\frac{M_{\eta }^{2}}{192\pi ^{2}}+\frac{M_{\pi }^{2}}{64\pi ^{2}}+\frac{9%
\overset{o}{m}_{\pi }^{2}}{64\pi ^{2}}+\frac{3\overset{o}{m}_{\eta }^{2}}{%
64\pi ^{2}} \\
B_{88}^{p} &=&\frac{1}{6}\overset{o}{m}_{\pi }^{2}\left(
J_{KK}^{r}(0)+2J_{\pi \pi }^{r}(0)+48L_{4}\right) +\frac{1}{2}\overset{o}{m}%
_{\eta }^{2}\left( J_{KK}^{r}(0)+16L_{4}\right)  \notag \\
&&+\frac{1}{3}\left( 4L_{2}-8L_{1}\right) (M_{\eta }^{2}+3M_{\pi }^{2}) \\
&&  \notag \\
C_{83}^{p} &=&0 \\
C_{33}^{p} &=&-\frac{1}{96\pi ^{2}}+\frac{1}{24}(J_{KK}^{r}(0)+8J_{\pi \pi
}^{r}(0)+192L_{1}+96L_{3}) \\
C_{88}^{p} &=&\frac{1}{24}(9J_{KK}^{r}(0)+32\left( 6L_{1}+L_{3}\right) ) \\
&&  \notag \\
D_{83}^{p} &=&0 \\
D_{33}^{p} &=&\frac{1}{24}(J_{KK}^{r}(0)+8J_{\pi \pi }^{r}(0)+96L_{2})+\frac{%
1}{192\pi ^{2}} \\
D_{88}^{p} &=&\frac{1}{24}(9J_{KK}^{r}(0)+96L_{2}+32L_{3})
\end{eqnarray}%
}

\subsection{Unitarity contribution}

{\normalsize 
\begin{eqnarray}
A_{83}^{u} &=&\frac{F^{4}}{F_{K}^{4}}\frac{{\Delta M_{K}^{2}}}{96\sqrt{3}}%
\frac{\left( {\bar{J}}_{KK}\left( 4M_{K}^{2}\right) -{\bar{J}}_{KK}\left(
s_{0}\right) \right) }{\left( 4M_{K}^{2}-s_{0}\right) }\left( 4\Delta _{K\pi
}+\Delta _{0}\right)  \notag \\
&&\times \left( 4\Sigma _{K\pi }-6s_{0}-5{\overset{o}{m}_{\pi }}^{2}-3{%
\overset{o}{m}_{\eta }}^{2}\right)  \notag \\
&&+\frac{F^{4}}{F_{K}^{4}}\frac{1}{24\sqrt{3}}{\bar{J}}_{KK}\left(
s_{0}\right) {\Delta M_{K}^{2}}\left( 3s_{0}-4M_{K}^{2}+4{\overset{o}{m}%
_{\pi }}^{2}\right)  \notag \\
&&+\frac{F^{4}}{F_{K}^{4}}\frac{1}{96\sqrt{3}R}{\bar{J}}_{KK}\left(
s_{0}\right) \Delta _{0}\left( 4{R\Delta M_{K}^{2}}+20\Delta _{K\pi }-\Delta
_{\eta \pi }-4{\overset{o}{m}_{\pi }}^{2}\right)  \notag \\
&&+\frac{F^{4}}{F_{K}^{4}}\frac{1}{32\sqrt{3}R}{\bar{J}}_{KK}\left(
s_{0}\right) \Delta _{0}^{2}  \notag \\
&&-\frac{F^{4}}{F_{\pi }^{2}F_{\eta }^{2}}\frac{1}{4\sqrt{3}R}{\bar{J}}_{\pi
\eta }\left( s_{0}\right) {\overset{o}{m}_{\pi }}^{2}\Delta _{0}-\frac{F^{4}%
}{F_{\eta }^{4}}\frac{1}{24\sqrt{3}R}{\bar{J}}_{\eta \eta }\left(
s_{0}\right) {\overset{o}{m}_{\pi }}^{2}\Delta _{0}  \notag \\
&&-\frac{F^{4}}{F_{\pi }^{4}}\frac{1}{24\sqrt{3}R}{\bar{J}}_{\pi \pi }\left(
s_{0}\right) \Delta _{0}\left( 11{\overset{o}{m}_{\pi }}^{2}+2\Delta _{\eta
\pi }\right) \\
&&  \notag \\
A_{33}^{u} &=&\frac{F^{4}}{F_{K}^{4}}\frac{1}{288}{\bar{J}}_{KK}\left(
s_{0}\right) \left( 8{\overset{o}{m}_{\pi }}^{2}+3\Delta _{0}-4\Delta _{K\pi
}\right)  \notag \\
&&\times \left( -4M_{K}^{2}+2\left( 4{\overset{o}{m}_{\pi }}^{2}+\Sigma
_{\eta \pi }\right) +3\Delta _{0}\right)  \notag \\
&&+\frac{F^{4}}{F_{\pi }^{2}F_{\eta }^{2}}\frac{1}{3}{\bar{J}}_{\pi \eta
}\left( s_{0}\right) {\overset{o}{m}_{\pi }}^{4}+\frac{F^{4}}{F_{\eta }^{4}}%
\frac{1}{6}{\bar{J}}_{\eta \eta }\left( s_{0}\right) {\overset{o}{m}_{\pi }}%
^{4}  \notag \\
&&+\frac{F^{4}}{F_{\pi }^{4}}\frac{1}{18}{\bar{J}}_{\pi \pi }\left(
s_{0}\right) {\overset{o}{m}_{\pi }}^{2}\left( 11{\overset{o}{m}_{\pi }}%
^{2}+2\Delta _{\eta \pi }\right) \\
&&  \notag \\
A_{88}^{u} &=&-\frac{F^{4}}{F_{K}^{4}}\frac{1}{36}{\bar{J}}_{KK}\left(
s_{0}\right) M_{K}^{2}\left( 16{\overset{o}{m}_{\pi }}^{2}+27s_{0}+3\Delta
_{0}\right)  \notag \\
&&+\frac{F^{4}}{F_{K}^{4}}\frac{1}{2}{\bar{J}}_{KK}\left( s_{0}\right)
M_{K}^{4}  \notag \\
&&+\frac{F^{4}}{F_{K}^{4}}\frac{1}{48}{\bar{J}}_{KK}\left( s_{0}\right)
\Delta _{0}\left( M_{\eta }^{2}+16{\overset{o}{m}_{\pi }}^{2}\right)  \notag
\\
&&+\frac{F^{4}}{F_{K}^{4}}\frac{1}{144}{\bar{J}}_{KK}\left( s_{0}\right)
M_{\pi }^{2}\left( 36M_{\eta }^{2}+56{\overset{o}{m}_{\pi }}^{2}+9\Delta
_{0}\right)  \notag \\
&&+\frac{F^{4}}{F_{K}^{4}}\frac{1}{18}{\bar{J}}_{KK}\left( s_{0}\right) {%
\overset{o}{m}_{\pi }}^{2}\left( M_{\eta }^{2}+4{\overset{o}{m}_{\pi }}%
^{2}\right)  \notag \\
&&+\frac{F^{4}}{F_{K}^{4}}\frac{1}{96}{\bar{J}}_{KK}\left( s_{0}\right)
(24M_{\pi }^{4}+11\Delta _{0}^{2})  \notag \\
&&+\frac{F^{4}}{F_{\pi }^{2}F_{\eta }^{2}}\frac{1}{3}{\bar{J}}_{\pi \eta
}\left( s_{0}\right) {\overset{o}{m}_{\pi }}^{4}  \notag \\
&&+\frac{F^{4}}{F_{\eta }^{4}}\frac{1}{18}{\bar{J}}_{\eta \eta }\left(
s_{0}\right) {\overset{o}{m}_{\pi }}^{2}\left( 3{\overset{o}{m}_{\pi }}%
^{2}+4\Delta _{0}\right)  \notag \\
&&+\frac{F^{4}}{F_{\pi }^{4}}\frac{1}{18}{\bar{J}}_{\pi \pi }\left(
s_{0}\right) {\overset{o}{m}_{\pi }}^{2}\left( 11{\overset{o}{m}_{\pi }}%
^{2}+2\Delta _{\eta \pi }\right) \\
&&  \notag \\
B_{83}^{u} &=&\frac{F^{4}}{F_{K}^{4}}\frac{{\Delta M_{K}^{2}}}{8\sqrt{3}}%
\frac{\left( {\bar{J}}_{KK}\left( 4M_{K}^{2}\right) -{\bar{J}}_{KK}\left(
s_{0}\right) \right) }{\left( 4M_{K}^{2}-s_{0}\right) }(3{\overset{o}{m}%
_{\pi }}^{2}+{\overset{o}{m}_{\eta }}^{2}+3s_{0}-4M_{K}^{2})  \notag \\
&&+\frac{F^{4}}{F_{K}^{4}}\frac{{\Delta M_{K}^{2}}}{96\sqrt{3}}\frac{\bar{J}%
_{KK}^{\prime }\left( s_{0}\right) \left( s_{0}-4M_{K}^{2}\right) -{\bar{J}}%
_{KK}\left( s_{0}\right) +{\bar{J}}_{KK}\left( 4M_{K}^{2}\right) }{\left(
s_{0}-4M_{K}^{2}\right) ^{2}}  \notag \\
&&\times \left( 4\Delta _{K\pi }+{\Delta }_{0}^{2}\right) \left( 4\Sigma
_{K\pi }-5{\overset{o}{m}_{\pi }}^{2}-3{\overset{o}{m}_{\eta }}%
^{2}-6s_{0}\right)  \notag \\
&&\frac{F^{4}}{F_{K}^{4}}\frac{1}{48\sqrt{3}R}\bar{J}_{KK}^{\prime }\left(
s_{0}\right) \left( \Delta _{0}-8{R\Delta M_{K}^{2}}\right) M_{K}^{2}  \notag
\\
&&+\frac{F^{4}}{F_{K}^{4}}\frac{1}{24\sqrt{3}}\bar{J}_{KK}^{\prime }\left(
s_{0}\right) {\Delta M_{K}^{2}}\left( 4{\overset{o}{m}_{\pi }}%
^{2}+3s_{0}\right)  \notag \\
&&+\frac{F^{4}}{F_{K}^{4}}\frac{1}{96\sqrt{3}R}\bar{J}_{KK}^{\prime }\left(
s_{0}\right) \Delta _{0}\left( 4{R\Delta M_{K}^{2}}-4{\overset{o}{m}_{\pi }}%
^{2}-\Sigma _{\eta \pi }\right)  \notag \\
&&-\frac{F^{4}}{F_{K}^{4}}\frac{1}{64\sqrt{3}R}\bar{J}_{KK}^{\prime }\left(
s_{0}\right) \Delta _{0}^{2}  \notag \\
&&+\frac{F^{4}}{F_{K}^{4}}\frac{1}{32\sqrt{3}R}{\bar{J}}_{KK}\left(
s_{0}\right) \left( 8{R\Delta M_{K}^{2}}-\Delta _{0}\right) -\frac{F^{4}}{%
F_{\eta }^{4}}\frac{1}{24\sqrt{3}R}\bar{J}_{\eta \eta }^{\prime }\left(
s_{0}\right) {\overset{o}{m}_{\pi }}^{2}\Delta _{0}  \notag \\
&&-\frac{F^{4}}{F_{\pi }^{4}}\frac{1}{24\sqrt{3}R}\bar{J}_{\pi \pi }^{\prime
}\left( s_{0}\right) 5\Delta _{0}\left( {\overset{o}{m}_{\pi }}^{2}+\Delta
_{\eta \pi }\right) -\frac{F^{4}}{F_{\pi }^{4}}\frac{5}{8\sqrt{3}R}{\bar{J}}%
_{\pi \pi }\left( s_{0}\right) \Delta _{0} \\
&&  \notag \\
B_{33}^{u} &=&\frac{F^{4}}{F_{K}^{4}}\frac{1}{288}\bar{J}_{KK}^{\prime
}\left( s_{0}\right) \left( -4\Delta _{K\pi }+8{\overset{o}{m}_{\pi }}%
^{2}+3\Delta _{0}\right)  \notag \\
&&\times \left( -4M_{K}^{2}+2\left( 4{\overset{o}{m}_{\pi }}^{2}+\Sigma
_{\eta \pi }\right) +3\Delta _{0}\right)  \notag \\
&&+\frac{F^{4}}{F_{\eta }^{4}}\frac{1}{6}\bar{J}_{\eta \eta }^{\prime
}\left( s_{0}\right) {\overset{o}{m}_{\pi }}^{4}+\frac{F^{4}}{F_{\pi }^{4}}%
\frac{5}{18}\bar{J}_{\pi \pi }^{\prime }\left( s_{0}\right) {\overset{o}{m}%
_{\pi }}^{2}\left( {\overset{o}{m}_{\pi }}^{2}+\Delta _{\eta \pi }\right) 
\notag \\
&&+\frac{F^{4}}{F_{K}^{4}}\frac{1}{24}{\bar{J}}_{KK}\left( s_{0}\right)
\left( -16M_{K}^{2}+6s_{0}+8{\overset{o}{m}_{\pi }}^{2}+3\Delta _{0}\right) 
\notag \\
&&+\frac{F^{4}}{F_{\pi }^{4}}\frac{1}{6}{\bar{J}}_{\pi \pi }\left(
s_{0}\right) \left( M_{\eta }^{2}-5M_{\pi }^{2}+10{\overset{o}{m}_{\pi }}%
^{2}\right)  \notag \\
&&+\frac{F^{4}}{F_{\pi }^{2}F_{\eta }^{2}}\frac{1}{2}{\bar{J}}_{\pi \eta
}\left( s_{0}\right) {\overset{o}{m}_{\pi }}^{2} \\
&&  \notag \\
B_{88}^{u} &=&\frac{F^{4}}{F_{K}^{4}}\frac{1}{12}\bar{J}_{KK}^{\prime
}\left( s_{0}\right) \Delta _{0}\left( 4{\overset{o}{m}_{\pi }}%
^{2}+3s_{0}-4M_{K}^{2}\right)  \notag \\
&&\frac{F^{4}}{F_{K}^{4}}\frac{1}{18}\bar{J}_{KK}^{\prime }\left(
s_{0}\right) {\overset{o}{m}_{\pi }}^{2}\left( \Delta _{\eta \pi }-8\Delta
_{K\pi }+4{\overset{o}{m}_{\pi }}^{2}\right)  \notag \\
&&\frac{F^{4}}{F_{K}^{4}}\frac{1}{12}\bar{J}_{KK}^{\prime }\left(
s_{0}\right) \Delta _{0}^{2}  \notag \\
&&+\frac{F^{4}}{F_{\eta }^{4}}\frac{1}{18}\bar{J}_{\eta \eta }^{\prime
}\left( s_{0}\right) {\overset{o}{m}_{\pi }}^{2}\left( 3{\overset{o}{m}_{\pi
}}^{2}+4\Delta _{0}\right)  \notag \\
&&+\frac{F^{4}}{F_{\pi }^{4}}\frac{5}{18}\bar{J}_{\pi \pi }^{\prime }\left(
s_{0}\right) {\overset{o}{m}_{\pi }}^{2}\left( {\overset{o}{m}_{\pi }}%
^{2}+\Delta _{\eta \pi }\right)  \notag \\
&&+\frac{F^{4}}{F_{K}^{4}}\frac{1}{6}{\bar{J}}_{KK}\left( s_{0}\right)
\left( 4{\overset{o}{m}_{\pi }}^{2}+3\Delta _{0}\right) +\frac{F^{4}}{F_{\pi
}^{4}}\frac{5}{6}{\bar{J}}_{\pi \pi }\left( s_{0}\right) {\overset{o}{m}%
_{\pi }}^{2} \\
&&  \notag \\
\frac{4R}{\sqrt{3}}C_{83}^{u} &=&-\frac{F^{4}}{F_{K}^{4}}\frac{{\Delta M}%
_{K}^{2}}{\left( 3s_{0}-12M_{K}^{2}\right) {}^{2}}\bar{J}^{\prime }\left(
M_{K}^{2},M_{K}^{2},s_{0}\right) \left( 7M_{\eta }^{2}+21M_{\pi }^{2}+20{%
\overset{o}{m}_{\pi }}^{2}+5\Delta _{0}\right)  \notag \\
&&\frac{F^{4}}{F_{K}^{4}}\frac{22{\Delta M}_{K}^{2}}{\left(
3s_{0}-12M_{K}^{2}\right) {}^{2}}\bar{J}^{\prime }\left(
M_{K}^{2},M_{K}^{2},s_{0}\right) M_{K}^{4}  \notag \\
&&+\frac{F^{4}}{F_{K}^{4}}\frac{1}{4}\frac{{\Delta M}_{K}^{2}}{\left(
3s_{0}-12M_{K}^{2}\right) {}^{2}}\bar{J}^{\prime }\left(
M_{K}^{2},M_{K}^{2},s_{0}\right) \Delta _{0}\left( 3M_{\eta }^{2}+M_{\pi
}^{2}+4{\overset{o}{m}_{\pi }}^{2}\right)  \notag \\
&&\frac{F^{4}}{F_{K}^{4}}\frac{1}{2}\frac{{\Delta M}_{K}^{2}}{\left(
3s_{0}-12M_{K}^{2}\right) {}^{2}}\bar{J}^{\prime }\left(
M_{K}^{2},M_{K}^{2},s_{0}\right)  \notag \\
&&\times \left( 4M_{\pi }^{2}\left( M_{\eta }^{2}+{\overset{o}{m}_{\pi }}%
^{2}\right) +4{\overset{o}{m}_{\pi }}^{2}M_{\eta }^{2}+M_{\eta }^{4}+7M_{\pi
}^{4}\right)  \notag \\
&&+\frac{F^{4}}{F_{K}^{4}}\frac{3}{8}\frac{{\Delta M}_{K}^{2}}{\left(
3s_{0}-12M_{K}^{2}\right) {}^{2}}\bar{J}^{\prime }\left(
M_{K}^{2},M_{K}^{2},s_{0}\right) \Delta _{0}^{2}  \notag \\
&&+\frac{F^{4}}{F_{K}^{4}}\frac{1}{18}\bar{J}^{\prime \prime }\left(
M_{K}^{2},M_{K}^{2},s_{0}\right) \left( \Delta _{0}-2{\Delta M}%
_{K}^{2}\right) M_{K}^{2}  \notag \\
&&+\frac{F^{4}}{F_{K}^{4}}\frac{1}{144}\bar{J}^{\prime \prime }\left(
M_{K}^{2},M_{K}^{2},s_{0}\right) \Delta _{0}\left( 4{\Delta M}%
_{K}^{2}-M_{\eta }^{2}-7M_{\pi }^{2}-4{\overset{o}{m}_{\pi }}^{2}\right) 
\notag \\
&&+\frac{F^{4}}{F_{K}^{4}}\frac{1}{36}\bar{J}^{\prime \prime }\left(
M_{K}^{2},M_{K}^{2},s_{0}\right) {\Delta M}_{K}^{2}\left( 4{\overset{o}{m}%
_{\pi }}^{2}+3s_{0}\right)  \notag \\
&&+\frac{F^{4}}{F_{K}^{4}}\frac{1}{48}\frac{{\Delta M}_{K}^{2}}{\left(
12M_{K}^{2}-3s_{0}\right) }\bar{J}^{\prime \prime }\left(
M_{K}^{2},M_{K}^{2},s_{0}\right) \left( 4\Delta _{K\pi }+\Delta _{0}\right) 
\notag \\
&&\times \left( -4M_{K}^{2}+2\left( 4{\overset{o}{m}_{\pi }}^{2}+\Sigma
_{\eta \pi }\right) +3\Delta _{0}\right)  \notag \\
&&+\frac{F^{4}}{F_{K}^{4}}\frac{1}{6}\bar{J}^{\prime }\left(
M_{K}^{2},M_{K}^{2},s_{0}\right) \left( 2{\Delta M}_{K}^{2}-\Delta
_{0}\right)  \notag \\
&&+\frac{F^{4}}{F_{K}^{4}}\frac{3}{8}\frac{{\Delta M}_{K}^{2}}{\left(
12M_{K}^{2}-3s_{0}\right) {}^{3}}\left( {\bar{J}}\left(
M_{K}^{2},M_{K}^{2},4M_{K}^{2}\right) -{\bar{J}}\left(
M_{K}^{2},M_{K}^{2},s_{0}\right) \right)  \notag \\
&&\times \left( 20M_{K}^{2}-4M_{\pi }^{2}+8{\overset{o}{m}_{\pi }}%
^{2}+3\Delta _{0}\right) \left( 20M_{K}^{2}-\Delta _{0}-2\Sigma _{\eta \pi
}\right)  \notag \\
&&-\frac{F^{4}}{F_{\pi }^{2}F_{\eta }^{2}}\frac{1}{18}\bar{J}^{\prime \prime
}\left( M_{\pi }^{2},M_{\eta }^{2},s_{0}\right) {\overset{o}{m}_{\pi }}%
^{2}\Delta _{0}-\frac{{F}^{4}}{F_{\eta }^{4}}\frac{1}{36}\bar{J}^{\prime
\prime }\left( M_{\eta }^{2},M_{\eta }^{2},s_{0}\right) {\overset{o}{m}_{\pi
}}^{2}\Delta _{0}  \notag \\
&&-\frac{{F}^{4}}{F_{\pi }^{4}}\frac{1}{72}\bar{J}^{\prime \prime }\left(
M_{\pi }^{2},M_{\pi }^{2},s_{0}\right) \Delta _{0}\left( 14{\overset{o}{m}%
_{\pi }}^{2}+8\Delta _{\eta \pi }\right)  \notag \\
&&-\frac{{F}^{4}}{F_{\pi }^{4}}\frac{2}{3}\bar{J}^{\prime }\left( M_{\pi
}^{2},M_{\pi }^{2},s_{0}\right) \Delta _{0}
\end{eqnarray}%
}

\begin{eqnarray}
\frac{4R}{\sqrt{3}}C_{33}^{u} &=&\frac{{F}^{4}}{F_{K}^{4}}\frac{1}{144\sqrt{3%
}}\bar{J}^{\prime \prime }\left( M_{K}^{2},M_{K}^{2},s_{0}\right) \left(
-4\Delta _{K\pi }+8{\overset{o}{m}_{\pi }}^{2}+3\Delta _{0}\right)  \notag \\
&&\times \left( -4M_{K}^{2}+2\left( 4{\overset{o}{m}_{\pi }}^{2}+\Sigma
_{\eta \pi }\right) +3\Delta _{0}\right)  \notag \\
&&+\frac{{F}^{4}}{F_{K}^{4}}\frac{1}{18\sqrt{3}}\bar{J}^{\prime }\left(
M_{K}^{2},M_{K}^{2},s_{0}\right) \left( 12M_{K}^{2}+3s_{0}+24{\overset{o}{m}%
_{\pi }}^{2}+9\Delta _{0}\right)  \notag \\
&&+\frac{{F}^{4}}{F_{\pi }^{2}F_{\eta }^{2}}\frac{2}{9\sqrt{3}}\bar{J}%
^{\prime \prime }\left( M_{\pi }^{2},M_{\eta }^{2},s_{0}\right) {\overset{o}{%
m}_{\pi }}^{4}+\frac{{F}^{4}}{F_{\eta }^{4}}\frac{1}{3\sqrt{3}}\bar{J}%
^{\prime \prime }\left( M_{\eta }^{2},M_{\eta }^{2},s_{0}\right) {\overset{o}%
{m}_{\pi }}^{4}  \notag \\
&&+\frac{{F}^{4}}{F_{\pi }^{4}}\frac{1}{18\sqrt{3}}\bar{J}^{\prime \prime
}\left( M_{\pi }^{2},M_{\pi }^{2},s_{0}\right) {\overset{o}{m}_{\pi }}%
^{2}\left( 14{\overset{o}{m}_{\pi }}^{2}+8\Delta _{\eta \pi }\right)  \notag
\\
&&+\frac{{F}^{4}}{F_{\pi }^{2}F_{\eta }^{2}}\frac{4}{3\sqrt{3}}\bar{J}%
^{\prime }\left( M_{\pi }^{2},M_{\eta }^{2},s_{0}\right) {\overset{o}{m}%
_{\pi }}^{2}  \notag \\
&&+\frac{{F}^{4}}{F_{\pi }^{4}}\frac{4}{9\sqrt{3}}\bar{J}^{\prime }\left(
M_{\pi }^{2},M_{\pi }^{2},s_{0}\right) \left( 12{\overset{o}{m}_{\pi }}%
^{2}+3s_{0}\right)  \notag \\
&&+\frac{{F}^{4}}{F_{K}^{4}}\frac{1}{6\sqrt{3}}{\bar{J}}\left(
M_{K}^{2},M_{K}^{2},s_{0}\right) +\frac{{F}^{4}}{F_{\pi }^{4}}\frac{4}{3%
\sqrt{3}}{\bar{J}}\left( M_{\pi }^{2},M_{\pi }^{2},s_{0}\right) \\
\frac{4R}{\sqrt{3}}C_{88}^{u} &=&\frac{{F}^{4}}{F_{K}^{4}}\frac{1}{6\sqrt{3}}%
\bar{J}^{\prime }\left( M_{K}^{2},M_{K}^{2},s_{0}\right)  \notag \\
&&\times \left( -12M_{K}^{2}+9s_{0}+16{\overset{o}{m}_{\pi }}^{2}+9\Delta
_{0}\right)  \notag \\
&&-\frac{{F}^{4}}{F_{K}^{4}}\frac{1}{36\sqrt{3}}\bar{J}^{\prime \prime
}\left( M_{K}^{2},M_{K}^{2},s_{0}\right) M_{K}^{2}  \notag \\
&&\times \left( 18s_{0}+32{\overset{o}{m}_{\pi }}^{2}+18\Delta _{0}\right) 
\notag \\
&&+\frac{{F}^{4}}{F_{K}^{4}}\frac{1}{3\sqrt{3}}\bar{J}^{\prime \prime
}\left( M_{K}^{2},M_{K}^{2},s_{0}\right) M_{K}^{4}  \notag \\
&&+\frac{{F}^{4}}{F_{K}^{4}}\frac{1}{144\sqrt{3}}\bar{J}^{\prime \prime
}\left( M_{K}^{2},M_{K}^{2},s_{0}\right) M_{\pi }^{2}  \notag \\
&&\times \left( 36M_{\eta }^{2}+112{\overset{o}{m}_{\pi }}^{2}+51\Delta
_{0}-12\Sigma _{\eta \pi }\right)  \notag \\
&&+\frac{{F}^{4}}{F_{K}^{4}}\frac{1}{144\sqrt{3}}\bar{J}^{\prime \prime
}\left( M_{K}^{2},M_{K}^{2},s_{0}\right)  \notag \\
&&\times \left( 16{\overset{o}{m}_{\pi }}^{2}M_{\eta }^{2}+15\Delta
_{0}M_{\eta }^{2}+36M_{\pi }^{4}+64{\overset{o}{m}_{\pi }}^{4}+96{\overset{o}%
{m}_{\pi }}^{2}\Delta _{0}+3\Delta _{0}\Sigma _{\eta \pi }+27\Delta
_{0}^{2}\right)  \notag \\
&&+\frac{{F}^{4}}{F_{\pi }^{2}F_{\eta }^{2}}\frac{2}{9\sqrt{3}}\bar{J}%
^{\prime \prime }\left( M_{\pi }^{2},M_{\eta }^{2},s_{0}\right) {\overset{o}{%
m}_{\pi }}^{4}  \notag \\
&&+\frac{{F}^{4}}{F_{\eta }^{4}}\frac{1}{9\sqrt{3}}\bar{J}^{\prime \prime
}\left( M_{\eta }^{2},M_{\eta }^{2},s_{0}\right) {\overset{o}{m}_{\pi }}%
^{2}\left( 3{\overset{o}{m}_{\pi }}^{2}+4\Delta _{0}\right)  \notag \\
&&+\frac{{F}^{4}}{F_{\pi }^{4}}\frac{1}{18\sqrt{3}}\bar{J}^{\prime \prime
}\left( M_{\pi }^{2},M_{\pi }^{2},s_{0}\right) {\overset{o}{m}_{\pi }}%
^{2}\left( 14{\overset{o}{m}_{\pi }}^{2}+8\Delta _{\eta \pi }\right)  \notag
\\
&&+\frac{{F}^{4}}{F_{\pi }^{4}}\frac{{8}}{3\sqrt{3}}\bar{J}^{\prime }\left(
M_{\pi }^{2},M_{\pi }^{2},s_{0}\right) {\overset{o}{m}_{\pi }}^{2}  \notag \\
&&+\frac{{F}^{4}}{F_{K}^{4}}\frac{\sqrt{3}}{2}{\bar{J}}\left(
M_{K}^{2},M_{K}^{2},s_{0}\right) \\
D_{83}^{u} &=&-\frac{F^{4}}{F_{K}^{4}}\frac{\sqrt{3}}{32R}\bar{J}%
_{KK}^{\prime }\left( s_{0}\right) \Delta _{0}  \notag \\
&&+\frac{F^{4}}{F_{K}^{4}}\frac{1}{128\sqrt{3}R}\bar{J}_{KK}^{\prime \prime
}\left( s_{0}\right) \Delta _{0}\left( 4\Delta _{K\pi }+\Delta _{0}\right) 
\notag \\
&&-\frac{F^{4}}{F_{\pi }^{2}F_{\eta }^{2}}\frac{1}{24\sqrt{3}R}\bar{J}_{\pi
\eta }^{\prime \prime }\left( s_{0}\right) {\overset{o}{m}_{\pi }}^{2}\Delta
_{0}  \notag \\
&&-\frac{F^{4}}{F_{\pi }^{4}}\frac{1}{48\sqrt{3}R}\bar{J}_{\pi \pi }^{\prime
\prime }\left( s_{0}\right) \Delta _{0}\left( 2{\overset{o}{m}_{\pi }}%
^{2}-\Delta _{\eta \pi }\right)  \notag \\
&&+\frac{F^{4}}{F_{\pi }^{4}}\frac{1}{8\sqrt{3}R}\bar{J}_{\pi \pi }^{\prime
}\left( s_{0}\right) \Delta _{0} \\
D_{33}^{u} &=&-\frac{F^{4}}{F_{K}^{4}}\frac{1}{72}\bar{J}_{KK}^{\prime
}\left( s_{0}\right) \left( 12M_{K}^{2}-3s_{0}\right)  \notag \\
&&+\frac{F^{4}}{F_{\pi }^{2}F_{\eta }^{2}}\frac{1}{18}\bar{J}_{\pi \eta
}^{\prime \prime }\left( s_{0}\right) {\overset{o}{m}_{\pi }}^{4}  \notag \\
&&+\frac{F^{4}}{F_{\pi }^{4}}\frac{1}{36}\bar{J}_{\pi \pi }^{\prime \prime
}\left( s_{0}\right) {\overset{o}{m}_{\pi }}^{2}\left( 2{\overset{o}{m}_{\pi
}}^{2}-\Delta _{\eta \pi }\right)  \notag \\
&&-\frac{F^{4}}{F_{\pi }^{4}}\frac{1}{9}\bar{J}_{\pi \pi }^{\prime }\left(
s_{0}\right) \left( 3\left( M_{\pi }^{2}+{\overset{o}{m}_{\pi }}^{2}\right)
-M_{\eta }^{2}\right)  \notag \\
&&-\frac{F^{4}}{F_{\pi }^{2}F_{\eta }^{2}}\frac{1}{6}\bar{J}_{\pi \eta
}^{\prime }\left( s_{0}\right) {\overset{o}{m}_{\pi }}^{2}+\frac{F^{4}}{%
F_{K}^{4}}\frac{1}{24}{\bar{J}}_{KK}\left( s_{0}\right) +\frac{F^{4}}{F_{\pi
}^{4}}\frac{1}{3}{\bar{J}}_{\pi \pi }\left( s_{0}\right) \\
D_{88}^{u} &=&\frac{F^{4}}{F_{K}^{4}}\frac{1}{192}\bar{J}_{KK}^{\prime
\prime }\left( s_{0}\right) \left( 4\Delta _{K\pi }+\Delta _{0}\right)
\left( 4M_{K}^{2}+\Delta _{0}-2\Sigma _{\eta \pi }\right)  \notag \\
&&+\frac{F^{4}}{F_{K}^{4}}\frac{1}{8}\bar{J}_{KK}^{\prime }\left(
s_{0}\right) \left( 3s_{0}-4M_{K}^{2}-\Delta _{0}\right) +\frac{F^{4}}{%
F_{\pi }^{2}F_{\eta }^{2}}\frac{1}{18}\bar{J}_{\pi \eta }^{\prime \prime
}\left( s_{0}\right) {\overset{o}{m}_{\pi }}^{4}  \notag \\
&&+\frac{F^{4}}{F_{\pi }^{4}}\frac{1}{36}\bar{J}_{\pi \pi }^{\prime \prime
}\left( s_{0}\right) {\overset{o}{m}_{\pi }}^{2}\left( 2{\overset{o}{m}_{\pi
}}^{2}-\Delta _{\eta \pi }\right)  \notag \\
&&-\frac{F^{4}}{F_{\pi }^{4}}\frac{1}{6}\bar{J}_{\pi \pi }^{\prime }\left(
s_{0}\right) {\overset{o}{m}_{\pi }}^{2}+\frac{F^{4}}{F_{K}^{4}}\frac{3}{8}{%
\bar{J}}_{KK}\left( s_{0}\right)
\end{eqnarray}%
Note that all the above observables are renormalization scale independent,
which can be verified by using the explicit $\mu $ dependence of the
chiral logarithms and $L_{i}$.

\section{Reparameterization of $A_{ab}^{p}$ and $B_{ab}^{p}$ \label{reparametrization_appendix}}

In this appendix, we list our final formulae for the Dalitz plot expansion
parameters $A_{ab}^{p}$ and $B_{ab}^{p}$ (for the definition see
appendix \ref{A_D_bare_expansion_appendix}), expressed in terms of the
physical masses and decay constants, the LECs $L_{1}$, $L_{2}$ and $L_{3}$,
the parameters $X$~$(Y)$, $Z$ and $r$ and the indirect remainders (namely $%
\delta _{F_{\pi }^{2}M_{\pi }^{2}}$, $~\delta _{F_{\pi }^{2}}$, $~\delta
_{F_{K}^{2}M_{K}^{2}}$, $~\delta _{F_{K}^{2}}$, $\delta _{F_{\eta
}^{2}M_{\eta }^{2}}$, $~\delta _{F_{\eta }^{2}}$, $~\delta _{\varepsilon FM}$
and $~\delta _{\varepsilon F}$). We use the notation 
\begin{eqnarray}
r_{2}^{\ast } &=&2\frac{F_{K}^{2}M_{K}^{2}}{F_{\pi }^{2}M_{\pi }^{2}}-1 \\
\epsilon (r) &=&2\frac{r_{2}^{\ast }-r}{r^{2}-1} \\
\eta (r) &=&\frac{2}{r-1}\left( \frac{F_{K}^{2}}{F_{\pi }^{2}}-1\right) \\
\Delta _{GMO} &=&\frac{3F_{\eta }^{2}M_{\eta }^{2}+F_{\pi }^{2}M_{\pi
}^{2}-4F_{K}^{2}M_{K}^{2}}{F_{\pi }^{2}M_{\pi }^{2}}.
\end{eqnarray}

{\normalsize 
\begin{eqnarray}
\frac{4R}{\sqrt{3}}A_{83}^{p} &=&-\frac{2}{9}X(r-1)F_{\pi }^{2}M_{\pi }^{2} 
\notag \\
&&-\frac{2}{9}\epsilon (r)r\frac{(r-1)(r+5)}{(r+2)}F_{\pi }^{2}M_{\pi }^{2}+%
\frac{1}{9}\eta (r)\frac{(r-1)}{(r+2)}F_{\pi }^{2}\left( (r+4)M_{\eta
}^{2}+3M_{\pi }^{2}r\right)  \notag \\
&&+\frac{2}{9}\Delta _{{GMO}}\frac{(r-4)}{(r-1)}F_{\pi }^{2}M_{\pi }^{2} 
\notag \\
&&+\frac{2}{9}(Z-1)\frac{(r-1)}{(r+2)}F_{\pi }^{2}(M_{\eta }^{2}-3M_{\pi
}^{2})+\frac{2}{9}(X-1)(r+8)\frac{(r-1)}{(r+2)}F_{\pi }^{2}M_{\pi }^{2} 
\notag \\
&&+\frac{1}{18}Y\frac{M_{\pi }^{2}M_{\eta }^{2}}{(r+2)}  \notag \\
&&\times \left( 2\left( 3r^{2}-8\right) J_{KK}^{r}(0)-(r+4)(2r+1)J_{\eta
\eta }^{r}(0)+((9-4r)r+20)J_{\pi \pi }^{r}(0)\right)  \notag \\
&&-\frac{1}{6}Y\frac{M_{\pi }^{4}r}{(r+2)}(2r(J_{KK}^{r}(0)+J_{\eta \eta
}^{r}(0)-2J_{\pi \pi }^{r}(0))+J_{\eta \eta }^{r}(0)-J_{\pi \pi }^{r}(0)) 
\notag \\
&&+\frac{1}{9}Y^{2}\frac{M_{\pi }^{4}}{(r+2)}  \notag \\
&&\times \left[ \left( 2(r(3r+4)+8)J_{KK}^{r}(0)-(r(9r+8)+16)J_{\pi \pi
}^{r}(0)\right) \right.  \notag \\
&&\left. +r(5r+4)J_{\eta \eta }^{r}(0)-2r(r+2)J_{\pi \eta }^{r}(0)\right] 
\notag \\
&&+\frac{M_{\pi }^{2}}{864\pi ^{2}}Y(r-1)\left( 5M_{\pi }^{2}\left(
22Y-9\right) -3M_{\eta }^{2}\right)  \notag \\
&&-\frac{2}{9}\frac{\left( (r+4)M_{\eta }^{2}+3M_{\pi }^{2}r\right) }{(r+2)}%
F_{K}^{2}\delta _{F_{K}}+\frac{2}{9}\frac{\left( (2r+3)M_{\eta }^{2}+3M_{\pi
}^{2}\right) }{(r+2)}F_{\pi }^{2}\delta _{F_{\pi }}  \notag \\
&&-\frac{2}{3}\frac{(r-4)}{(r-1)}F_{\eta }^{2}M_{\eta }^{2}\delta _{F_{\eta
}M_{\eta }}-\frac{2}{9}\frac{(r(r(r+3)+3)-16)}{\left( r^{2}+r-2\right) }%
F_{\pi }^{2}M_{\pi }^{2}\delta _{F_{\pi }M_{\pi }}  \notag \\
&&+\frac{8}{9}\frac{(2r+1)\left( r^{2}+r-8\right) }{\left(
r^{3}+2r^{2}-r-2\right) }F_{K}^{2}M_{K}^{2}\delta _{F_{K}M_{K}} \\
&&  \notag \\
\frac{4}{\sqrt{3}}A_{33}^{p} &=&\frac{4}{3\sqrt{3}}XF_{\pi }^{2}M_{\pi }^{2}+%
\frac{4}{\sqrt{3}}\epsilon (r)\frac{r}{(r+2)}F_{\pi }^{2}M_{\pi }^{2}-\frac{2%
}{\sqrt{3}}\eta (r)\frac{r}{(r+2)}F_{\pi }^{2}s_{0}  \notag \\
&&-\frac{4}{3\sqrt{3}}\frac{F_{\pi }^{2}}{(r+2)}M_{\pi }^{2}(X-1)(r+8) 
\notag \\
&&+\frac{4}{\sqrt{3}}\frac{F_{\pi }^{2}}{(r+2)}(Z-1)s_{0}  \notag \\
&&-\frac{1}{9\sqrt{3}}\left( J_{KK}^{r}(0)+8\left( J_{\pi \pi
}^{r}(0)+8L_{1}+8L_{2}+4L_{3}\right) \right) M_{\eta }^{2}\left( M_{\eta
}^{2}-3M_{\pi }^{2}\right)  \notag \\
&&+\frac{1}{\sqrt{3}}Y\frac{r}{\left( r^{2}+r-2\right) }  \notag \\
&&\times (2r(J_{KK}^{r}(0)+J_{\eta \eta }^{r}(0)-2J_{\pi \pi
}^{r}(0))+J_{\eta \eta }^{r}(0)-J_{\pi \pi }^{r}(0))M_{\pi }^{2}s_{0}  \notag
\\
&&-\frac{2}{\sqrt{3}}Y^{2}\frac{r^{2}}{\left( r^{2}+r-2\right) }%
(2J_{KK}^{r}(0)+J_{\eta \eta }^{r}(0)-3J_{\pi \pi }^{r}(0))M_{\pi }^{4} 
\notag \\
&&+\frac{1}{144\sqrt{3}\pi ^{2}}Y\left( M_{\pi }^{2}(45-74Y)+15M_{\eta
}^{2}\right) M_{\pi }^{2}  \notag \\
&&+\frac{4}{\sqrt{3}}\frac{1}{\left( r^{2}+r-2\right) }s_{0}\left(
rF_{K}^{2}\delta _{F_{K}}-F_{\pi }^{2}\delta _{F_{\pi }}\right)  \notag \\
&&-\frac{16}{\sqrt{3}}\frac{r}{\left( r^{3}+2r^{2}-r-2\right) }%
F_{K}^{2}M_{K}^{2}\delta _{F_{K}M_{K}}  \notag \\
&&-\frac{4}{3\sqrt{3}}\frac{\left( r^{2}+r-8\right) }{\left(
r^{2}+r-2\right) }F_{\pi }^{2}M_{\pi }^{2}\delta _{F_{\pi }M_{\pi }} \\
&&  \notag \\
\frac{4}{\sqrt{3}}A_{88}^{p} &=&\frac{4}{3\sqrt{3}}XF_{\pi }^{2}M_{\pi }^{2}+%
\frac{4}{3\sqrt{3}}\epsilon (r)\frac{r(2r+1)}{(r+2)}F_{\pi }^{2}M_{\pi }^{2}
\notag \\
&&+\frac{2}{9\sqrt{3}}\eta (r)\frac{\left( (4-7r)M_{\eta }^{2}+3M_{\pi
}^{2}(r-4)\right) }{(r+2)}F_{\pi }^{2}  \notag \\
&&+\frac{8}{\sqrt{3}}\Delta _{{GMO}}\frac{1}{(3-3r)}F_{\pi }^{2}M_{\pi }^{2}
\notag \\
&&-\frac{4}{9\sqrt{3}}(Z-1)\frac{\left( M_{\eta }^{2}(2r-5)-3M_{\pi
}^{2}(2r+1)\right) }{(r+2)}F_{\pi }^{2}  \notag \\
&&-\frac{4}{3\sqrt{3}}(X-1)\frac{(5r+4)}{(r+2)}F_{\pi }^{2}M_{\pi }^{2} 
\notag \\
&&-\frac{1}{9\sqrt{3}}\left( 9J_{KK}^{r}(0)+32\left( 2\left(
L_{1}+L_{2}\right) +L_{3}\right) \right) M_{\eta }^{2}\left( M_{\eta
}^{2}-3M_{\pi }^{2}\right)  \notag \\
&&+\frac{1}{9\sqrt{3}}Y\frac{M_{\pi }^{2}}{\left( r^{2}+r-2\right) }  \notag
\\
&&\times \left[ 3M_{\pi }^{2}\left( 2\left( 8-5r^{2}\right)
J_{KK}^{r}(0)+((7-2r)r+4)J_{\eta \eta }^{r}(0)+(r(12r-7)-20)J_{\pi \pi
}^{r}(0)\right) \right.  \notag \\
&&\left. -M_{\eta }^{2}\left( 2(r(r+12)-16)J_{KK}^{r}(0)+\left(
-14r^{2}+r+4\right) J_{\eta \eta }^{r}(0)+(r(12r-25)+28)J_{\pi \pi
}^{r}(0)\right) \right]  \notag \\
&&+\frac{2}{9\sqrt{3}}Y^{2}\frac{M_{\pi }^{4}}{\left( r^{2}+r-2\right) }%
\left[ 6(r(3r-2)-4)J_{KK}^{r}(0)+4\left( r^{2}+r-2\right) J_{\pi \eta
}^{r}(0)\right.  \notag \\
&&\left. +(5(r-2)r-4)J_{\eta \eta }^{r}(0)+9((2-3r)r+4)J_{\pi \pi }^{r}(0) 
\right]  \notag \\
&&+\frac{M_{\pi }^{2}}{432\sqrt{3}\pi ^{2}}Y\left[ 3(19-4r)M_{\eta
}^{2}\right.  \notag \\
&&\left. +M_{\pi }^{2}\left( 9(4r+11)-2(44r+67)Y\right) \right]  \notag \\
&&-\frac{4}{9\sqrt{3}}\frac{\left( (4-7r)M_{\eta }^{2}+3M_{\pi
}^{2}(r-4)\right) }{\left( r^{2}+r-2\right) }F_{K}^{2}\delta _{F_{K}}  \notag
\\
&&+\frac{4}{9\sqrt{3}}\frac{\left( 3M_{\pi }^{2}\left( 2r^{2}-5\right)
-\left( 2r^{2}+1\right) M_{\eta }^{2}\right) }{\left( r^{2}+r-2\right) }%
F_{\pi }^{2}\delta _{F_{\pi }}  \notag \\
&&-\frac{16}{3\sqrt{3}}\frac{(r(4r+7)+4)}{(r-1)(r+1)(r+2)}%
F_{K}^{2}M_{K}^{2}\delta _{F_{K}M_{K}}  \notag \\
&&-\frac{4}{3}\frac{((r-5)r-8)}{\sqrt{3}\left( r^{2}+r-2\right) }F_{\pi
}^{2}M_{\pi }^{2}\delta _{F_{\pi }M_{\pi }}+\frac{8}{\sqrt{3}}\frac{1}{(r-1)}%
F_{\eta }^{2}M_{\eta }^{2}\delta _{F_{\eta }M_{\eta }} \\
&&  \notag \\
\frac{4}{\sqrt{3}}B_{83}^{p} &=&-\frac{1}{3}\eta (r)\frac{r(r-1)}{(r+2)}%
F_{\pi }^{2}  \notag \\
&&+\frac{2}{3}(Z-1)\frac{(r-1)}{(r+2)}F_{\pi }^{2}  \notag \\
&&+\frac{1}{6}Y\frac{M_{\pi }^{2}}{(r+2)}\left[ 2r^{2}J_{KK}^{r}(0)+2\left(
-2r^{2}+r+3\right) J_{\pi \pi }^{r}(0)\right.  \notag \\
&&\left. +(2r+1)rJ_{\eta \eta }^{r}(0)-3(r+2)J_{\pi \eta }^{r}(0)\right] 
\notag \\
&&+\frac{5}{96\pi ^{2}}Y(r-1)M_{\pi }^{2}  \notag \\
&&+\frac{2}{3}\frac{1}{(r+2)}\left( rF_{K}^{2}\delta _{F_{K}}-F_{\pi
}^{2}\delta _{F_{\pi }}\right) \\
&&  \notag \\
\frac{4}{\sqrt{3}}B_{33}^{p} &=&\frac{4}{\sqrt{3}}ZF_{\pi }^{2}+\frac{4}{%
\sqrt{3}}\eta (r)\frac{r}{(r+2)}F_{\pi }^{2}  \notag \\
&&-\frac{4}{\sqrt{3}}(Z-1)\frac{(r+4)}{(r+2)}F_{\pi }^{2}  \notag \\
&&-\frac{16}{\sqrt{3}}\left( (2L_{1}-L_{2})+L_{3}\right) s_{0}  \notag \\
&&-\frac{2}{\sqrt{3}}Y\frac{M_{\pi }^{2}r}{\left( r^{2}+r-2\right) }\left[
2r(J_{KK}^{r}(0)+J_{\eta \eta }^{r}(0)-2J_{\pi \pi }^{r}(0))\right.  \notag
\\
&&\left. +J_{\eta \eta }^{r}(0)-J_{\pi \pi }^{r}(0)\right]  \notag \\
&&+\frac{s_{0}}{16\sqrt{3}\pi ^{2}}-\frac{5M_{\pi }^{2}}{8\sqrt{3}\pi ^{2}}Y
\notag \\
&&-\frac{8}{\sqrt{3}}\frac{r}{\left( r^{2}+r-2\right) }F_{K}^{2}\delta
_{F_{K}}-\frac{4}{\sqrt{3}}\frac{\left( r^{2}+r-4\right) }{\left(
r^{2}+r-2\right) }F_{\pi }^{2}\delta _{F_{\pi }} \\
&&  \notag \\
\frac{4}{\sqrt{3}}B_{88}^{p} &=&-\frac{8}{3}\eta (r)F_{\pi }^{2}  \notag \\
&&-\frac{8}{3\sqrt{3}}(Z-1)F_{\pi }^{2}-\frac{16}{\sqrt{3}}\left(
2L_{1}-L_{2}\right) s_{0}  \notag \\
&&+\frac{4}{3\sqrt{3}}Y\frac{M_{\pi }^{2}}{(r-1)}  \notag \\
&&\times ((r+1)J_{KK}^{r}(0)+(2r+1)J_{\eta \eta }^{r}(0)-(3r+2)J_{\pi \pi
}^{r}(0))  \notag \\
&&-\frac{M_{\pi }^{2}}{12\sqrt{3}\pi ^{2}}Y(r+2)  \notag \\
&&+\frac{16}{3\sqrt{3}}\frac{1}{(r-1)}F_{K}^{2}\delta _{F_{K}}  \notag \\
&&-\frac{8}{3\sqrt{3}}\frac{(r+1)}{(r-1)}F_{\pi }^{2}\delta _{F_{\pi }}
\end{eqnarray}%
}

\section{$\pi^0$-$\eta$ mixing at $O(p^{4})$ \label{Op4_mixing}}

In this appendix, we discuss the interrelation between the
safe observable $G\left( s,t,;u\right) $ and the scattering
amplitude in the presence of the $\pi^0$-$\eta$ mixing in more detail. 
We can write the generating functional in the form
\begin{eqnarray}
Z[a] &=&\frac{1}{2\,}\int \mathrm{d}^{4}x\Big[Z_{33}(\partial \pi
_{3}[a]-a_{3}F_{0})^{2} \notag \\
&&+Z_{88}(\partial \eta
_{8}[a]-a_{8}F_{0})^{2}+2Z_{38}(\partial \pi _{3}[a]-a_{3}F_{0})\cdot
(\partial \eta _{8}[a]-a_{8}F_{0})  \notag \\
&&-\mathcal{M}_{33}\pi _{3}[a]^{2}-\mathcal{M}_{88}\eta _{8}[a]^{2}-2%
\mathcal{M}_{38}\pi _{3}[a]\eta _{8}[a]\Big]+\ldots +O(p^{6}),
\end{eqnarray}%
where we have restricted ourselves to the $\pi _{3}$-$\eta _{8}$ sector. Let us
remind that we treat the mixing parameters $Z_{38},\mathcal{M}_{38}$ at the
leading order in the $1/R$ expansion. The corresponding equations of motion are equivalent, up
to higher order corrections, to a stationarity condition for the generating
functional itself, namely 
\begin{equation}
\frac{\delta Z[a]}{\delta \phi }=0,
\end{equation}%
or explicitly 
\begin{eqnarray}
Z_{33}\partial ^{2}\pi _{3}[a]+Z_{38}\partial ^{2}\eta _{8}[a]+\mathcal{M}%
_{33}\pi _{3}[a]+\mathcal{M}_{38}\eta _{8}[a] &=&F_{0}Z_{33}\partial\cdot
a_{3}+F_{0}Z_{38}\partial\cdot a_{8}+\ldots \notag \\ \\
Z_{38}\partial ^{2}\pi _{3}[a]+Z_{88}\partial ^{2}\eta _{8}[a]+\mathcal{M}%
_{38}\pi _{3}[a]+\mathcal{M}_{88}\eta _{8}[a] &=&F_{0}Z_{38}\partial\cdot
a_{3}+F_{0}Z_{88}\partial\cdot a_{8}+\ldots \notag \\
\end{eqnarray}%
We can diagonalize the kinetic terms by means of an orthogonal transformation $%
O^{T}O=\mathbf{1}$ 
\begin{equation}
{\bold{Z}}=O^{T}\cdot\mathrm{diag}(Z_{3},Z_{8})\cdot O,
\end{equation}%
where the eigenvalues satisfy $Z_{3},Z_{8}>0$. After rescaling, we get 
\begin{equation}
{\bold{Z}}=O^{^{\prime }T}\cdot\mathbf{1}\cdot O^{^{\prime }},
\end{equation}%
with
\begin{equation}
O^{^{\prime }}=\mathrm{diag}(Z_{3}^{1/2},Z_{8}^{1/2})\cdot O.
\end{equation}%
Subsequently, we diagonalize the transformed mass terms $(O^{^{\prime
}-1})^{T}\mathcal{M}O^{^{\prime }-1}$ with another orthogonal transformation 
$O^{^{\prime \prime }T}O^{^{\prime \prime }}=\mathbf{1}$, which does not
effect the kinetic term 
\begin{equation}
(O^{^{\prime }-1})^{T}\cdot\mathcal{M}\cdot O^{^{\prime }-1}=O^{^{\prime \prime }T}\cdot%
\mathrm{diag}(M_{\pi ^{0}}^{2},M_{\eta }^{2})\cdot O^{^{\prime \prime }}.
\end{equation}%
As a result, we can write%
\begin{equation}
\mathcal{M}=G^{T}\cdot\mathrm{diag}(M_{\pi ^{0}}^{2},M_{\eta }^{2})\cdot G,
\end{equation}%
where the matrix $G$ reads%
\begin{equation}
G=O^{^{\prime \prime }}\cdot O^{^{\prime }}=O^{^{\prime \prime }}\cdot\mathrm{diag}%
(Z_{3}^{1/2},Z_{8}^{1/2})\cdot O.
\end{equation}%
Then we can rewrite the generating functional in terms of the physical
fields $\pi ^{0}$ and $\eta $, according to 
\begin{equation}
Z[a]=\frac{1}{2\,}\int \mathrm{d}^{4}x[(\partial \pi
^{0}[a]-a^{0}F_{0})^{2}+(\partial \eta \lbrack a]-a^{\eta }F_{0})^{2}-M_{\pi
^{0}}^{2}\pi ^{0}[a]^{2}-M_{\eta }^{2}\eta \lbrack a]^{2}]+\ldots +O(p^{6}),
\end{equation}%
with 
\begin{eqnarray}
\left( 
\begin{array}{l}
\pi ^{0}[a] \\ 
\eta \lbrack a]%
\end{array}%
\right) &=&G\left( 
\begin{array}{l}
\pi _{3}[a] \\ 
\eta _{8}[a]%
\end{array}%
\right) \\
\left( 
\begin{array}{l}
a^{\pi ^{0}} \\ 
a^{\eta }%
\end{array}%
\right) &=&G\left( 
\begin{array}{l}
a_{3} \\ 
a_{8}%
\end{array}%
\right).
\end{eqnarray}%
In terms of this fields, the equations of motion has become diagonal 
\begin{eqnarray}
(\partial ^{2}+M_{\pi ^{0}}^{2})\pi ^{0}[a] &=&F_{0}\partial \cdot a^{\pi
^{0}}+\ldots \\
(\partial ^{2}+M_{\eta }^{2})\eta \lbrack a] &=&F_{0}\partial \cdot a^{\eta
}+\ldots
\end{eqnarray}%
and therefore the functional derivatives of their solutions with respect to $%
a_{3}$ are 
\begin{eqnarray}
\frac{\delta \pi ^{0}[a]}{\delta a_{3}} &=&\frac{F_{0}}{\partial ^{2}+M_{\pi
^{0}}^{2}}\frac{\delta \partial \cdot a^{\pi ^{0}}}{\delta a_{3}}+\ldots
=F_{0}G^{\pi ^{0}3}\frac{\partial }{\partial ^{2}+M_{\pi ^{0}}^{2}}+\ldots \\
\frac{\delta \eta \lbrack a]}{\delta a_{3}} &=&\frac{F_{0}}{\partial
^{2}+M_{\eta }^{2}}\frac{\delta \partial \cdot a^{\eta }}{\delta a_{3}}%
+\ldots =F_{0}G^{\eta 3}\frac{\partial }{\partial ^{2}+M_{\eta }^{2}}+\ldots
,
\end{eqnarray}%
or more generally 
\begin{equation}
\frac{\delta P[a]}{\delta a_{i}}=F_{0}G^{Pi}\frac{\partial }{\partial
^{2}+M_{P}^{2}}+\ldots .
\end{equation}%
For the second functional derivatives, we therefore get 
\begin{equation}
\frac{1}{i}\frac{\delta ^{2}Z[a]}{\delta a_{i}\delta a_{j}}%
|_{a=0}=iF_{0}^{2}\sum_{P}\frac{\partial \partial }{\partial ^{2}+M_{P}^{2}}%
G^{Pi}G^{Pj}+\ldots
\end{equation}%
and thus 
\begin{equation}
\langle 0|j_{\mu 5}^{i}(0)|p,P\rangle =ip_{\mu }F_{0}G^{Pi}.
\end{equation}%
We can identify the elements of the matrix $G$ to be 
\begin{equation}
G^{Pi}=\frac{F^{Pi}}{F_{0}}.
\end{equation}%
The entries of the matrix $F$ in terms of the physical decay constants are 
\begin{equation}
{\bold{F}}=\left( 
\begin{array}{cc}
F^{\pi ^{0}3} & F^{\pi ^{0}8} \\ 
F^{\eta 3} & F^{\eta 8}%
\end{array}%
\right) =\left( 
\begin{array}{cc}
F_{\pi } & \varepsilon _{\pi }F_{\pi } \\ 
-\varepsilon _{\eta }F_{\eta } & F_{\eta }%
\end{array}%
\right),
\end{equation}%
where $\varepsilon _{i}=O(1/R)$ are the mixing angles. The inverse matrix to
the first order in the isospin breaking then takes the form 
\begin{equation}
{\bold{F}}^{-1}=\left( 
\begin{array}{ll}
F_{\pi }^{-1} & -\varepsilon _{\pi }F_{\eta }^{-1} \\ 
\varepsilon _{\eta }F_{\pi }^{-1} & F_{\eta }^{-1}%
\end{array}%
\right) +O\left( \frac{1}{R^{2}}\right) .
\end{equation}%
The LSZ formulae give, for $p^{2}\rightarrow M_{P}^{2}$ 
\begin{eqnarray}
\langle 0|\widetilde{j_{\mu 5}^{i}}(p)\ldots |0\rangle &=&\frac{i}{%
p^{2}-M_{P}^{2}}\langle 0|j_{\mu 5}^{i}(0)|p,P\rangle \langle p,P|\cdots
|0\rangle +reg.  \notag \\
&=&\frac{i}{p^{2}-M_{P}^{2}}iF_{0}G^{Pi}p_{\mu }\langle p,P|\cdots |0\rangle
+reg.
\end{eqnarray}%
Writing the generating functional in the physical basis 
\begin{equation}
Z[a]=\sum_{n}\int d^{4}x_{1}\ldots d^{4}x_{n}P_{1}[a](x_{1})\ldots
P_{n}(x_{n})V_{n}(x_{1},\ldots ,x_{n})[a],
\end{equation}%
we symbolically get (tilde denotes a Fourier transform here) 
\begin{eqnarray}
\langle 0|\widetilde{j_{\mu 5}^{i}}(p)\ldots |0\rangle &=&\frac{i}{%
p^{2}-M_{P}^{2}}p_{\mu }F_{0}G^{Pi}\widetilde{V}_{n}+reg. \\
&=&\frac{i}{p^{2}-M_{P}^{2}}p_{\mu }F^{Pi}\widetilde{V}_{n}+reg.
\end{eqnarray}%
This means that in order to extract the physical amplitudes from the generating
functional, we can diagonalize the $O(p^{4})$ kinetic and mass terms and then
use the generating functional as a non-local Lagrangian. The
diagonalization is achieved by the substitution 
\begin{eqnarray}
\pi _{3} &=&\frac{F_{0}}{F_{\pi }}\pi ^{0}-\varepsilon _{\pi }\frac{F_{0}}{%
F_{\eta }}\eta \\
\eta _{8} &=&\frac{F_{0}}{F_{\eta }}\eta +\varepsilon _{\eta }\frac{F_{0}}{%
F_{\pi }}\pi ^{0}
\end{eqnarray}%
in the generating functional $Z[a]$. 

Alternatively, one can work in the $\pi _{3},\eta _{8}$ basis. The following relation between the safe observable $G$, in terms of the original fields, and the physical amplitude is then obtained
\begin{equation}
G^{i_{1}i_{2}i_{3}i_{4}}(s,t;u)=\sum_{P_{1}P_{2}P_{3}P_{4}}F^{i_{i}P_{i}}%
\ldots F^{i_{4}P_{4}}A_{P_{1}P_{2}P_{3}P_{4}}(s,t;u).
\end{equation}%

Solving this relation algebraically with respect to the amplitude $A_{P_{1}P_{2}P_{3}P_{4}}(s,t;u)$ is equivalent to using the diagonalization procedure in the first approach.

\newpage

\bibliographystyle{utphys}
\bibliography{Bibliography}
{}

\end{document}